# Extracting Signal out of Chaos: Advancements on MAGI for Bayesian Analysis of Dynamical Systems

A THESIS PRESENTED
BY
SKYLER WU
TO
THE DEPARTMENTS OF STATISTICS AND MATHEMATICS

IN PARTIAL FULFILLMENT OF THE REQUIREMENTS
FOR THE DEGREE OF
BACHELOR OF ARTS (HONORS)
IN THE SUBJECT OF
STATISTICS AND MATHEMATICS

HARVARD UNIVERSITY
CAMBRIDGE, MASSACHUSETTS
MAY 2024

# Extracting Signal out of Chaos: Advancements on MAGI for Bayesian Analysis of Dynamical Systems

## Abstract

Dynamical systems, and the systems of ordinary differential equations (ODEs) that govern them, are ubiquitous in many scientific disciplines for modeling natural phenomena, from the Fitzhugh-Nagumo model for neuron action potentials [14], to the Lotka-Volterra model for predator-prey dynamics [30], and to the SIR model for infectious disease outbreaks [20], among countless other use cases.

Given noisy and sparse data from an ODE-based dynamical system, there are many questions that scientists may hope to ask, including the following. First, what are the original parameters governing the system? Second, can we reconstruct what the underlying ground-truth, unnoised trajectory would have looked like? Third, can we infer, hopefully with uncertainty quantification, whether our system is chaotic or stable, based on just a few data points? Fourth, can we predict the future trajectory of this system, even in chaotic regimes?

This senior thesis builds off of Yang et al.'s 2021 paper [55] introducing the manifold-constrained Gaussian process inference (MAGI) method for Bayesian parameter inference and trajectory reconstruction of ODE-based dynamical systems. We are interested in answering the above questions under sparse and noisy data conditions, using four regimes from the Lorenz system as test beds. Specifically, we provide the following novel contributions. First, we introduce Pilot MAGI (pMAGI), a novel methodological upgrade on the base MAGI method that confers significantly-improved numerical stability, parameter inference, and trajectory reconstruction. Second, we demonstrate, for the first time to our knowledge, how one can combine MAGI-based methods with dynamical systems theory to provide probabilistic classifications of whether a system is stable or chaotic. Third, we demonstrate how pMAGI performs favorably in many settings against much more computationally-expensive and overparameterized methods such as Particle Swarm Evolution, Differential Evolution, and Physics-Informed Neural Networks (PINNs) on the task of parameter inference and trajectory reconstruction. Fourth, we introduce Pilot MAGI Sequential Prediction (PMSP), a novel method building upon pMAGI that allows one to predict the trajectory of ODE-based dynamical systems multiple time steps into the future, given only sparse and noisy observations. We show that PMSP can output accurate future predictions even on chaotic dynamical systems and significantly outperform PINN-based methods.

Overall, we contribute to the literature two novel methods, pMAGI and PMSP, that serve as Bayesian, uncertainty-quantified competitors to the Physics-Informed Neural Network. We hope that researchers in any discipline that uses ODEs for modeling phenomena can benefit from our methodology.

# Contents

# Listing of figures

For my loving mom and dad: Zheng and Sumin.

# Acknowledgments

FIRST, I would like to express my deepest gratitude to Professor Samuel Kou for taking a chance on me as his student, and for his endless support, patience, mentoring, guidance, and reassurance over the past two years. The research process was certainly nonlinear and I have made more mistakes and oversights than I can even recall. And yet, each time, Professor Kou has always addressed these shortcomings with his patience, gently guiding me back on track and teaching me how to prevent such errors in the future. The experience of working under Professor Kou was also one of my primary reasons for applying to graduate school in statistics — the excitement, anticipation, and learning I experienced in each of our meetings will continue to inspire me for decades to come. In the same vein, I would like to express my deepest gratitude to Professor Shihao Yang, our close collaborator and a trusted mentor. I do not think I would have found the confidence nor developed the intuition and technical ability to stay the course on this project had it not been for Professor Yang's generous donation of his time to mentor me over the past two years. Professor Kou and Professor Yang, thank you.

I WOULD ALSO LIKE TO THANK Professor Jun Liu for serving as my second reader and for introducing me to the world of Monte Carlo. I also thank Dr. Alex Young for coordinating the entire STAT 99R operation and community, and for providing some much-needed calm and reassurance during the thesis-writing process.

MANY OF THE EXPERIMENTS performed for this senior thesis were run on the FASRC cluster supported by the FAS Division of Science Research Computing Group at Harvard University. To the FASRC staff, thank you.

NEXT, I must thank Professor Joseph Blitzstein. From inspiring me to pursue statistics, to always having his office door open when I start overthinking and/or panicking, to being one of the first "Ivy League" professors I dared reach out to as a frightened first year in the middle of a pandemic due to his friendliness and approachability, to generously advising me and always checking in on me on basically all things statistics-related (and many not), and to being a role model of the type of teacher I wish to become one day, Professor Blitzstein, thank you.

I WOULD ALSO LIKE TO THANK Dr. Edward Raff and Fred Lu for their encouragement over the years as my bosses at Booz Allen Hamilton. Thank you for giving this wide-eyed undergraduate a chance on your research team, for taking me under your wings, for putting up with my countless questions and bugs and glitches, for teaching me how to ask questions, recalibrate, and circle the wagons, and for your support on ... all the things in all the times. Our work together was one of the main impetuses for my desire to pursue a career in research.

To my first research bosses in college, Professor Lucas M. Stolerman and Professor Mauricio Santillana, thank you for taking a chance on me, for your constant encouragement and guidance, and for patiently shepherding me through our first manuscript together, as we endeavored to robustly and accurately forecast Dengue fever in hundreds of locations around the world.

To Professor Himabindu Lakkaraju and Professor Jiaqi Ma, thank you for your patience and guidance on our first project together in the land of LLMs and generative AI, and for inspiring me to think not only about increasing performance, but also about asking some crucial questions — how do we know that a model is working the way that it should be? How can we probe what the model actually is doing under-the-hood?

To Professor Finale Doshi-Velez, thank you for introducing me to my lifelong passion of machine learning and for taking a chance on me and allowing me to join your course staff. To Dr. Weiwei Pan, from mentoring me on how to become a better teacher, to guiding me on how to become a more effective writer and communicator, thank you. Serving on the COMPSCI 181 course staff was one of the most powerful experiences that furthered my dream of hopefully one day becoming a professor.

I would also like to thank Yash, Ginnie, and Jason as some of my closest upperclassman mentors. Had it not been for Yash's encouragement and mentoring, I would have likely dropped out of COMPSCI 181 and stayed far away from the realms of statistics and machine learning. Had it not been for Ginnie's gentle reassurances, I do not think I would have come out of senior fall in one piece. Had it not been for my long-time battle buddy in the world of academics, Jason, I think life would not have been nearly as great.

To my (extended) suitemates Kevin, Mincheol, Nithin, Oliver, Ryan, Sunera, Wittmann, and Yiting, thank you — it's been a great past three years. I will forever treasure both the most serious of our conversations and the … not-so-most-serious of our conversations and will always be grateful for your support and partnership on the math and statistics fronts. It's been a good run, gentlemen. To my long-time battle buddies Alex Rojas and Will Nickols, and to my long-time co-TFs Cat and Charu, working together with you all was incredibly formative towards shaping me as a learner, a doer, a teacher, and a friend — both inside and outside of the classroom.

To my PhD student mentors Adela, Alan, Anvit, Benedikt, Benjamin, Biyonka, Buyu, Hans, Kyla, Matthew, Nathan, Nic, Nowell, Philip, Ritwik, Sarah, Souhardya, Wenqi, Xiaodong, Yuanchuan, Yuzhou, and Zeyang, thank you for all of your mentoring, jokes, and encouragement over the last two years and for graciously allowing me to work within this wonderful community on the 6th floor. To more (winner winner) chicken dinners!

Last, but certainly not least, I would like to thank my mom and dad, Zheng and Sumin. Thank you for your sacrifice, for starting a new life in a foreign country, for keeping our family as one unwavering unit with a love that transcends geographical boundaries, and for your unconditional support my entire life — all, so that I could have a better life and education. Words can not hope to do justice to what I wish to convey, so I will just end with the following six words: thank you and I love you.

# 1
# Introduction

## 1.1 Motivation

Dynamical systems, and the systems of ordinary differential equations (ODEs) that govern them, have become ubiquitous in many scientific disciplines for modeling natural phenomena, from the Fitzhugh-Nagumo model for neuron action potentials [14], to the Lotka-Volterra model for predator-prey dynamics [30], and to the SIR model for infectious disease outbreaks [20], among countless other use cases. For the purposes of this thesis, we will restrict the term *dynamical system* to refer to a collection of variables (e.g., particles) whose values throughout time are governed by a system of ODEs. Suppose $\mathbf{x}(t) = (x_1(t), x_2(t), \ldots, x_n(t)) \in \mathbb{R}^n$ are variables whose behaviors over time are described by an ODE with parameters $\boldsymbol{\theta}$. Then, $\mathbf{f}(\mathbf{x}(t), t; \boldsymbol{\theta})$, the vector containing the derivatives of each variable over time can be written as follows, at any time point $t$:

$$\mathbf{f}(\mathbf{x}(t), t; \boldsymbol{\theta}) = \dot{\mathbf{x}}(t) = \frac{d\mathbf{x}(t)}{dt} = \left(\frac{dx_1(t)}{dt}, \frac{dx_2(t)}{dt}, \ldots, \frac{dx_n(t)}{dt}\right).$$

In theory, with infinite numerical precision, given initial conditions $\mathbf{x}(0)$ and parameter values $\boldsymbol{\theta}$, one can integrate over an interval $[0, T]$ to obtain the values of $\mathbf{x}(t)$ at any time point $t \in [0, T]$. In practice, such numerical integration is performed using an algorithm such as Runge-Kutta[23] or LSODA[37], as implemented in a scientific computing library like `SciPy`[50].

With an ODE system, there are many learning tasks that one may wish to perform:

1. *Parameter Inference*: given observations at a set of time points of an ODE-based dynamical system, **can we recover the original parameters $\boldsymbol{\theta}$ that govern the ODE equations?** For example, parameter inference in the SIR model[20] allows one to compute the *basic reproduction number* that governs whether or not and how fast an infectious disease outbreak will spread[4]. This information could be very helpful, if not life-saving if provided to public health officials.

   A natural question to ask is — what happens if the observations we are provided are noisy, for example, due to faulty sensors or imperfect public health surveillance? How would our parameter inference performance change? What if we only receive noisy observations at very few and infrequent time steps – how would our performance change under *sparse* and *noisy* conditions?

2. *Trajectory Reconstruction*: given sparse and noisy observations from an ODE system, can we **reconstruct what the underlying ground-truth, unnoised trajectory would have looked like?** Put another way, can we cut through the noise, extract the true signal, and infer within our interval of observation?

3. *Probabilistic Stability Classification:* some dynamical systems will eventually converge to a fixed point. Others will continue to oscillate erratically for perpetuity. Given sparse and noisy observations from an ODE system, can we probabilistically infer whether the system will eventually converge to a fixed point or not?

4. *(Future) Prediction:* given sparse and noisy observations from an ODE system, can we predict how the system's components will evolve in the future? For example, under the SIR model, how many cases of COVID-19 will we see in a year's time?

Throughout this thesis, we will refer to the above enumerated tasks as the *four learning tasks*.

## 1.2 Objectives and Contributions

One method that shows promise for performing at least the first two tasks above is the manifold-constrained Gaussian process inference method, or MAGI, introduced by Yang et al.[55] At a very high level, MAGI is a Bayesian method that a) models sparse, noisy observations from an ODE system as components of a Gaussian Process (GP); b) restricts the Gaussian Process to lie on a manifold that respects the governing equations of the ODE system; and c) facilitates parameter inference and component sampling with principled uncertainty quantification via Hamiltonian Monte Carlo[35]. We elaborate on each aspect of the MAGI method in Chapter 3.

Building on Yang et al.'s MAGI, the objectives and novel contributions of this senior thesis are fourfold. First we introduce *Pilot MAGI* (pMAGI), a novel numerically-stabler and higher-performance improvement on the base MAGI method. Second, we quantify pMAGI's parameter inference and trajectory reconstruction abilities on four ODE test beds governed by the famous Lorenz system. Third, we demonstrate how and to what extent pMAGI can be novelly adapted towards probabilistic stability classification of Lorenz system trajectories. Fourth, we introduce *Pilot MAGI Sequential Prediction* (PMSP), a novel method for probabilistic prediction of ODE dynamical systems multiple steps into the future, with natural uncertainty quantification.

While working through each of these objectives, we will compare pMAGI and PMSP's performances against three competitor methods as appropriate: Particle Swarm Evolution (PSO), Differential Evolution (DE), and Physics-Informed Neural Networks (PINNs). We invite the interested reader to explore Chapter 2 for details on these competitor methods. As alluded to above, we will use four systems' trajectories governed by the famous Lorenz equations as our evaluation test beds. As we will soon make clear below, the Lorenz system was selected as our evaluation benchmark due to its uniquely exciting, if not pathological, dynamics.

## 1.3 Reproducibility and Data Availability

The source code for reproducing the main experiments contained in this thesis can be found at the following GitHub link: https://github.com/skbwu/pMAGI-and-PMSP/tree/main. Each folder in this repository corresponds to one major experiment and/or model-class involved in this thesis. The contents of this repository reproduce the raw model outputs (e.g., Hamiltonian Monte Carlo samples, mean trajectories, point estimates of parameters, etc.) of all models discussed in this thesis. This repository also contains the datasets used to benchmark various competitor methods against pMAGI and PMSP.

## 1.4 Chaotic Dynamics and the Lorenz System

One of the most famous chaotic ordinary differential equations of all time is the Lorenz system, first introduced by Lorenz et al. in 1963 for modeling atmospheric convection and weather processes. Since its first formulation, the Lorenz system has found many applications in many fields, including biology, circuit theory, mechanics, and metereology [19,45]. The Lorenz system is famous for being a deterministic system, but having the potential to output "extremely erratic dynamics" [46].

Mathematically, the Lorenz system has three components $\mathbf{x} = (x, y, z)$, all of which are functions of time $t$ alone, governed by the following non-linear system of ordinary differential equations (ODEs), parameterized by three presumably-positive parameters $\beta, \rho,$ and $\sigma$[29]:

$$\frac{dx}{dt} = \sigma(y - x), \quad \frac{dy}{dt} = x(\rho - z) - y, \quad \frac{dz}{dt} = xy - \beta z.$$

Formal analysis[45] of the governing ODEs reveals that, for all parameter values, the origin is a stationary point; moreover, the origin is a stable global attractor if $0 < \rho < 1$. For $\rho > 1$, the system possesses two additional stationary points at[45]

$$(x, y, z) = (\pm\sqrt{\beta(\rho - 1)}, \pm\sqrt{\beta(\rho - 1)}, \rho - 1).$$

Examining the eigenvalues of the dynamical system, one finds that if

$$1 < \rho < \sigma\frac{\sigma + \beta + 3}{\sigma - \beta - 1} = \rho_H,$$

then the two additional stationary points are stable[45]. If $\rho$ is greater than the threshold $\rho_H$ (Lorenz refers to $\rho_H$ as the "critical value of $\rho$ for the instability of steady convection"[29]), then these two additional stationary points will lose their stability, and the trajectory of the system may (but not always) exhibit chaotic behavior in various forms[45]. Within the context of this thesis, we will refer to a Lorenz system trajectory generated by a certain parameterization of $(\beta, \rho, \sigma)$ as "stable" if it converges to a stationary point. While there is no universally-accepted definition of chaos[46], for this paper, we will borrow Strogatz's definition of chaos as "aperiodic long-term behavior in a deterministic system that exhibits sensitive dependence on initial conditions"[46].

Visual identification of whether a Lorenz system will converge to a stable fixed point or demonstrate chaotic behavior in perpetuity is extremely difficult. Such difficulty is exacerbated by the phenomena of certain Lorenz system trajectories exhibiting "preturbulence" (also known as "transient chaos"[46]) — when the trajectories behave like a strange attractor (i.e., butterfly) for a long period of time before eventually converging to a stable fixed point[45]. Other Lorenz trajectories will exhibit "intermittent chaos," where the system alternates between chaotic and stable-looking behavior[45]. Because of the Lorenz system's potential chaotic nature, it is very difficult to nu-merically predict the future state of a chaotic system for long periods of time, due to the extreme sensitivity to initial conditions, and errors introduced through floating point operations and rounding[46].

In summary, the Lorenz system is an excellent, descriptive benchmark for our ODE learning tasks due to its numerical pathologies, chaotic potential, and unpredictable behavior.

## 1.5 Testbeds

We will evaluate the performance of pMAGI and MAGI, alongside their appropriate competitor methods, on four test beds that are all governed by the Lorenz system but differing in their governing parameter values of $\theta = (\beta, \rho, \sigma)$ and their initial conditions of $\mathbf{x}(0) = (x(0), y(0), z(0))$. From established sources[19], we know that the governing parameter values of $\beta = \frac{8}{3}, \rho = 28, \sigma = 10$ yield chaotic regimes. We also know via the analyses presented in Sparrow[45], that if we hold $\beta = \frac{8}{3}$ and $\sigma = 10$, then the parameter settings of $\rho = 6$ and $\rho = 23$ will correspond to stable regimes.

We present the four test beds alongside their governing parameter values, initial conditions, and names that we will use to reference them throughout this thesis in the figure below. To generate the ground-truth for these test beds, we numerically-integrate from $t = 0$ to $t = 10$ using the solve_ivp module from SciPy[50], working under the assumption that integration on such a short interval should not accrue meaningful numerical imprecision.

The *Stable (Canonical)* regime is named as such because it converges to a fixed point ("stable") and because such convergence is very patently clear even at $t = 10$ ("canonical"). While the *Stable (Transient Chaos)* regime also eventually converges (see next figure) to a fixed point ("stable"), from $t = 0$ to $t = 10$, it looks visually indistinguishable from the chaotic regimes, and hence the "Transient Chaos" descriptor. The *Chaotic (Butterfly)* regime is named as such because it is mathematically chaotic as referenced above, and because even from $t = 0$ to $t = 10$, one can already see the dual-wing butterfly attractor pattern. Finally, the *Chaotic (No Butterfly)* regime is named as such because while its parameters also dictate chaos, its initial conditions cause it to not show the two wings of the butterfly from $t = 0$ to $t = 10$, instead opting for a spiral pattern on this interval.

One particularly salient question is — if given only sparse, short, and visually-similar looking intervals of observation of the Stable (Transient Chaos) and Chaotic (Butterfly) regimes, can models predict which system will be chaotic and which will be stable? We will explore this question in the following chapters.

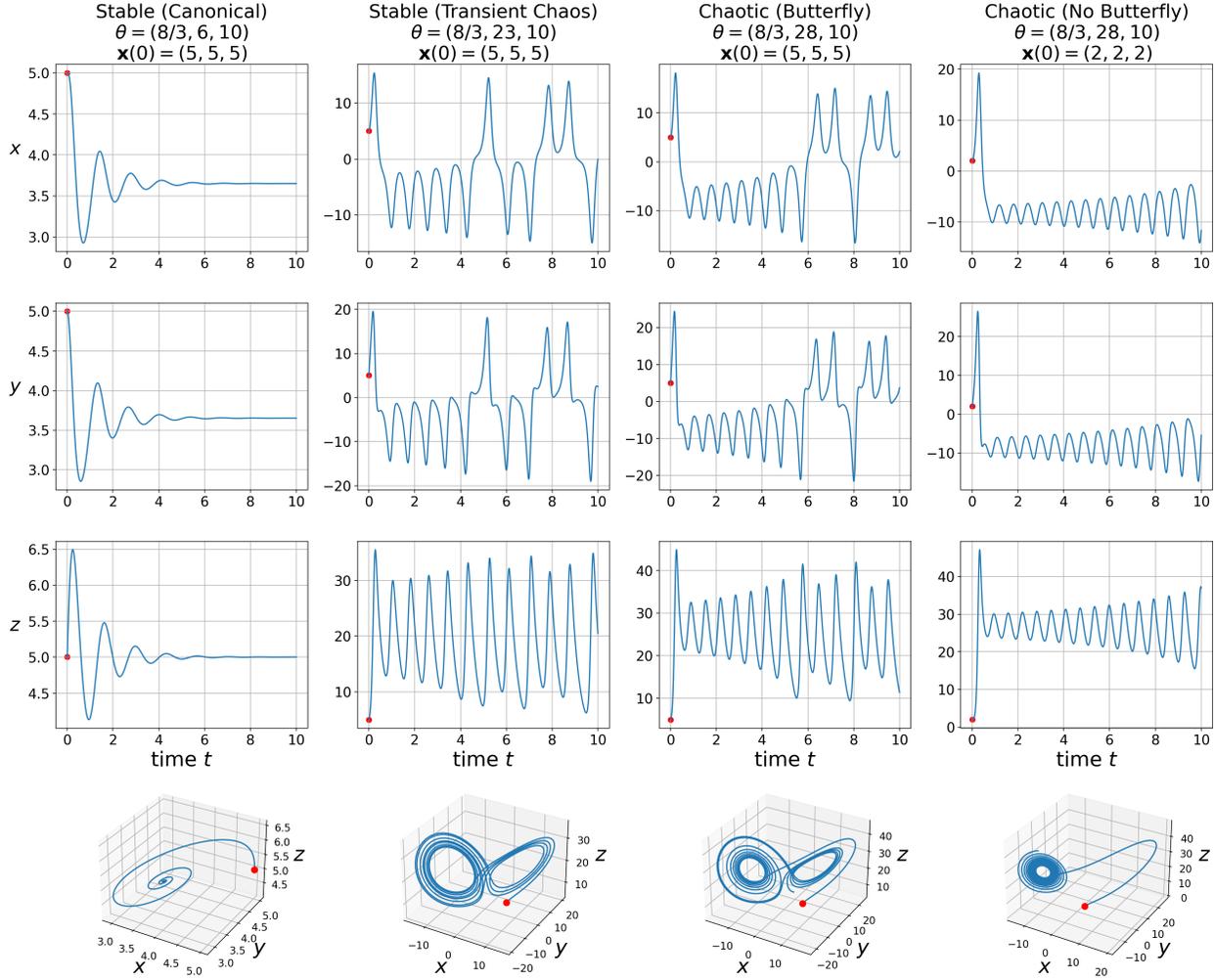

**Figure 1.1:** The four testbed regimes that we will use to evaluate pMAGI, MAGI, and their competitor methods on the four learning tasks. Each column corresponds to a regime, while the first three rows show the $x$, $y$, and $z$ components of each regime from $t = 0$ to $t = 10$. The fourth row shows the 3D trajectory of the system. For all subplots, the red dot marks the initial conditions of the system.

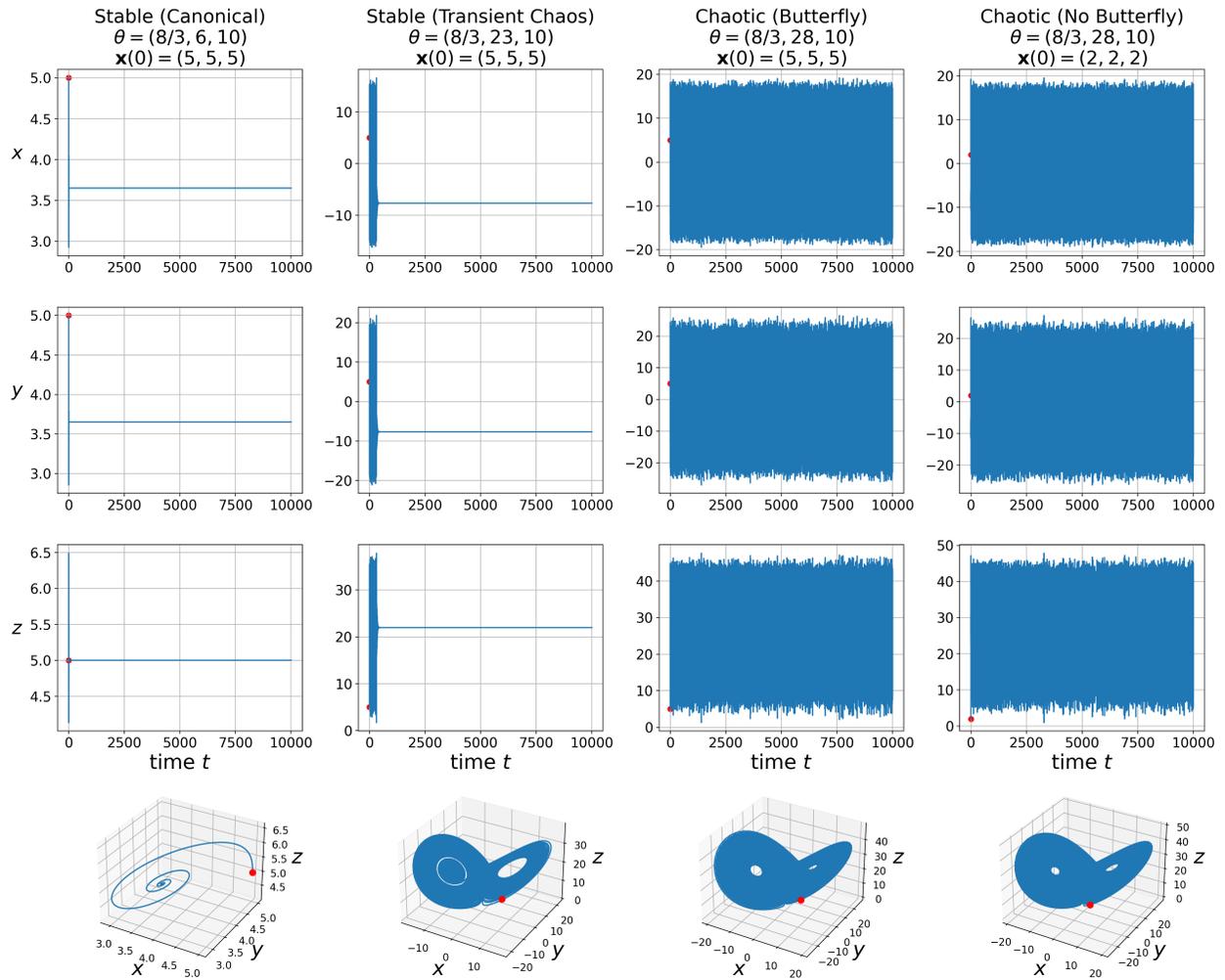

**Figure 1.2:** The exact same four testbed regimes from the previous figure, except integrated from $t = 0$ to $t = 10000$, with the same organization and axes labeling as earlier. Here, it is very clear that the two stable regimes — Stable (Canonical) and Stable (Transient Chaos) — eventually converge to a fixed point, while the two chaotic regimes — Chaotic (Butterfly) and Chaotic (No Butterfly) — continue to oscillate unpredictably at least on this very long time interval.

Because we are interested in model performance under noisy and sparse observation settings, after performing numerical integration to obtain our ground truth, we incorporate the following data operations:

1. Interval truncation: we truncate our data to smaller intervals of $t \in [0, T_{max}]$, for $T_{max} \in [2, 4, 6, 8, 10]$. An active research question is — how much data and/or signal is necessary for successful parameter infer-ence, stability classification, and prediction?

2. Injecting noise: for each component $x, y, z$, we will add Gaussian noise to all of our integrated trajectory points' components with mean 0 and standard deviation governed by a noise level parameter $\alpha$. For example, for the $x$ component, we will set the noise standard deviation $\sigma_x$ as follows:

$$\sigma_x = \alpha \cdot \left( \max_{t \in [0,10]} x(t) - \min_{t \in [0,10]} x(t) \right),$$

repeating this process analogously for our $y, z$ components. This formulation allows us to have equal *relative* noise levels on all three components. For this thesis, we will try $\alpha$ levels of $1.5 \cdot 10^{-1}, 1.5 \cdot 10^{-2}, 1.5 \cdot 10^{-3}, 1.5 \cdot 10^{-4}$, and $1.5 \cdot 10^{-5}$ to test model performance across a wide range of noise intensities.

3. Density of observations: we will sparsify our integrated and noised trajectories to generate datasets with either $d_{obs} = 5, 10, 20$, or $40$ dense observations per unit time.

4. Random seeds: for reproducibility and to quantify model variance, we will generate $n = 10$ datasets per combination of the above settings, for each of our four testbed regimes. We save all of these datasets as .csv files to have a common set of datasets for all models on all learning tasks.

With our datasets generated, we proceed in the next chapter to describe some existing methods in the literature for parameter inference, trajectory reconstruction, and future prediction of ODE-based dynamical systems.

# 2
# Related Work

The purpose of this chapter is to present an overview of existing methods for parameter inference and/or trajectory reconstruction of ODE-based dynamical systems, along with their strengths and weaknesses. We begin with a discussion of parameter-inference-only methods (Particle Swarm Optimization and Differential Evolution), before moving on to neural network architectures that are capable of parameter inference, trajectory reconstruction, and prediction. Finally, we close with some less easily-categorizable methods.

## 2.1 Particle Swarm Optimization

Particle Swarm Optimization (PSO) methods, also known as multiple-learner methods, are, by design, only capable of parameter inference. In general, these methods formulate the ODE parameter inference problem in terms of finding the best set of parameters θ that minimizes the error between the predicted (via numerical integration) and observed values of the systems' states [2,6,18,47]. For example, the idea behind Bavafa et al.'s system, specifically, called Teaching Learning-Based Optimization (TLBO), is to a) initialize a set of "learners" (each representing one possible set of parameters); b) select the "learner" that achieved the best mean squared error (MSE) as the "teacher"; c) update the remaining learners based on the new "teacher" and each individual learner's performance and condition; and d) run the next iteration of the algorithm, repeating until convergence [6]. A similar algorithm approach is the Artificial Honey Bee Colony Algorithm, which mimics the swarm behavior of a colony of bees towards finding an optimal set of parameters in trying to balance exploration and exploitation of the large parameter space [26]. Others have also combined such Particle Swarm methods together with other optimization methods for increased performance [25,56]. Bavafa et al. demonstrate that their TLBO algorithm is able to precisely recover the governing parameters of a chaotic Lorenz system in both offline (parameters fixed) and online (parameters can vary over time) settings [6]. However, it is unclear from Bavafa et al.'s work on how many observations the system must encounter for robust parameter inference [6]. One major disadvantage of PSO methods is that they are incapable of providing uncertainty quantification of their outputted estimates. Another major disadvantage of PSO methods is that they require extensive numerical integration to produce the predicted trajectories at each iteration, which is extremely computationally expensive and prone to finite precision errors. Another potential disadvantage is that the papers discussed above do not contain any discussion on how their methods perform when provided *noisy* data from a system. In the following chapters, we will evaluate the PSO method as a benchmark under the same noisy data conditions as our pMAGI methodologies.

## 2.2 Differential Evolution Algorithms

Similar to PSO methods, differential evolution algorithms (DE) are also, by design, only capable of parameter inference. The main idea behind DE algorithms towards ODE parameter inference is to begin with a large set of candidate solutions (sets of parameters), and, at each stage, combine parent candidate solutions together (potentially with some modifications, called mutations) to form new child candidate solutions, with a child candidate solution only replacing its parent solution if it is more optimal, as measured through an objective function such as MSE on inferred versus observed trajectories [26]. This approach was also shown by Li et al. to succeed in recovering the parameters governing the chaotic Lorenz system [26]. Variants of evolutionary algorithms, such as the cuckoo search algorithm, have also shown success in recovering chaotic Lorenz parameters from noiseless observations [27]. However, such evolutionary algorithms are vulnerable to being trapped inside local optima and/or converge too slowly for practical use [26]. Given their non-probabilistic, optimization-based nature, like PSO, DE algorithms are also incapable of providing any form of uncertainty quantification on their outputted estimates. Like PSO, DE algorithms also require expensive numerical integration at each iteration – a major disadvantage. Again, the papers referenced above do not contain any discussion regarding the performance of their methods under noisy data settings. In the following chapters, we will also include the DE method as a benchmark against our pMAGI methods.

## 2.3 Deep Learning and Neural Networks

Neural network architectures have also demonstrated excellent performance towards modeling dynamical systems, including on both the forward problem of generating solutions for ODEs and PDEs [7,39,40], and the reverse problem of learning the governing equations [5,39,40,38]. Indeed, neural network architectures such as LSTMs and autoencoders can learn to interpolate within the time ranges of observed data very accurately [1,5]. Some neural network architectures can even accurately forecast the trajectories of the chaotic Lorenz system for some steps into the future, albeit under conditions with no external noise injected [12,13,17,44]. How these architectures per-form when trained on noisy data is unclear from the referenced papers.

One specific class of neural networks for modeling dynamical systems are physics-informed neural networks, or PINNs, which generally are formulated with loss functions that incorporate expert knowledge about the characteristics of a dynamical system as soft constraints[21,24,39].

For concreteness, suppose we have an ODE dynamical system containing three components $(x(t), y(t), z(t))$ with derivatives $\frac{dx}{dt} = f_X(x, y, z, t; \theta)$, $\frac{dy}{dt} = f_Y(x, y, z, t; \theta)$, and $\frac{dz}{dt} = f_Z(x, y, z, t; \theta)$, where $\theta$ represents the parameters governing the ODE equations $f_X, f_Y, f_Z$. To keep notation concise, let $\mathbf{f} \in \mathbb{R}^3$ be defined as

$$\mathbf{f}(x, y, z, t; \theta) = (f_X(x, y, z, t; \theta), f_Y(x, y, z, t; \theta), f_Z(x, y, z, t; \theta)).$$

Let $\mathbf{W}$ represent the feed-forward weights of our PINN (i.e., the weights used in the forward-pass of the network), $\hat{\theta}$ represent the PINN's current estimates of $\theta$, and $\mathrm{NN}(t; \mathbf{W}) \in \mathbb{R}^3 = (\hat{x}(t), \hat{y}(t), \hat{z}(t))$ represent the PINN's prediction of our system's trajectory at time $t$ (i.e., the output of the forward pass). Note that $\hat{\theta}$ itself, which is indeed an autograd-enabled parameter attribute of the PINN module, is not involved in the forward-pass — it is only estimated via enforcement of the soft physics-based constraint that we will soon discuss below. Suppose we have (potentially noisy) observations $\{(x_{obs}(t), y_{obs}(t), z_{obs}(t))\}_{t \in \mathcal{T}}$, where the set $\mathcal{T}$ contains our observation time points. Using backpropagation and automatic differentiation, it is possible to compute, for any time point $t$,

$$\mathbf{f}_{\mathrm{NN}}(t; \mathbf{W}) = \frac{d\mathrm{NN}(t; \mathbf{W})}{dt} \in \mathbb{R}^3.$$

Under this concrete example, the loss function $L(\mathbf{W}, \hat{\theta})$ of our PINN would look as follows:

$$L(\mathbf{W}, \hat{\theta}) = \frac{\lambda}{|\mathcal{T}|} \sum_{t \in \mathcal{T}} ||\mathrm{NN}(t; \mathbf{W}) - (x_{obs}(t), y_{obs}(t), z_{obs}(t))||_2^2 +$$

$$\frac{1}{|\mathcal{T}|} \sum_{t \in \mathcal{T}} ||\mathbf{f}_{\mathrm{NN}}(t; \mathbf{W}) - \mathbf{f}(\hat{x}(t), \hat{y}(t), \hat{z}(t), t; \hat{\theta})||_2^2.$$

The first additive term in this PINN loss function above is called the *reconstruction loss*, which captures the MSE between the predicted and observed values of the systems' states, like a standard regression neural network. The second additive term, called the *physics-based loss*, captures how closely the derivatives obtained via automatic differentiation of the neural networks' functional representation of the system's trajectory matches the true governing equation derivatives. The $\lambda$ hyperparameter governs how much the PINN training regime prioritizes reconstruction error versus physics-based error. Because it is effectively impossible to drive both loss components to zero, PINNs are said to incorporate domain knowledge (i.e., the physics-based term) only as soft constraints. As a consequence, the learned neural network function may not sufficiently-faithfully obey the governing laws of the system of interest, and may produce undesirable and unrealistic outputs[21]. Nonetheless, PINNs have demonstrated potential to perform both trajectory reconstruction and prediction (forward problem), and parameter inference (inverse problem) of the underlying dynamical systems[31,39,10,33,34,39,40].

In practice, to enforce our physics-based structure more forcefully, one might evaluate the physics-based loss at a finer-grained set of time points $\mathcal{T}_{physics}$, which could look something like inserting 2-3 time points between each observed time point, or randomly sampling, at each epoch, time points within our observation interval to compute the physics-loss. For the purposes of this thesis, to present our competitor methods in the best light possible, unless specified otherwise, we will evaluate the physics-based loss at a uniform grid of $d = 40$ time points per unit time (i.e., at 400 evenly-spaced points if we observe data on an interval of length-10). With this setup, the combined loss function of our PINN would look as follows:

$$L(\mathbf{W}, \hat{\boldsymbol{\theta}}) = \frac{\lambda}{|\mathcal{T}|} \sum_{t \in \mathcal{T}} ||\text{NN}(t; \mathbf{W}) - (x_{obs}(t), y_{obs}(t), z_{obs}(t))||_2^2 +$$

$$\frac{1}{|\mathcal{T}_{physics}|} \sum_{t \in \mathcal{T}_{physics}} ||\mathbf{f}_{\text{NN}}(t; \mathbf{W}) - \mathbf{f}(\hat{x}(t), \hat{y}(t), \hat{z}(t), t; \hat{\boldsymbol{\theta}})||_2^2.$$

Nonetheless, even with such modifications, in general, many neural networks (including PINNs) are vulnerable to the following disadvantages: a) their learning abilities are oftentimes limited to very simple problems with small parameter values; b) their loss functions, by virtue of being neural networks, are topographically complex with many, many local minima; and c) they are inherently uninterpretable (without extensive modification) due to the complexity of the function encoded by the black-box neural network[21]. For PINNs specifically, as we will detail in the later chapters, their performance is heavily dependent on the $\lambda$ hyperparameter and quite volatile — another major disadvantage. In contrast, pMAGI is robust to a wide range of specified hyperparameter values. Another disadvantage of the black-box architecture of neural networks is that, in general, it is effectively infeasible to do any form of uncertainty quantification on the learned dynamics and outputed predictions. Finally, neural network architectures generally are very computationally-expensive to train per epoch and require many epochs of training to learn useful function approximations.

In full transparency, related work by Yang et al. proposes Bayesian PINNs which can solve both the forward and inverse problems of PDE dynamical systems, while using Hamiltonian Monte Carlo and/or variational inference methods to construct a posterior and perform uncertainty quantification, under the assumption that observations of the system are distributed as Gaussians centered around the true values[54]. However, the Bayesian PINN methodology requires that the true noise levels be specified as input into the neural network architecture — a severe limitation. In the following chapters, we include standard PINNs as the main competitor method against pMAGI and **do not provide the true noise levels as input to any model.**

## 2.4 Other Methods

Other approaches for computationally-efficiently discovering the equations governing PDEs and ODEs include modeling the governing equations as a sparse regression problem[3,42,43] and applying Galerkin projections or other dimension-reduction approximations[9,15,28] to a high-dimensional PDE system to reduce the state and parameter space to facilitate Monte Carlo sampling.

However, like other black-box methods, these approaches are disadvantaged by the fact that they abandon the first-principles functional forms of the equations governing the system, and rely on surrogate functions to approximate the original governing equations. The consequence of this property is that one cannot recover the original parameters governing the first-principles equations. In contrast, Bongard et al. explore using symbolic equations composed of various combinations of mathematical operations and fitting various sets of parameters on each of these symbolic equations until one symbolic equation is found which, upon numerical integration, produces trajectories that are closest to the ground truth noisy observations of the system[8]. The main disadvantage with this method is that it requires frequent numerical integration, which is very computationally expensive.

Yet another method for Lorenz system parameter estimation is introduced by Ng, which combines data assimilation (inserting observations into an numerically-integrated approximate solution) with an iterative parameter inference update step based on setting the interpolation step differences between the observed and simulated solutions to 0[36]. But, Ng's approach is limited to a noiseless environment. Collage coding, a conceptually-similar approach introduced by Kunze & Heidler, converts the Lorenz parameter estimation problem in terms of a contractive map on a complete metric space, discretizes the ODE integration into update equations, and then solves the collage distance minimization problem inspired by Collage's theorem to recover system parameters[22]. However, such a method may be vulnerable to the choice of discretization step size.

One method somewhat similar in concept to the MAGI method is Adaptive Gradient Matching, which bypass computationally expensive numerical integration by setting up posterior distributions with Gaussian processes that aim to minimize the derivatives predicted from the ODE governing equations and the derivatives approximated by calculating slopes of an interpolant of the observed data[11,51]. However, such approximated gradients may be heavily-dependent on the quality of the interpolation. In contrast, the MAGI method does not involve interpolation of the observed data to approximate gradients, but rather directly uses the fact that the gradient of a Gaussian process is still a Gaussian process.

In the next chapter, we will introduce the manifold-constrained Gaussian process inference (MAGI) method.

# 3
# Overview of the MAGI Method

In this chapter, we will provide an overview of the original MAGI methodology as introduced by Yang et al.[55] We will give a high-level, undergraduate-friendly description focusing on big ideas, and refer readers interested in the full mathematical details to the original *PNAS* MAGI paper[55]. But, before we introduce MAGI, we ought to first introduce some preliminary concepts of Gaussian Processes and Hamiltonian Monte Carlo.

## 3.1 Gaussian Processes (GP)

Formally, in the context of time-indexed random variables, a Gaussian process (GP) is a (potentially uncountably infinite) collection of random vectors $\mathbf{X}(t) = (X_1(t), \ldots, X_d(t)) \in \mathbb{R}^d$ indexed by time $t$, such that any finite subset of them follows a joint multivariate Gaussian distribution[41]. Analogous to a standard Gaussian distribution, any Gaussian Process is fully specified by a mean function $\mathbf{m}(t) = \mathbb{E}[\mathbf{X}(t)]$ and covariance function $k(t, t') = \mathrm{Cov}(\mathbf{X}(t), \mathbf{X}(t'))$, with the notation $\mathbf{X}(t) \sim \mathcal{GP}(\mathbf{m}(t), k(t, t'))$[41]. The choice of covariance function also allows us to encode prior assumptions such as our data being periodic or close to periodic[16].

Mathematically, GPs are elegant for a few reasons of particular interest to our use cases. First, the sums and products of Gaussian processes are also Gaussian processes[16]. Second, the derivatives of a Gaussian Process (gradients, in the multivariate setting) are also Gaussian processes due to differentiation being a linear operator, assuming that the covariance function is at least twice-differentiable (see Yang et al. and Ch. 9 of Rasmussen et al.)[41,55]. Third, the posterior distribution of $\mathbf{X}(t)$ conditional on crystallized observations $\mathbf{X}(t')$ for some set of observed timesteps $t' \in \tau$ is still a Gaussian process, bestowing us with convenient Gaussian conjugacy properties, albeit with updated mean and covariance functions, and naturally fitting into a Bayesian modeling paradigm.

More practically, GPs are typically deployed as data-fitting and/or interpolation procedures with natural uncertainty quantification properties. In other words, placing a GP model on our data allows us to sample smooth mean functions that interpolate a set of observed datapoints and allow us to quantify the uncertainty at any given set of time points via variances and covariances. Given that GPs are fully defined by their mean and covariance functions, through GPs, we have access to a method of data-fitting that achieves the best of both worlds with respect to both parameter efficiency and non-linear expressiveness, with uncertainty quantification.

## 3.2 Hamiltonian Monte Carlo (HMC)

Given a posterior distribution $p(\theta \mid \mathbf{y})$ defined up to a proportionality constant, the objective of a Markov Chain Monte Carlo method, in general, is to sample draws of $\theta$ from this target posterior. Unfortunately, if $\theta$ contains many components (i.e., high-dimensional), samples drawn from traditional methods like the Metropolis algorithm converge very slowly to the target distribution because these traditional methods typically explore the state space using local random walks [16]. For a Bayesian method like MAGI that requires sampling from relatively high-dimensional posterior distributions, we need something that is more powerful and converges faster — Hamiltonian Monte Carlo.

Hamiltonian Monte Carlo (HMC) bypasses the local random walk mechanism in the proposal distribution of the Metropolis and/or Metropolis-Hastings algorithms, and instead leverages concepts from physics (i.e., Hamiltonian dynamics) to substantially increase the rate of exploration of the state space and *decrease the autocorrelation between subsequent samples*. A scaling factor $\varepsilon$ governing how close numerically we replicate the Hamiltonian dynamics (analogous to a learning rate). Specifically, at a high level, for each component $\theta_j$ of our parameter vector $\theta$, we define a corresponding *momentum* variable $\phi_j$ that we will use to simulate the Hamiltonian dynamics via a procedure called *leapfrog steps* [16]. While Metropolis algorithms involve a proposal distribution on $\theta$ itself (albeit, a conditional one given the previous sample), HMC adopts a *marginal* proposal distribution on $\phi$ that we use at each sampling step, usually $\phi \sim \mathcal{N}(0, M)$ for some (usually diagonal) covariance matrix $M$ [16]. Also, in contrast to a Metropolis setup, HMC requires that the user input not only $p(\theta \mid \mathbf{y})$ (up to a proportionality constant), but also $\nabla_\theta \log p(\theta \mid \mathbf{y})$ — the gradient of the log-posterior with respect to $\theta$ [16].

Below, we provide algorithmic overviews of the leapfrog step and overall HMC algorithms [16]. The interested reader is invited to read Ch. 12 of Gelman et al.'s *Bayesian Data Analysis* for a more comprehensive treatment.

## Algorithm 1 Leapfrog

**Input:** $\theta$ - initialized parameter value, $\phi$ - initialized momentum vector, $M$ - mass matrix, $L$ - number of leapfrog steps, $\varepsilon$ - scaling factor.
**Output:** Proposal values $\theta^*$ and $\phi^*$.

1: **function** LEAPFROG($\theta, \phi; M, L, \varepsilon$)
2:     **for** $l \leftarrow 1$ to $L$ **do**
3:         $\phi \leftarrow \phi + \frac{1}{2}\varepsilon \nabla_\theta \log p(\theta \mid \mathbf{y})$     ▷ Update momentum vector $\phi$ using gradient
4:         $\theta \leftarrow \theta + \varepsilon M^{-1} \phi$     ▷ Update position of $\theta$
5:         $\phi \leftarrow \phi + \frac{1}{2}\varepsilon \nabla_\theta \log p(\theta \mid \mathbf{y})$     ▷ Update momentum vector $\phi$ again using gradient
6:     **end for**
7:     $\theta^* \leftarrow \theta$ and $\phi^* \leftarrow \phi$     ▷ Encode our outputs to avoid notational confusion
8:     **Return** $\theta^*, \phi^*$.
9: **end function**

## Algorithm 2 Hamiltonian Monte Carlo (HMC)

**Input:** $\theta^{(0)}$ - initialized parameter value, $L$ - number of leapfrog steps, $\varepsilon$ - scaling factor, $M$ - mass matrix (conventionally diagonal), $N$ - number of samples.
**Output:** Samples $\{\theta^{(1)}, \ldots, \theta^{(N)}\}$

1: **function** HMC($\theta^{(0)}; L, \varepsilon, M, N$)
2:     **for** $t \leftarrow 1$ to $N$ **do**
3:         Draw $\phi \sim \mathcal{N}(0, M)$     ▷ Sample starting momentum vector
4:         $\theta^*, \phi^* \leftarrow$ Leapfrog($\theta^{(t-1)}, \phi; \varepsilon, L$)     ▷ Perform $L$ leapfrog steps
5:         $r \leftarrow \frac{p(\theta^* \mid y) p(\phi^*)}{p(\theta^{(t-1)} \mid y) p(\phi)}$     ▷ Compute acceptance probability
6:         $\theta^{(t)} \leftarrow \theta^*$ with probability $\min(r, 1)$, else $\theta^{(t)} \leftarrow \theta^{(t-1)}$     ▷ Add to our list of samples
7:     **end for**
8:     **Return** $\{\theta^{(1)}, \ldots, \theta^{(N)}\}$.
9: **end function**

Big picture, HMC provides us with a mechanism to sample efficiently from high-dimensional posterior distributions, including those governed by Gaussian processes. Having introduced both GPs and HMC — the two main components of the MAGI method — we now provide a high-level overview of the MAGI method.

## 3.3 Manifold-Constrained Gaussian Process Inference (MAGI)

Suppose we are interested in a dynamical system $\mathbf{x}(t) \in \mathbb{R}^d$ governed by a set of ordinary differential equations $\mathbf{f}$ parameterized by $\boldsymbol{\theta} \in \mathbb{R}^p$. Mathematically, we have

$$\mathbf{f}(\mathbf{x}(t), t; \boldsymbol{\theta}) = \dot{\mathbf{x}}(t) = \frac{d\mathbf{x}(t)}{dt} = \left( \frac{dx_1(t)}{dt}, \frac{dx_2(t)}{dt}, \ldots, \frac{dx_d(t)}{dt} \right).$$

Let $\tau_{obs}$ represent a set of times $t$ for where we have noisy observations $\mathbf{x}_{obs}(t) \in \mathbb{R}^d$ from this system. Next, define $\nabla_{\boldsymbol{\theta}} \mathbf{f}(\mathbf{x}(t), t; \boldsymbol{\theta}) \in \mathbb{R}^{d \times p}$ as the matrix of partial derivatives of the ODEs with respect to the components of $\boldsymbol{\theta}$. Also, define $\nabla_{\mathbf{x}} \mathbf{f}(\mathbf{x}(t), t; \boldsymbol{\theta}) \in \mathbb{R}^{d \times d}$ as the matrix of partial derivatives of the ODEs with respect to the components of $\mathbf{x}$. Let $\tau_{inf}$ denote a set of times $t$ where we wish to infer the unnoised true values $\mathbf{x}(t) \in \mathbb{R}^d$. In practice, to achieve a more "continuous" understanding of the system, we will select $\tau_{inf}$ as a finely-spaced grid of time steps between $t = T_{\min}$ and $t = T_{\max}$, where $[T_{\min}, T_{\max}]$ denotes our interval of interest.

Introduced by Yang et al. in 2021[55], the manifold-constrained Gaussian process inference method (MAGI) is a Bayesian method for inferring the parameters $\boldsymbol{\theta} \in \mathbb{R}^p$ governing the ODE system $\mathbf{f}$ and the values of the system's components $\mathbf{x}(t) \in \mathbb{R}^d$ at a set of desired time points $t \in \tau_{inf}$, given noisy observations $\mathcal{X}_{obs} = \{\mathbf{x}_{obs}(t) \mid t \in \tau_{obs}\}$ from the system and the functions $(\mathbf{f}, \nabla_{\boldsymbol{\theta}} \mathbf{f}, \nabla_{\mathbf{x}} \mathbf{f})$. Compared to other methods like PSO, DE, and PINNs, MAGI a) neither requires computationally-expensive numerical integration nor hardware-intensive automatic differentiation; b) possesses natural uncertainty quantification abilities via its Bayesian paradigm; and c) employs a statistically-principled Bayesian method for accounting for noisy and/or missing data.

Mathematically, MAGI models noisy observations $\mathbf{x}_{obs}(t)$ as the crystallized values of a Gaussian process $\mathbf{X}(t) \in \mathbb{R}^d$[55,53]. However, if we just use a Gaussian process (GP) directly out of the box, given that standard GPs are just a form of statistically-principled interpolation without any dynamical system information, we will only be able to reasonably interpolate estimates of $\mathbf{x}(t)$ within the interval containing our noisy observations, and will neither be able to perform any uncertainty-quantified parameter inference on $\boldsymbol{\theta}$ nor out-of-interval prediction of the trajectories of $\mathbf{x}(t)$. To allow for Bayesian uncertainty quantification of our parameter inference, MAGI models $\theta$ as realizations from a distribution $\pi(\theta)$. Now, with probabilistic frameworks on both $\theta$ and $\mathbf{X}(t)$, we can compute the joint distribution of $(\mathbf{X}(t), \theta)$[55,53].

However, we still have not encoded any information about the underlying dynamical system in our modelling setup. Fortunately, as alluded to during our discussion on GPs, we know that the derivatives of a GP are still GPs. Let $\dot{\mathbf{X}}(t) \in \mathbb{R}^d$ denote the GP representing the derivatives of $\mathbf{X}(t)$ with respect to time $t$. Thus, we can obtain a joint distribution over $(\mathbf{X}(t), \dot{\mathbf{X}}(t), \theta)$. Intuitively, we want $(\mathbf{X}(t), \dot{\mathbf{X}}(t), \theta)$ to respect the ODEs governing our dynamical system at all time steps $t$. Letting $\mathbf{X}_i(t)$ refer to the $i^{th}$ component of $\mathbf{X}(t) \in \mathbb{R}^d$, and defining analogous notation by extension, mathematically, we desire our draws of $(\mathbf{X}(t), \dot{\mathbf{X}}(t), \theta)$ to satisfy the following constraint for all time steps $t$ and components $i$: $\dot{\mathbf{X}}_i(t) = \mathbf{f}(\mathbf{X}(t), t, \theta)_i$. Indeed, this constraint is called a *manifold constraint* by Yang et al.[55] because it forces our draws from our GP to lie on a manifold. If this constraint were satisfied completely, then the distribution over $(\mathbf{X}(t), \dot{\mathbf{X}}(t), \theta)$ would capture the dynamical system perfectly.

Unfortunately, in reality, this is not computationally possible. In practice, we just hope that this constraint is approximately satisfied for times $t \in \tau_{inf}$. Specifically, define an error variable $W_{\tau_{inf}}$ as follows:

$$W_{\tau_{inf}} = \max_{t \in \tau_{inf}, i \in \{1,\ldots,d\}} |\dot{\mathbf{X}}_i(t) - \mathbf{f}(\mathbf{X}(t), t, \theta)_i|.$$

To encode both our dynamical system information and the information from our noisy observations, we can construct a manifold-constrained joint posterior distribution over $(\mathbf{X}(t), \theta)$ for times $t \in \tau_{inf}$, conditioning on our data $\mathcal{X}_{obs}$ (for data interpolation) and our manifold constraint $W_{\tau_{inf}} = 0$ (to respect the ODE system). The $\dot{\mathbf{X}}(t)$ can be omitted because it is a deterministic function of $(\mathbf{X}(t), \theta)$ when conditioning on these two entities. Yang et al. demonstrates that this joint posterior $p(\mathbf{X}(t), \theta \mid \mathcal{X}_{obs}, W_{\tau_{inf}} = 0)$, for times $t \in \tau_{inf}$, can be written in a closed-form expression, and we invite the interested reader to look at their *PNAS* paper and Supplementary Information for additional details and the full closed-form formulas[55]. If the true noise levels $\sigma = (\sigma_1, \ldots, \sigma_d)$ for each component are unspecified, these variables can also be incorporated into the posterior to yield a full posterior $p(\mathbf{X}(t), \sigma, \theta \mid \mathcal{X}_{obs}, W_{\tau_{inf}} = 0)$, for times $t \in \tau_{inf}$.

With this closed-form posterior distribution, assuming the covariance function of our GP is fully-specified (more details soon), we can sample from this posterior distribution using Hamiltonian Monte Carlo to obtain draws of $\mathcal{X}_{inf} = \{\mathbf{X}(t) \mid t \in \tau_{inf}\}$ and $\theta$ [55]. Practically, with these samples, we can compute the posterior means, standard deviations, and any desired summary statistics of $\mathcal{X}_{inf} = \{\mathbf{X}(t) \mid t \in \tau_{inf}\}$, $\sigma$, and $\theta$ for use in downstream analyses. With these samples, depending on our choice of $\tau_{inf}$, we can perform trajectory reconstruction and future prediction as well, with natural uncertainty quantification on these tasks simply by computing summary statistics of our sampled draws. **For the purposes of this thesis, to output point estimates of $\theta$, inferred trajectories, or future predicted trajectories, we will use the posterior means of our relevant samples.**

By default, MAGI uses the Matern kernel $k(t, t')$ with degrees of freedom $\nu = 2.01$ (see Rasmussen et al. Ch. 4 [41]) as the covariance function of the GP, with hyperparameters $(\phi_1, \phi_2)$ set independently for each component $X_i(t)$ of our GP and $B_\nu$ representing the modified Bessel function of the second kind [55]:

$$k(t, t') = \phi_1 \frac{2^{1-\nu}}{\Gamma(\nu)} \left(\sqrt{2\nu}\frac{|t-t'|}{\phi_2}\right)^\nu B_\nu\left(\sqrt{2\nu}\frac{|t-t'|}{\phi_2}\right).$$

In practice, for each component $i \in \{1, \ldots, d\}$, $(\phi_1, \phi_2)$ are fit using an optimization routine based on its marginal likelihood before the start of the posterior sampling phase (please see Yang et al. [55] for details). **Heuristically, $\phi_1$ should reflect the "overall variance level of the GP" (i.e., how much the trajectory of this component fluctuates over time vs. simply flat-lining) and $\phi_2$ should reflect approximately half the "period" of the system (i.e., how often do we see "cycles", very loosely speaking)** [55]. The choices of $(\phi_1, \phi_2)$ will be very salient in the next chapter.

When using the MAGI R package itself [52], there are a few relevant settings that should be discussed. First, MAGI allows the user the option to manually specify fixed values of $(\phi_1, \phi_2)$ and $\sigma$ for each component, as opposed to letting the algorithm fit these values. MAGI also gives users the option to warm-start its HMC sampling process with initial values for $\mathcal{X}_{inf}$ and $\theta$ to hopefully speed up convergence. Given its Bayesian GP structure,

MAGI can also innately handle missing data — specifically, if $\mathbf{X}(t)$ for $t \in \tau_{obs}$ is only observed for a subset of the components, MAGI can still sample over the missing components' posterior distributions. In fact, if for every point $t \in \tau_{obs}$, a subset of components is completely unobserved, MAGI can still, in theory, sample over said components' posterior distributions. Practically, for the tasks of parameter inference and trajectory reconstruction, we will construct $\tau_{inf}$ by taking $\tau_{obs}$ (assuming it is sorted sequentially and evenly-spaced) and inserting $2^k - 1$ evenly-spaced time points between each observation time point in $\tau_{obs}$. We denote this nonnegative integer $k$-value as the *discretization* level. **Larger values of $k$ correspond to finer-resolution time grids, causing our manifold constraint to become more strict. However, this comes at the expense of an increased number of variables to sample from in our posterior distribution, leading to slower convergence.** Note that a discretization level of 0 entails setting $\tau_{inf}$ equal to $\tau_{obs}$. In later chapters, we will also experiment with different discretization levels for different subintervals of interest.

For the remainder of this thesis, let $\mathcal{S}$ encapsulate all of MAGI's runtime-specific settings like the number of HMC steps $N$ (16001, unless otherwise indicated), the burn-in ratio of our HMC sampler (in practice, 0.5 unless otherwise indicated), and any exogenously-specified quantities. With this convention, let us define a MAGISolver function that encapsulates the entire MAGI operation, parameterized and outputting as follows, to reduce notational clutter in future chapters:

$$(\hat{\mathcal{X}}_{inf}, \hat{\theta}, \hat{\sigma}, (\hat{\phi}_1, \hat{\phi}_2)) = \text{MAGISolver}(\mathcal{X}_{obs}, \tau_{inf}; (\mathbf{f}, \nabla_\theta \mathbf{f}, \nabla_\mathbf{x} \mathbf{f}), \mathcal{S}),$$

where $\hat{\mathcal{X}}_{inf}$, $\hat{\theta}$, and $\hat{\sigma}$ represent our size-$N$ posterior HMC samples of $\mathcal{X}_{inf}$, $\theta$, and $\sigma$, and $(\hat{\phi}_1, \hat{\phi}_2)$ refers to the two $d$-dimensional vectors containing our estimated Matern kernel hyperparameters for each component. For notational consistency, if noise levels $\sigma$ and/or the Matern kernel hyperparameters $(\phi_1, \phi_2)$ have already been pre-specified as exogenous quantities in $\mathcal{S}$, then $\hat{\sigma} = \sigma$ and/or $(\hat{\phi}_1, \hat{\phi}_2) = (\phi_1, \phi_2)$, respectively.

In the next chapter, we will introduce Pilot MAGI (pMAGI), a novel methodological improvement on MAGI towards increasing numerical stability, parameter inference accuracy, and trajectory reconstruction fidelity.

# 4

# Methodological Improvements on MAGI

In this chapter, we will introduce *Pilot MAGI* (pMAGI), a novel methodological improvement on MAGI towards increasing numerical stability, parameter inference accuracy, and trajectory reconstruction fidelity. We begin by concretely stating the numerical instability problem and identifying some potential causes. Next, we propose the pMAGI method and present our statistical intuition for why it should yield improvements over base MAGI. We end with extensive benchmarks demonstrating pMAGI's substantial performance advantages over base MAGI on both parameter inference and trajectory reconstruction across a wide range of test bed settings.

Before we proceed into technical details, we provide the following figures to visually emphasize the performance differences between MAGI and pMAGI. Figure 4.1 shows an instance where standard MAGI fails to reconstruct the ground truth trajectory of a Stable (Canonical) regime, with the resultant numerically-unstable reconstructed trajectory including values on the order of $10^{16}$, which is certainly incorrect. In contrast, Figure 4.2 demonstrates pMAGI's clear success on the **exact same data simulation settings** that caused standard MAGI to fail in Figure 4.1. In the sections to come, we will provide explanations regarding the details and nuances in each of these two figures.

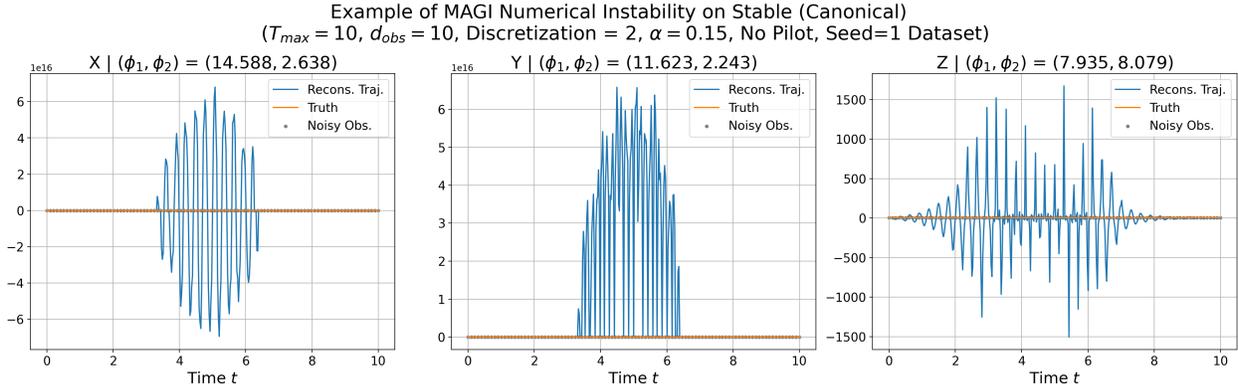

**Figure 4.1:** Example of numerical instability on MAGI on trajectory reconstruction for the Stable (Canonical) regime with interval length $T_{max} = 10.0$, noisy observation density $d_{obs} = 10$ observations per unit time, discretization $D = 2$, and noise level multiplier $\alpha = 0.15$. The noisy observations are shown as grey dots, the unnoised ground truth trajectory is shown as an orange line, and the reconstructed trajectory is shown as a blue line. **Note that the y-axis of the $X$ and $Y$ components is on the order of $10^{16}$.**

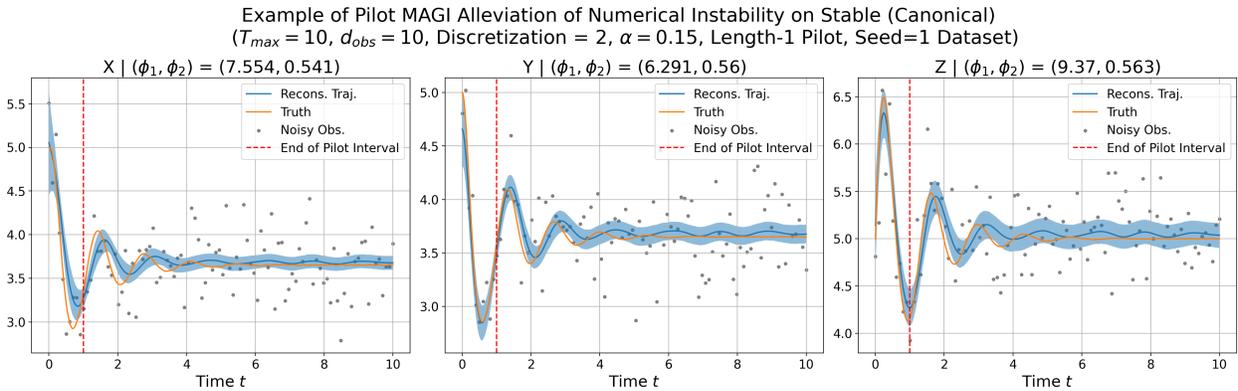

**Figure 4.2:** Example of Pilot MAGI (pMAGI) resolving numerical instability on the Stable (Canonical) regime with a length-$1$ pilot module under the **exact same data simulation settings** as Figure 4.1. The noisy observations are shown as grey dots, the unnoised ground truth trajectory is shown as an orange line, and the reconstructed trajectory is shown as a blue line. The interval band shows our $95\%$ predictive interval obtained via summary statistics on our posterior samples. The dotted red line marks the end of the pilot module.

## 4.1 Numerical Problems with MAGI

From extensive empirical testing, we observe that especially for long observation intervals with sparse observation density (i.e., few observations per 1-unit of time), the base MAGI algorithm sometimes has difficulty selecting appropriate values of the $(\phi_1, \phi_2)$ hyperparameters. The consequence of this misspecification of hyperparameter values is that the HMC sampling procedure in MAGI will become numerically unstable, generating trajectories that "blow-up" by multiple orders of magnitude. In Figure 4.1 from above, we demonstrated base MAGI's failed trajectory reconstruction on a dataset from the Stable (Canonical) regime, passing in as input sparse and noisy observations on the interval $t \in [0.0, T_{max}]$ for $T_{max} = 10.0$, with $d_{obs} = 10$ noisy observations per unit time, for a total of 100 noisy observations on this interval. The noise level multiplier used for this trial was $\alpha = 0.15$.

Examining Figure 4.1 more closely, we see a potential diagnosis for the cause of MAGI's numerical instability. In the titles of each subplot, we print the inferred $(\phi_1, \phi_2)$ hyperparameters for each component of the Lorenz system. As discussed in the previous chapter, the $\phi_2$ hyperparameter should reflect approximately half the "period" of the system. In particular, the $Z$ component having a fitted $\phi_2 = 8.079$ is very concerning, because the approximate "period" of the stable Lorenz system is definitely not on the order of 16 units of time. In general, from extensive numerical testing, we do not need $\phi_2$ to be very precise, but it should be within an order of magnitude of the optimal value to yield numerical instability. Furthermore, from traditional understandings of dynamical systems, the $(\phi_1, \phi_2)$ hyperparameter values for each component inferred for our choices of $T_{max} \in \{2.0, 4.0, 6.0, 8.0, 10.0\}$ should be relatively constant and/or convergent, and, at the minimum, within an order of magnitude of each other.

To follow up on this diagnosis, in Figure 4.3, we plot MAGI's estimated $(\phi_1, \phi_2)$ values for many possible intervals $t \in [0.0, T_{max}]$ for $T_{max} \in \{2.0, 4.0, 6.0, 8.0, 10.0\}$, across ten randomly-seeded datasets apiece. Assuming that $(\phi_1, \phi_2)$ estimation is consistent in the sense that taking $\alpha \to 0$ and $d_{obs} \to \infty$ should send $(\phi_1, \phi_2)$ to their optimal values, we can take the lines corresponding to $d_{obs} = 40$ as most reflective of the ground truth optima.

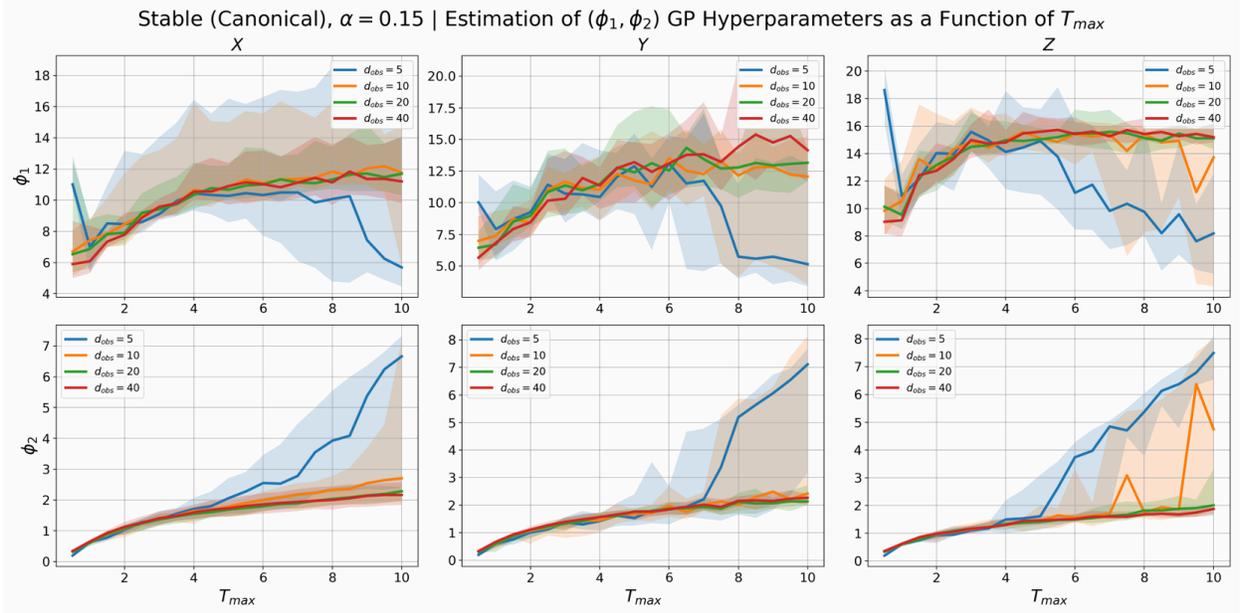

**Figure 4.3:** Estimated $(\phi_1, \phi_2)$ values (y-axis) for each Lorenz component as a function of $T_{max}$ (x-axis) on the Stable (Canonical) regime with noise level multiplier $\alpha = 0.15$, averaged across ten randomly-seeded datasets. Specifically, we are showing estimated $(\phi_1, \phi_2)$ values on the interval $t \in [0, T_{max}]$ as points on the curve. The colors indicate the density of noisy observations $d_{obs}$ per unit time. The correspondingly-colored intervals demonstrate the minimum and maximum estimated hyperparameter values across the ten randomly-seeded datasets.

Under this assumption, we do observe from Figure 4.3 above that, especially at larger $T_{max}$ values, our estimated hyperparameter values tend to diverge from the ground truth optima for low observation densities (i.e., $d_{obs} = 5$ and $d_{obs} = 10$), with the $Z$ component's deviance being most apparent. These results also hold across our other three regimes — the plots of which we relegate to the Appendix A.1. Thus, we hypothesize that one direction towards improving MAGI's numerical instability is to improve its $(\phi_1, \phi_2)$ estimation routine.

## 4.2 Proposed Solution: Pilot MAGI (pMAGI)

From the $(\phi_1, \phi_2)$ estimation figures above, we note that MAGI tends to have stable hyperparameter estimation for small intervals of $t \in [0, T_{max}]$. As such we propose the *Pilot MAGI* (pMAGI) method below.

**Algorithm 3** Pilot MAGI (pMAGI)
---
**Input:** $T_{\text{max,pilot}}$ - pilot interval length, $D_{\text{pilot}}$ - pilot discretization, $\mathcal{X}_{\text{obs}}$ - noisy observations, $\tau_{inf}$ - desired discretized timesteps .
**Output:** Samples of trajectories $\hat{\mathcal{X}}_{\text{inf}}$ at discretized times $\tau_{inf}$, and samples of parameters $\hat{\theta}$.
 1: **function** PMAGI($T_{\text{max,pilot}}, D_{\text{pilot}}, \mathcal{X}_{obs}, \tau_{inf}$)
 2: $\quad \tau_{\text{obs, pilot}} \leftarrow \{t \in \tau_{\text{obs}} \mid t \leq T_{\text{max,pilot}}\}$ $\quad\quad\quad\quad\quad\quad\quad\quad\quad\quad\quad$ ▷ Construct pilot interval
 3: $\quad$ Construct $\mathcal{X}_{\text{obs, pilot}}$ accordingly
 4: $\quad \tau_{\text{inf, pilot}} \leftarrow$ discretize $\tau_{\text{obs, pilot}}$ with level $D_{\text{pilot}}$.
 5: $\quad (\hat{\phi}_1, \hat{\phi}_2), \hat{\sigma} \leftarrow$ MAGISolver($\mathcal{X}_{\text{obs, pilot}}, \tau_{\text{inf, pilot}}; (\mathbf{f}, \nabla_\theta \mathbf{f}, \nabla_\mathbf{x} \mathbf{f}), \mathcal{S}$). $\quad$ ▷ Execute pilot module
 6: $\quad$ Pass $(\hat{\phi}_1, \hat{\phi}_2)$ and $\hat{\sigma}$ as fixed, exogenous inputs into MAGISolver again and run MAGISolver on the full interval of observed data. Obtain posterior samples

$$\hat{\mathcal{X}}_{\text{inf}}, \hat{\theta} \leftarrow \text{MAGISolver}(\mathcal{X}_{\text{obs}}, \tau_{\text{inf}}; (\mathbf{f}, \nabla_\theta \mathbf{f}, \nabla_\mathbf{x} \mathbf{f}), \mathcal{S} \cup \{(\hat{\phi}_1, \hat{\phi}_2), \hat{\sigma}\}).$$

 7: $\quad$ **Return** samples $\hat{\mathcal{X}}_{\text{inf}}, \hat{\theta}$.
 8: **end function**
---

**In short, pMAGI operates by fitting our $(\phi_1, \phi_2)$ and $\sigma$ on a short interval of our data (as opposed to the full observation interval) and passing the fitted values as fixed, exogenous inputs into the full MAGI routine on the full observation interval.** We will define the procedures performed before the full MAGI run as the *pilot module with length $T_{max,pilot}$*. The reason for inferring $\sigma$ during the pilot module is because the `MAGI` package itself currently requires noise levels to be exogenously specified if $(\phi_1, \phi_2)$ are exogenously specified. In practice, when running experiments, we will hyperparameter tune over $T_{\text{max,pilot}} \in \{1, 2\}$ and $D_{\text{pilot}} \in \{0, 1, 2\}$, using 4001 HMC steps for the pilot call to MAGISolver and 16001 HMC steps for the subsequent full MAGISolver call. To be clear, a setting of $D_{\text{pilot}} = 0$ entails that $\tau_{\text{inf, pilot}}$ is set equal to $\tau_{\text{obs, pilot}}$.

Using the exact data simulation settings from Figure 4.1, we applied the pMAGI method with a length-1 pilot module and showed our reconstructed trajectories in Figure 4.2. Indeed, our reconstructed trajectories match the ground truth much more closely, with our predictive intervals almost always capturing the ground truth. The printed $(\phi_1, \phi_2)$ hyperparameter values are also much more reasonable. Our proposed method worked!

## 4.3 pMAGI vs. MAGI Full Benchmark Results

Having established the potential of our pMAGI method, we proceed to benchmark pMAGI vs. MAGI on a wide range of Lorenz settings, focusing on parameter inference and trajectory reconstruction. Specifically, we will try all four Lorenz testbed regimes, over $T_{max} \in \{2.0, 4.0, 6.0, 8.0, 10.0\}$ and $d_{obs} \in \{5, 10, 20, 40\}$, with 10 randomly-seeded datasets for reproducibility. For both pMAGI and MAGI, we will set overall discretization equal to $\log_2(40/d_{obs})$ to ensure an equal number of discretized timesteps in $\tau_{inf}$ regardless of $d_{obs}$. For the results in this chapter, we will hyperparameter tune pMAGI over $T_{max,pilot} \in \{1, 2\}$ and $D_{\text{pilot}} \in \{0, 1, 2\}$ and show the metrics corresponding to the best pMAGI variants in the subsequent figures. Though, one major advantage of pMAGI is that it is quite robust to the choice of hyperparameters, as we will show in later chapters when comparing against PSO, DE, and PINNs.

On notation, let the true parameter values for each regime be denoted $\theta = (\beta, \rho, \sigma)$ and let our MAGISolver output be denoted $\hat{\theta} = (\hat{\beta}, \hat{\rho}, \hat{\sigma})$, where $\hat{\beta} = (\hat{\beta}_1, \ldots, \hat{\beta}_N)$ refers to our posterior draws of $\beta$, with $\hat{\rho}$ and $\hat{\sigma}$ defined analogously. We will quantify parameter inference using *scaled L1 error*, defined for $\beta$ as

$$L(\hat{\beta}; \beta) = \frac{1}{N} \sum_{i=1}^{N} \frac{|\hat{\beta}_i - \beta|}{|\beta|},$$

with errors on $\rho$ and $\sigma$ defined analogously. For trajectory reconstruction, let $\mathcal{X}_{true} = (\mathbf{X}_{true}, \mathbf{Y}_{true}, \mathbf{Z}_{true})$ refer to the true, unnoised trajectories of the relevant regime, where $\mathbf{X}_{true} = \{X(t) \mid t \in \tau_{inf}\}$ and $\mathbf{Y}_{true}, \mathbf{Z}_{true}$ are defined accordingly. In practice, we compute these ground truth unnoised trajectories using high-precision numerical integration. Analogously, we can unpack our MAGISolver output as $\hat{\mathcal{X}}_{inf} = (\hat{\mathbf{X}}_{inf}, \hat{\mathbf{Y}}_{inf}, \hat{\mathbf{Z}}_{inf})$, where $\hat{\mathbf{X}}_{inf} = \{\hat{X}_{inf}(t) \mid t \in \tau_{inf}\}$ and $\hat{\mathbf{Y}}_{inf}, \hat{\mathbf{Z}}_{inf}$ are defined accordingly. Specifically, for each time $t \in \tau_{inf}$, we let $\hat{X}_{inf}(t)$ be the posterior mean computed over all HMC samples. We will quantify trajectory reconstruction performance using *scaled mean absolute error* (sMAE), defined for component $X$ as

$$L_X(\hat{\mathbf{X}}_{inf}; \mathbf{X}_{true}) = \frac{1}{|\tau_{inf}|} \sum_{t \in \tau_{inf}} \frac{|\hat{X}_{inf}(t) - X(t)|}{|X(t)|},$$

with errors on components $Y$ and $Z$ defined accordingly.

### 4.3.1　Parameter Inference

For the main text, we will show the parameter inference scaled L1 errors of pMAGI vs. MAGI on the Chaotic (Butterfly) regime for three main reasons. First, it is representative of the general performance differences between pMAGI vs. MAGI. Second, according to traditional dynamical systems theory, it should be the most difficult regime to perform parameter inference on due to its chaotic nature and sensitivity to slight perturbations in initial conditions and parameter settings. Third, compared to the Chaotic (No Butterfly) regime, it has more visually erratic behavior which should present a bigger challenge for GP-based fitting methods. We invite the reader to explore Appendix A.2 to see our results for the other three regimes, which are quite similar.

From Figure 4.4, we see that on inferring the $\beta$ parameter, pMAGI significantly outperforms MAGI, especially at $d_{obs} = 10$ and $d_{obs} = 20$. By $d_{obs} = 40$, the two methods' performances on $\beta$ become relatively equal, which makes sense given that the $(\phi_1, \phi_2)$ hyperparameter estimation routine for base MAGI has likely recovered and stabilized due to the abundance of signal. On the critical $\rho$ parameter that effectively governs the stability vs. chaos of the system, we see that at $d_{obs} = 10$ and $d_{obs} = 20$, pMAGI considerably outperforms base MAGI at nearly all $T_{max}$ values. For the $\sigma$ parameter, again, pMAGI considerably outperforms MAGI at $d_{obs} = 10$ and $d_{obs} = 20$, though it should be noted that both pMAGI and MAGI have considerable bias. In general, pMAGI and MAGI have comparable performance at the extremes of $d_{obs} = 5$ and $d_{obs} = 40$, but pMAGI wins considerably at the intermediate settings of $d_{obs} = 10$ and $d_{obs} = 20$. Having explained the $d_{obs} = 40$ phenomena, one explanation for the parity in performance at $d_{obs} = 5$ is that this setting simply has too little signal for both pMAGI and MAGI to capture, and thus both do not perform well, especially on $\beta$ and $\sigma$.

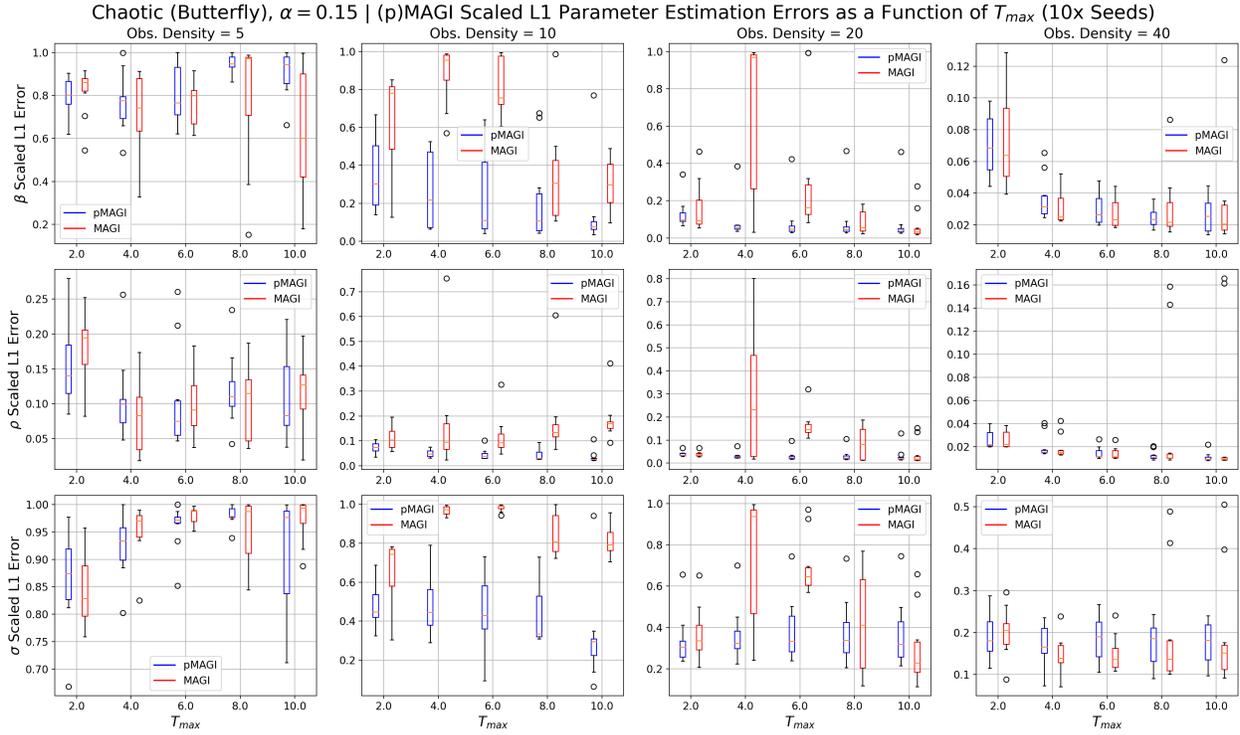

**Figure 4.4:** pMAGI vs. MAGI Parameter Inference on the Chaotic (Butterfly) regime. Each row corresponds to one parameter in $(\beta, \rho, \sigma)$ and each column corresponds to the density of noised observations per unit time $d_{obs} \in \{5, 10, 20, 40\}$. The $x$-axis for each subplot corresponds to the length of the observation interval, i.e. $t \in [0.0, T_{max}]$. The box-and-whisker plots show the distribution of the scaled L1 parameter estimation errors for the best pMAGI vs. MAGI variants across the 10 random-seeded datasets. Blue (always on the left) corresponds to pMAGI while red (always on the right) corresponds to MAGI. **Lower errors indicate better performance.**

Overall, however, it is clear that pMAGI has a decisive advantage over MAGI on parameter inference performance, especially at the intermediate ranges of $d_{obs} = 10$ and $d_{obs} = 20$.

### 4.3.2 Trajectory Reconstruction

For the same reasons provided in the previous subsection, we will focus our discussion on trajectory reconstruction performance of pMAGI vs. MAGI on the Chaotic (Butterfly) regime. From Figure 4.5, we see that with only a few exceptions, pMAGI achieves considerably lower trajectory reconstruction errors on all three components $X, Y, Z$ at $d_{obs} = 10$ and $d_{obs} = 20$. Just as in the parameter inference discussion, we note that the perfor-mance between pMAGI and MAGI is roughly comparable at $d_{obs} = 40$, likely due to there being sufficient signal to overcome the volatility of the base MAGI hyperparameter fitting routine. At $d_{obs} = 5$, both methods' errors are quite large, indicating that both MAGI and pMAGI are struggling with too sparse of a signal.

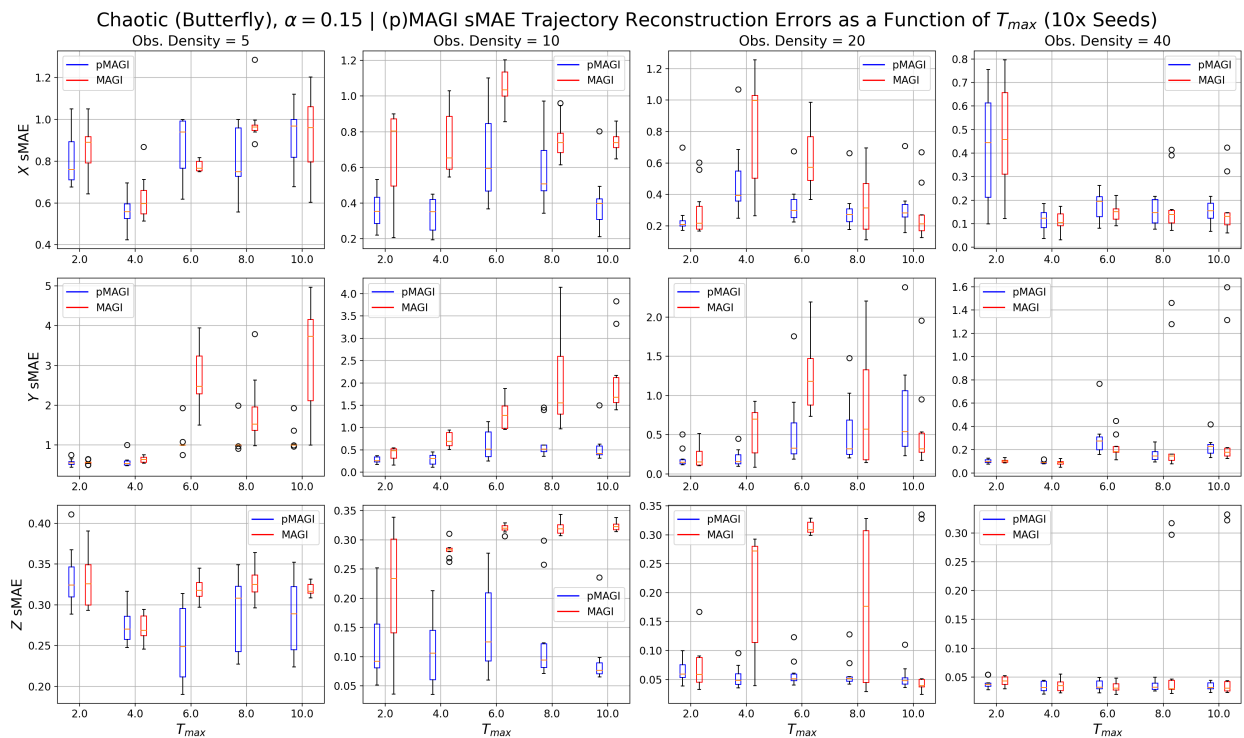

**Figure 4.5**: pMAGI vs. MAGI Trajectory Reconstruction on the Chaotic (Butterfly) regime. Each row corresponds to one component in $(X, Y, Z)$ and each column corresponds to the density of noised observations per unit time $d_{obs} \in \{5, 10, 20, 40\}$. The *x*-axis for each subplot corresponds to the length of the observation interval, i.e. $t \in [0.0, T_{max}]$. The box-and-whisker plots show the distribution of the sMAE errors for the best pMAGI vs. MAGI variants across the 10 random-seeded datasets. Blue (always on the left) corresponds to pMAGI while red (always on the right) corresponds to MAGI. **Lower errors indicate better performance.**

Nonetheless, it is clear that especially at the intermediate settings of $d_{obs} = 10$ and $d_{obs} = 20$, pMAGI is a considerable improvement over MAGI when it comes to trajectory reconstruction.

## 4.4 Main Takeaways

From all the visualizations and benchmarks presented, it is clear that our newly-proposed pMAGI method comes with **three significant advantages over the base MAGI.** First, pMAGI is much more numerically-stable than MAGI. Second, pMAGI possesses significantly-increased parameter inference accuracy compared to MAGI. Third, pMAGI also possesses much-improved trajectory reconstruction abilities compared compared to base MAGI, especially at intermediate observation densities $d_{obs}$.

# 5
# Parameter Inference and Identifiability

In this chapter, we evaluate pMAGI's parameter inference performance against two optimization-based methods, Particle Swarm Optimization (PSO) and Differential Evolution (DE), and one deep-learning method, Physics-Informed Neural Networks (PINNs). We demonstrate that, especially after taking computational costs into account, pMAGI presents itself favorably against all three competitor methods. A deeper investigation of pMAGI's error curves also poses some exciting new questions regarding parameter identifiability in chaotic vs. stable settings. Specifically, we conjecture that parameter identifiability in Lorenz systems under noisy and sparse data conditions is not so much determined by mathematical notions of stability vs. chaos, but rather on the shapes of the observed components of the trajectories themselves.

## 5.1 EXPERIMENTAL SETUPS

We will evaluate pMAGI and all three of its competitor methods on the four Lorenz testbed regimes described in Section 1.5. We will truncate our observation intervals to be on $t \in [0.0, T_{max}]$ for $T_{max} \in \{2.0, 4.0, 6.0, 8.0, 10.0\}$. For each of these trajectories, we feed into our models noisy and sparse observations with noise level multiplier $\alpha = 0.15$ with observation densities $d_{obs} \in \{5, 10, 20, 40\}$. As a concrete example, a setting with $d_{obs} = 10$ and $T_{max} = 6.0$ entails giving the models 60 noisy observations between $t = 0.0$ and $t = 6.0$ as input. To increase confidence in our results, we will repeat each setting across ten randomly-seeded datasets, with the datasets common across all models.

For pMAGI, we will set overall discretization equal to $\log_2(40/d_{obs})$ to ensure an equal number of discretized timesteps in $\tau_{inf}$ regardless of $d_{obs}$. We will hyperparameter tune over pilot interval lengths $T_{max,\text{pilot}} \in \{0.0, 1.0, 2.0\}$ and pilot discretizations $D_{pilot} \in \{0, 1, 2, 3\}$. Note that $T_{max,\text{pilot}} = 0.0$ corresponds to no pilot module (i.e., standard MAGI). We use the `MAGI R` package as described in Wong et al[52]. We use 4001 HMC steps for the pilot module and 16001 steps for the main MAGISolver call. We also set a burn-in ratio of 0.5, which governs the proportion of samples we discard. This is a precautionary measure against lack of convergence in early sampling steps.

For Particle Swarm Optimization (PSO), we use the `PySwarms` Python package implemented by Miranda et al.[32], using $n_{\text{particles}} = 120$ and cognitive and social parameters $c_1 = c_2 = 2$, extrapolating from discussions by He et al.[18]. We hyperparameter tune over the inertia parameter $w \in \{0.9, 0.8, 0.7, 0.6, 0.5, 0.4\}$. We invite the interested reader to explore Miranda et al.[32] for a deeper discussion on PSO.

For Differential Evolution (DE), we use the implementation from the `SciPy.optimize`[50] library. We fix a population size of 45 and hyperparameter tune over the mutation parameter interval $F \in (F_{lb}, F_{ub})$ for $(F_{lb}, F_{ub}) \in$ $\{(0.5, 1.0), (0.5, 1.5), (1.0, 1.5)\}$. We also hyperparameter tune over the crossover probability $CR \in \{0.7, 0.5, 0.3\}$. We invite the interested reader to explore the `scipy.optimize.differential_evolution` documentation.

For Physics-Informed Neural Networks (PINN), we use the codebase written by van Herten et al.[49] with 3 dense-layers and 32 nodes in each layer. We hyperparameter tune over $\lambda$, which governs the relative weights of the reconstruction and physics-based loss components, trying $\lambda \in \{0.1, 1.0, 10.0, 100.0, 1000.0\}$. We train each PINN for 60000 epochs with an Adam optimizer using learning rate $\eta = 0.01$, following van Herten et al.[49].

It is important to note that PSO, DE, and PINN are incapable of uncertainty quantification in the sense that they can only output a final point estimate for $\theta$. In contrast, pMAGI outputs a set of HMC posterior samples $\hat{\theta}$, to facilitate uncertainty quantification. To keep our comparisons apples-to-apples, we will take the posterior mean of our pMAGI HMC samples as our pMAGI point estimate for $\theta$.

Let the point estimates of any given model be $\hat{\theta} = (\hat{\beta}, \hat{\rho}, \hat{\sigma})$ and the ground truth values be $\theta = (\beta, \rho, \sigma)$. We define the L1 parameter estimation error on $\beta$ as $L(\hat{\beta}; \beta) = \frac{|\beta - \hat{\beta}|}{|\beta|}$, and define the errors on $\rho, \sigma$ accordingly. In the accompanying figures for this thesis, we will focus on the best variants of each model.

## 5.2   pMAGI vs. PSO and DE

The performance trends between pMAGI, PSO, and DE do not differ significantly across regimes. As such, to avoid redundancy and dilution of the main message of this text, we will focus on a most representative regime — the Chaotic (No Butterfly). We invite the reader to explore the corresponding figures for the other regimes in Appendix A.4. From Figure 5.1, we see that on inference of the $\beta$ parameter, pMAGI incurs much larger errors than both PSO and DE at $d_{obs} = 5$, though the performance gap shrinks significantly as observation density $d_{obs}$ increases, while not completely disappearing. On the critical $\rho$ parameter that, in effect, governs the stability and/or chaos of the system, we see that all three methods pMAGI, PSO, and DE incur very, very low errors at all observation densities, though it seems like pMAGI is, on average, a few percentage points behind.

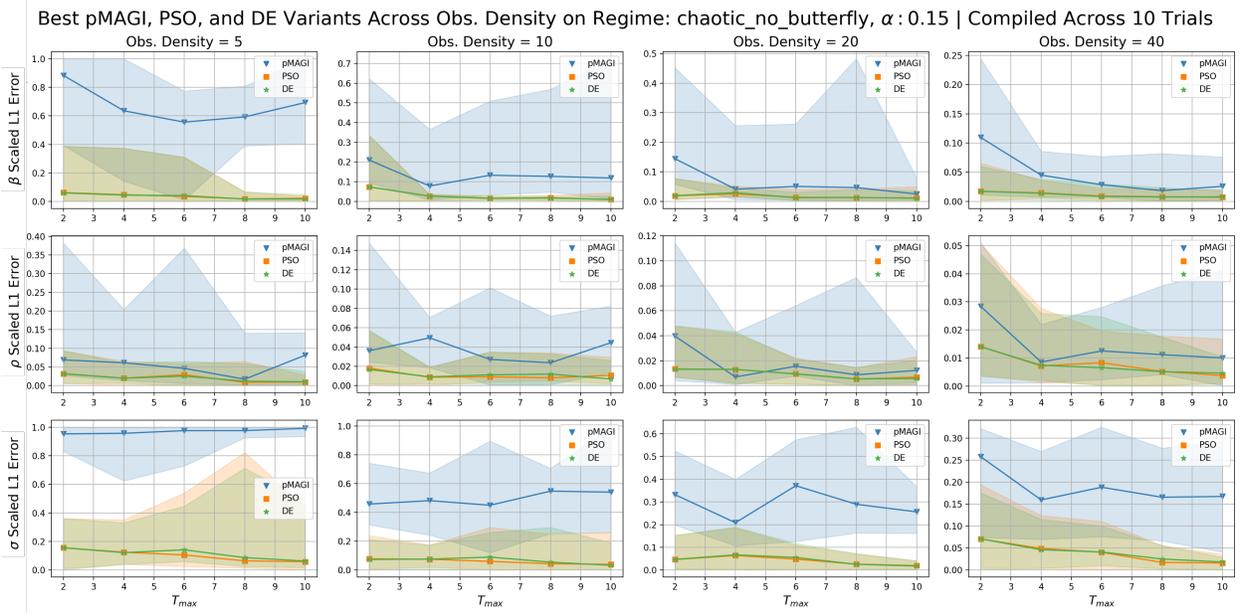

**Figure 5.1:** pMAGI vs. PSO and DE Parameter Inference on the Chaotic (No Butterfly) regime. The x-axis of each subplot is $T_{max}$, the length of our noisy observation interval $t \in [0, T_{max}]$. Blue represents pMAGI, orange represents PSO, and green represents DE. The solid lines indicate the average errors across all ten randomly-seeded trials. The colored confidence bands represent the maximum and minimum errors accrued on the given setting across ten randomly-seeded trials. **Lower errors indicate better performance.**

Finally, on the $\sigma$ parameter, pMAGI consistently incurs higher errors than PSO and DE, though by $d_{obs} = 40$, the gap has been closed significantly, but not completely. Indeed, these observations lead us to suspect that pMAGI and MAGI may have lingering biases on $\sigma$ due to the nature of the Bayesian prior used. Upon further investigation, we note that pMAGI tends to underestimate $\sigma$, which could be due to the Bayesian prior drawing the parameter values towards 0. It is possible that $\beta$ and $\rho$ are a lot more identifiable from the data alone and thus pMAGI can overpower the priors on these two parameters. However, much more extensive investigation will be necessary to draw any definitive conclusions.

Overall, the mere fact that pMAGI can perform in the same performance envelope against optimization-based methods like PSO and DE is already a major win, if we remember that PSO and DE require repeated execution of computationally-expensive numerical integration at each iteration. Also, recall that PSO and DE are both incapable of uncertainty quantification.

## 5.3  pMAGI vs. PINN

The performance trends between pMAGI and PINN can be roughly clustered into two groups — Stable (Canonical) on one end, and the three other regimes on the other end. As such, in the main text, we will focus our analyses on the Stable (Canonical) and Chaotic (Butterfly) regimes. The interested reader is invited to explore Appendix A.5 for corresponding figures on the other two regimes.

From Figure 5.2, we see that on the $\beta$ and $\rho$ parameters for the Stable (Canonical) regime, pMAGI and PINN incur very similar errors across all observation densities $d_{obs}$. On the $\sigma$ parameter, however, echoing the PSO and DE discussions earlier, we observe that pMAGI incurs significantly larger errors than PINN across virtually all settings on the Stable (Canonical) regime. Indeed, we even observe that pMAGI's errors on $\sigma$ tend to increase as $T_{max}$ increases, which is counterintuitive because one would expect that the more data we have over more time, the more accurate our parameter inference becomes. The increasing (absolute) bias on $\sigma$ is especially concerning when we observe that pMAGI's errors on $\beta$ and $\rho$ do not seem to increase with $T_{max}$. Nonetheless, both pMAGI and PINN do an excellent job inferring $\beta$ and $\rho$ on the Stable (Canonical) regime, even with such a high noise level multiplier of $\alpha = 0.15$.

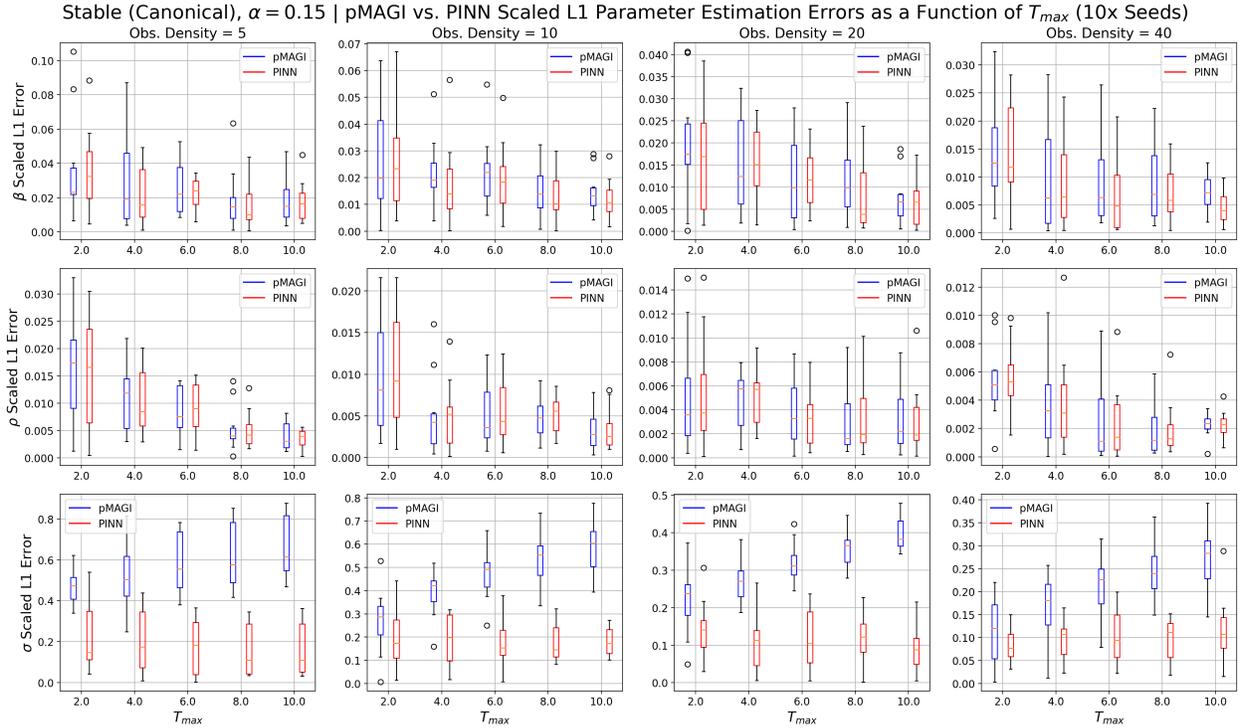

**Figure 5.2:** pMAGI vs. PINN Parameter Inference on the Stable (Canonical) regime. Blue represents pMAGI, while red represents PINN. The x-axis of each subplot is $T_{max}$, the length of our noisy observation interval $t \in [0, T_{max}]$. **Lower error values indicate better performance.**

From Figure 5.3, on the Chaotic (Butterfly) regime, we observe that by $d_{obs} = 10$, the performance gap between PINN and pMAGI has narrowed considerably, despite an initial lead of PINN over pMAGI at $d_{obs} = 5$. Indeed, at larger $T_{max}$ values, pMAGI begins to noticeably outperform PINN on inference of the $\beta$ and $\rho$ parameters at $d_{obs} = 20$ and $d_{obs} = 40$. Nonetheless, across all observation densities $d_{obs}$, pMAGI retains a considerable bias on $\sigma$.

However, when comparing pMAGI vs. PINN parameter inference performance, it is important to note that we are only running pMAGI for $20K$ HMC steps (if we account for the pilot module), while we are training the PINN for $60K$ epochs. Furthermore, the PINN contains hundreds, if not thousands, more parameters than pMAGI. The fact that pMAGI can perform in a similar performance envelope as PINN, and even outperform it on certain settings, is, by any definition, a victory for end users who value computational efficiency.

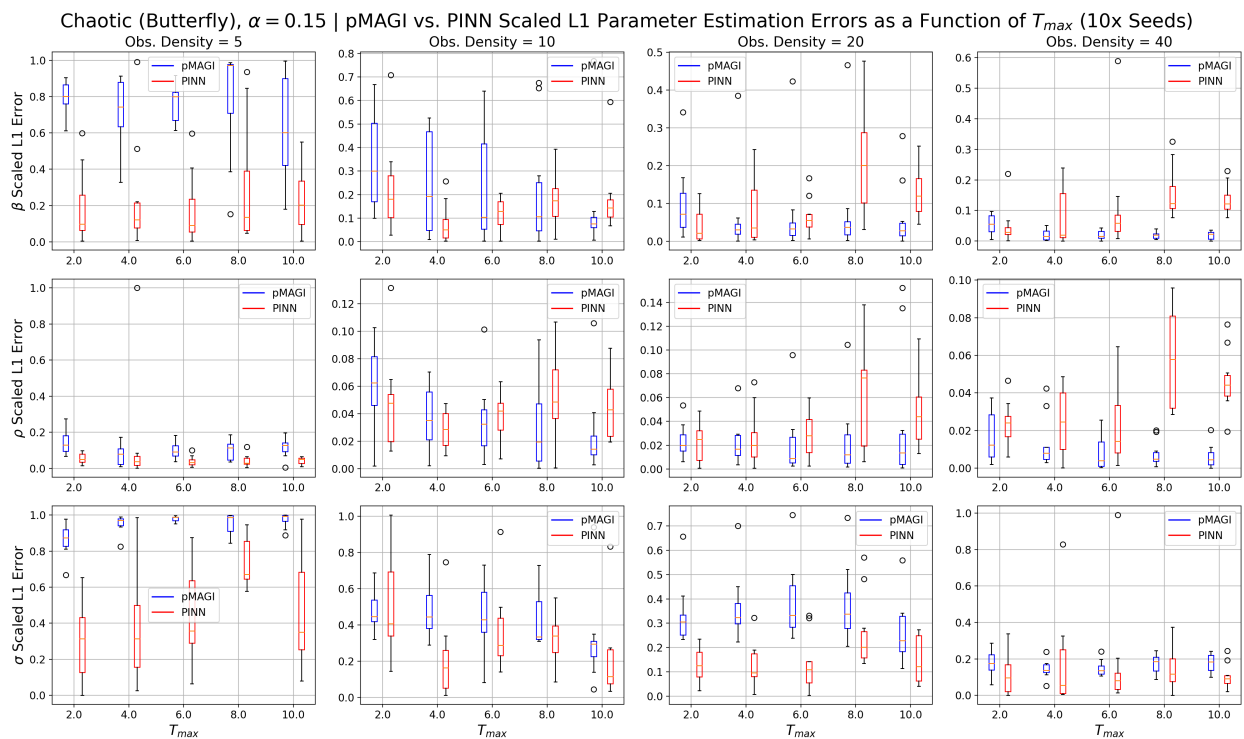

**Figure 5.3:** pMAGI vs. PINN Parameter Inference on the Chaotic (Butterfly) regime. Blue represents pMAGI, while red represents PINN. The x-axis of each subplot is $T_{max}$, the length of our noisy observation interval $t \in [0, T_{max}]$. **Lower error values indicate better performance.**

## 5.4 Insights on Identifiability

From Figures 5.1 - 5.3, with the exception of pMAGI's occasional increasing bias on $\sigma$ as a function of $T_{max}$, we observe that, holding regime and $d_{obs}$ fixed, all three methods' incurred errors on each parameter remain relatively constant across $T_{max}$ values. In other words, increasing the length of the observation interval does not lead to higher-accuracy parameter inference. This is indeed a counterintuitive finding as one would expect that the longer time we observe a system, the more information we have about the system, and thus the more accurately we can infer the parameters governing the system. Yet, this appears to not be the case, at least using our four vastly-different parameter inference methods pMAGI, PINN, PSO, and DE.

Another natural question one may ask is — holding all else fixed, which of stable vs. chaotic regimes have more identifiable parameters under noisy and sparse observation conditions? This is an intriguing question because according to dynamical systems theory, chaotic systems are infamous for outputting significantly-deviating trajectories when initial conditions and/or parameters are only slightly perturbed. To gain some insight into this question, we repeat the experimental setup from Section 5.1, but now try not only noise level multiplier $\alpha = 0.15$, but also $\alpha \in \{1.5 \times 10^{-5}, 1.5 \times 10^{-4}, 1.5 \times 10^{-3}, 1.5 \times 10^{-2}\}$. We focus on pMAGI in this section due to it being more statistically-principled than the other methods, and use the same hyperparameter tuning procedure discussed in Section 5.1.

Because we are working only with the statistically-principled pMAGI, we can compute errors leveraging the Bayesian sampling nature of our method. Specifically, let $\hat{\theta}^{(i)} = (\hat{\beta}^{(i)}, \hat{\rho}^{(i)}, \hat{\sigma}^{(i)})$ represent the $i^{th}$ HMC sample of $\theta$ for a given run of pMAGI. Accounting for the burn-in ratio which governs the proportion of samples we discard due to lack of convergence, we have $N_{\text{samp}} = 8000$ HMC samples. Let $\hat{\beta} = \{\hat{\beta}^{(1)}, \ldots, \hat{\beta}^{(N_{\text{samp}})}\}$. We define mean absolute percent error (MAPE) on $\beta$ as follows, with the corresponding errors on $\rho, \sigma$ defined analogously:

$$MAPE(\hat{\beta}; \beta) = \frac{1}{N_{\text{samp}}} \sum_{i=1}^{N_{\text{samp}}} \frac{|\beta - \hat{\beta}^{(i)}|}{|\beta|}.$$

This MAPE metric is more useful because it allows us to better consider the impact of each HMC posterior sample $\hat{\theta}^{(i)}$, giving us potentially more granular insights.

All in all, the error curves generated from these experiments did not look significantly visually-different across noise level multiplier $\alpha$ values, so we will focus our discussion on $\alpha = 0.015$ as a representative example, and relegate the other $\alpha$-values' figures to Appendix A.6.

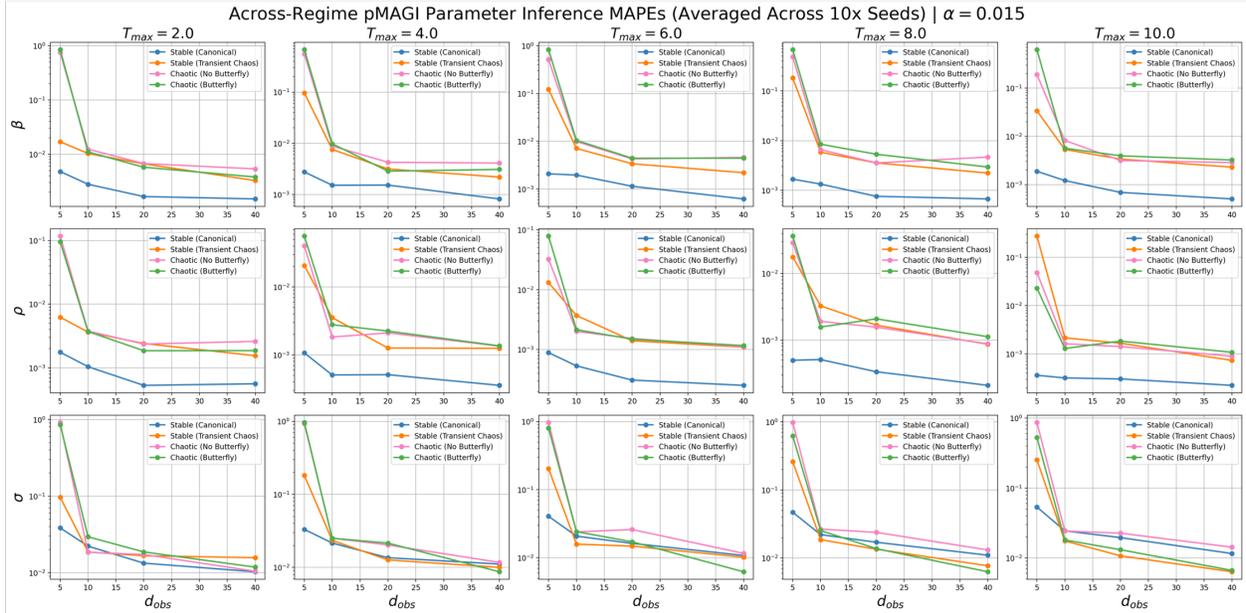

**Figure 5.4:** Lorenz parameter identifiability at $\alpha = 0.015$ as quantified with MAPE. The x-axis of each subplot shows the observation density $d_{obs}$. The y-axis of each subplot, rendered on the log-scale for readability, shows the MAPE on the given parameter. The solid lines correspond to the mean errors averaged across ten randomly-seeded datasets. The error curves corresponding to the Stable (Canonical) regime are shown in blue, the Stable (Transient Chaos) in orange, the Chaotic (No Butterfly) in pink, and the Chaotic (Butterfly) in green. **Lower error values indicate better performance, which imply stronger identifiability.**

From Figure 5.4, we observe that for all regimes on all parameters and across all $T_{max}$ values, MAPE decreases with increasing observation density $d_{obs}$. Combining this finding with our insights in the previous sections, we conclude that Lorenz system identifiability (i.e., how accurately can we identify the parameters governing the system) is *not a function of how long we observe our system (as governed by $T_{max}$)*, but rather a function of *how frequently do we observe the trajectories of our system (as governed by $d_{obs}$)*. Furthermore, we note that on the $\beta$ and $\rho$ parameter inference tasks, the Stable (Canonical) regime incurs an MAPE error curve that is significantly lower (even on the log-scale) than that of the other three regimes, which have curves that are effectively on top of each other. The first implication of this observation is that we can deduce that the Stable (Canonical) regime's parameters are the most identifiable compared to the other three regimes.

The even more interesting implication is that as measured via MAPE, the Stable (Transient Chaos) regime's parameters are equally identifiable as those of the Chaotic (Butterfly) and Chaotic (No Butterfly) regimes'. This is a very pivotal observation. Visually, we know that the Stable (Transient Chaos) regime looks quite similar to the Chaotic (Butterfly) regime from $t = 0$ to $t = 10$. Yet, mathematically, we also know, as their names imply, that one regime is mathematically stable, while the other regime is mathematically chaotic. Putting the pieces together, we hypothesize that **Lorenz system identifiability is *not* necessarily tied to the mathematical stability or chaos of a system, but rather the shape of the system's observed trajectory itself.** Indeed, the Stable (Canonical) regime looks much different visually than the other three regimes, in the sense that it converges to a fixed point very clearly by $t = 10$. Though, in the spirit of solid science, we must emphasize that this statement is only a hypothesis, and much more extensive testing and ablation studies across a much larger set of test bed regimes will be necessary to further explore this statement.

## 5.5 Main Takeaways

In this chapter, we demonstrated that pMAGI can achieve comparable parameter inference performance to that of PSO, DE, and PINN. This is especially remarkable because pMAGI neither requires computationally-expensive numerical integration nor automatic differentiation. From a compute efficiency standpoint, pMAGI certainly presents itself in a very favorable light. In addition, a more thorough examination of pMAGI's error curves also raises some exciting new inquiries abut parameter identifiability at the intersection of Bayesian inverse problem methods and traditional dynamical systems theory.

In the next chapter, we will explore how pMAGI's outputed HMC samples can be used to probabilistically infer whether a Lorenz system is stable or chaotic, based only on noisy and sparse observations.

# 6
# Probabilistic Classification of Stability

In the previous chapter, we explored pMAGI's parameter inference capabilities, benchmarked against three non-probabilistic competitor methods. One natural question to ask is — what is so good about having a probabilistic, sampling-based method like pMAGI? One obvious answer is that instead of point estimates of our parameters $\theta$, we can provide an uncertainty-quantified estimate of $\theta$ using a posterior mean and standard deviation computed from our samples. But even such uncertainty-quantified estimates are not unique to a probabilistic, sampling-based method like pMAGI. Variational Bayes [48] methods that approximate the posterior distribution of $\theta$ can also provide uncertainty-quantified parameter estimates.

In this chapter, we demonstrate that one can combine the posterior HMC $\hat{\theta}$ samples from pMAGI with dynamical systems theory about the Lorenz system to construct an estimator of the probability that an underlying trajectory is stable, given only noisy and sparse observations from the system.

## 6.1 A Bayesian Stability Probability Estimator

Specifically, recall from our discussion in Chapter 1 that, for the Lorenz system, if $0 < \rho < 1$, then the origin is a stable global attractor[45]. Also, if the following condition is satisfied, then our system will have two stable stationary points[45]:

$$1 < \rho < \sigma \frac{\sigma + \beta + 3}{\sigma - \beta - 1}.$$

Using our straightforward definition of stability, we can conclude that a given Lorenz system will be stable if

$$\rho < \sigma \frac{\sigma + \beta + 3}{\sigma - \beta - 1}.$$

Suppose we have $N_{\text{samp}}$ HMC samples $\{\hat{\theta}^{(1)}, \ldots, \hat{\theta}^{(N_{\text{samp}})}\}$ from an application of pMAGI on some set of sparse and noisy observations $\mathcal{X}_{obs}$ and times $\tau_{obs}$, which we know came from some Lorenz system with unknown parameters $\theta = (\beta, \rho, \sigma)$. Specifically, let $\hat{\theta}^{(i)} = (\hat{\beta}^{(i)}, \hat{\rho}^{(i)}, \hat{\sigma}^{(i)})$ represent the $i^{th}$ HMC sample of $\theta$ from this run of pMAGI. To investigate whether our underlying system is stable, we could compute point estimates $(\hat{\beta}, \hat{\rho}, \hat{\sigma})$ by taking the posterior mean of our HMC samples and then checking whether

$$\hat{\rho} < \hat{\sigma} \frac{\hat{\sigma} + \hat{\beta} + 3}{\hat{\sigma} - \hat{\beta} - 1}.$$

However, using this method, we ignore the uncertainty on our prediction of whether the system is stable or chaotic — we only have a binary answer. An alternative is to approach the problem from a Bayesian, probabilistic setup. Using our HMC samples, we can construct the following estimator:

$$\Pr(\text{stable} \mid \mathcal{X}_{obs}, \tau_{obs}) = \mathbb{E}\left(\mathbf{I}\left(\rho < \sigma\frac{\sigma+\beta+3}{\sigma-\beta-1}\right) \mid \mathcal{X}_{obs}, \tau_{obs}\right) \approx \frac{1}{N_{\text{samp}}} \sum_{i=1}^{N_{\text{samp}}} \mathbf{I}\left(\hat{\rho}^{(i)} < \hat{\sigma}^{(i)}\frac{\hat{\sigma}^{(i)}+\hat{\beta}^{(i)}+3}{\hat{\sigma}^{(i)}-\hat{\beta}^{(i)}-1}\right),$$

where $\mathbf{I}(\dots)$ is an indicator function. Technically speaking, there may be some pathological Lorenz regime that is indeed stable, but whose $(\beta, \rho, \sigma)$ do not satisfy the aforementioned condition, but for the purposes of this proof-of-concept, we will use the equal sign conditional on the understanding that, technically, the equal sign should be an less-than-or-equal-to sign. This estimator for $\Pr(\text{stable} \mid \mathcal{X}_{obs}, \tau_{obs})$ demonstrates the utility of the HMC sampling in our pMAGI method — PSO, DE, and PINN are all incapable of joining this framework.

Intuitively, this estimator is useful because we are working with sparse and noisy data. Heuristically, it is possible that there are multiple possible regimes that could serve as the underlying ground truth of our noisy data. The above Bayesian estimator naturally captures this uncertainty. However, we still need to investigate to what extent this estimator actually works well in reality and agrees with the underlying ground truth. We do this below.

## 6.2 Experimental Validation

We retain the pMAGI experimental setup and data from Section 5.4, but focus only on $D_{pilot} = 1$ and $T_{max,\text{pilot}} = 1.0$ to avoid repetitiveness, given that pMAGI performance does not change significantly with our hyperparameter choices (we will elaborate on this more in the next chapter). For each pMAGI run, we will compute our Bayesian estimate of the stability probability of the system. In the figures below, we will plot the means of these stability probabilities across ten randomly-seeded datasets at each noise level multiplier $\alpha$.

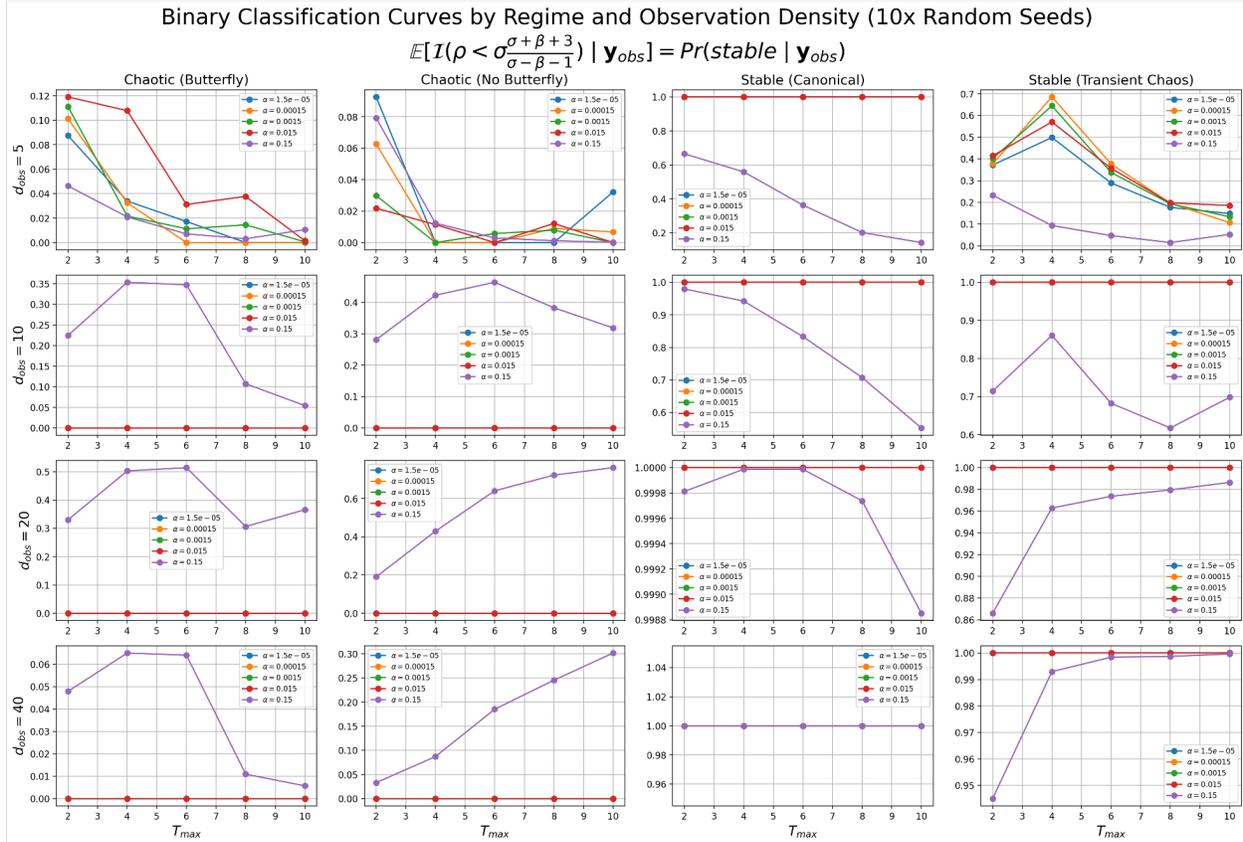

**Figure 6.1:** pMAGI mean estimates of stability probability by regime (column) and $d_{obs}$ (row). The x-axis is the $T_{max}$, the length of our noisy observation interval $t \in [0, T_{max}]$. The y-axis is the mean estimated stability probability averaged across ten randomly-seeded trials. To reduce notational clutter, $(\mathcal{X}_{obs}, \tau_{obs})$ was abbreviated to $\mathbf{y}_{obs}$. $\alpha = 0.15$ corresponds to purple, $\alpha = 0.015$ to red, $\alpha = 1.5 \times 10^{-3}$ to green, $\alpha = 1.5 \times 10^{-4}$ to orange, and $\alpha = 1.5 \times 10^{-5}$ to blue.

Looking at Figure 6.1, if our Bayesian stability probability estimate implied a perfect classifier, then we should see all curves in the Chaotic (Butterfly) and Chaotic (No Butterly) regime subplots at 0.0 (i.e., the estimated probability of these regimes being stable should be 0.0). Analogously, under these ideal conditions, we should see all curves for the Stable (Canonical) and Stable (Transient Chaos) regimes at 1.0 (i.e., the estimated probability of these regimes being stable should be 1.0). With this intuition, we observe that for all noise level multiplier levels $\alpha < 0.15$, their resultant curves seem to match this ideal behavior in 13 out of the 16 possible combinations of regime and observation density $d_{obs}$. The only outliers to this behavior are the Chaotic (Butterfly), Chaotic (No Butterfly), and Stable (Transient Chaos) regimes at observation density $d_{obs} = 5$. Intuitively, it makes sense that these three settings would manifest the most deviant behavior because $d_{obs} = 5$ entails having the least signal possible, and we know from our discussions in the previous chapter that the Stable (Canonical) regime is the most identifiable with respect to parameter inference.

Looking more closely at the y-axes, for all $\alpha$ values at $d_{obs} = 5$ on the Chaotic (Butterfly) and Chaotic (No Butterfly) regimes, we observe that the estimated stability probability curves are all quite close to 0.0 already, and that the estimated stability probability curves, for the most part, seem to decrease as $T_{max}$ increases. Based on these observations, one hypothesis is that when pMAGI is uncertain about whether a regime is stable or chaotic (as opposed to having probability values all at 0.0 or 1.0), increasing the observation interval length via $T_{max}$ increases our estimated stability classification confidence, and correctly so.

However, this hypothesis seems to encounter some opposition when we look at the $\alpha < 0.15$ stability probability curves for the Stable (Transient Chaos) case at $d_{obs} = 5$. Here, we would ideally like all curves to have value 1.0, but it appears that our $\alpha < 0.15$ curves increase in estimated stability probability when increasing $T_{max} = 2$ to $T_{max} = 4$ (which would make sense), but then decrease uniformly from $T_{max} = 4$ to $T_{max} = 10$. Before analyzing too much into this observation, it is possible that we simply do not have enough signal to accurately classify the Stable (Transient Chaos) regime based on only $d_{obs} = 5$ observations per unit time. However, it is also possible that our findings suggest that some intervals are more informative of stability or chaos than others. Specifically, it is possible that the interval $t \in [0, 4.0]$ is just the most informative towards classifying the noisy observed trajectories from the Stable (Transient) Chaos regime as indeed being stable. If this is the case, then there is much promising future work in using pMAGI to augment theoretical dynamical systems work on analyzing stable vs. chaotic dynamics — specifically, in the identification of the most informative intervals and/or features of a system that give off chaotic or stable signatures. Nonetheless, these discussions are merely hypotheses — hypotheses that need much more extensive testing and evaluation across a much larger set of test bed systems.

Now, we turn our attention to the $\alpha = 0.15$ curves (our highest noise level multiplier setting), shown via the purple lines in each subplot. On the Chaotic (Butterfly) regime, for $d_{obs} \in \{10, 20, 40\}$, we observe that the general shape of the $\alpha = 0.15$ curve seems relatively fixed. The peaks and plateaus at $T_{max} = 4$ and $T_{max} = 6$ potentially suggest that these two intervals provide the least useful information regarding whether the system is stable or chaotic, if we believe our previous hypotheses to be true. For the Chaotic (No Butterfly) regime, we observe that, for $d_{obs} \in \{10, 20, 40\}$, our purple estimated stability probability curves increase steadily for $d_{obs} \in \{20, 40\}$ *in the wrong direction*, leading us to believe that the "no butterfly" portion of the Chaotic (No Butterfly) regime may actually mislead pMAGI. At $d_{obs} = 10$, we see an increase in the estimated stability probability from $T_{max} = 2$ to $T_{max} = 6$, followed by an equal-magnitude decrease — this is likely an artifact of $\alpha = 0.15$ being too high of a noise level for pMAGI to extract useful signal from. On the Stable (Canonical) regime, for $d_{obs} \in \{5, 10, 20\}$, we see that, in general, estimated stability probability actually decreases with increasing $T_{max}$. Taking into account what we know about the Stable (Canonical) regime, it is likely that noise level multiplier $\alpha = 0.15$ degrades too much of the already-sparse signal at the effectively-flatlined portions of the trajectory. However, this hypothesis encounters some opposition at $d_{obs} = 40$, where the $\alpha = 0.15$ matches the ideal: this is likely because by doubling our observation density, we can also evaluate our error term $W$ across more time points and cut through the noise more effectively.

Finally, on the Stable (Transient Chaos) regime, we observe that for $\alpha = 0.15$ at $d_{obs} \in \{20, 40\}$, a quick inspection of the y-axis values suggests that our classification accuracy is quite good, and that **pMAGI was able to recognize that the Stable (Transient Chaos) regime, despite its initial trajectory looking visually-similar to the Chaotic (Butterfly) regime, was indeed stable, with very high confidence.** At $d_{obs} = 10$, our classification accuracy (as measured through confidence) oscillates around 70%, which is still pretty strong, given the sparse observations. At $d_{obs} = 5$, pMAGI becomes too confident ... in the wrong direction. It is possible that pMAGI actually become confused by the chaotic-looking initial trajectories of the Stable (Transient Chaos) regime.

One final curiosity we had was that perhaps pMAGI's systemic bias on underestimating $\sigma$ was a root cause of the observed deviations from the ideal curves. To test this hypothesis, we recomputed the stability probability estimates for each setting, but manually hard-coded $\sigma$ to its true value of 10.0 for each sample. We relegate the resultant curves to Appendix A.7 because such $\sigma$-correction did not always have a corrective effect on our probabilistic binary classification, with overall insignificant effect. In fact, sometimes $\sigma$-correction would only exacerbate uncertainty ... in the wrong direction.

## 6.3 Main Takeaways

Taking all of these observations into consideration, we can make a few conclusions and/or hypotheses. First, overall, for $\alpha < 0.15$, our Bayesian stability probability estimate is quite well-aligned and reflective of the ground truth. For many not-already-ideal settings at $\alpha < 0.15$, we also observe that increasing our observation length $T_{max}$ increases both accuracy and confidence. The uncertainty as indicated through the estimated stability probability curves of $\alpha = 0.15$ also matches our intuition that $\alpha = 0.15$ is simply too noisy and degrading too much of the signal. Our analyses above also suggest potential hypotheses that some observation intervals are more indicative of chaotic and/or stable signatures than others, making pMAGI a potentially useful tool to deploy in complement with theoretical dynamical systems analyses.

In any case, this chapter serves as an overall successful proof-of-concept of the utility of sampling-based methods like pMAGI for probabilistic stability classification, based only on noisy and sparse observations. In the next chapter, we will explore pMAGI's ability to reconstruct the unnoised trajectory of a system, compared to the overparameterized Physics-Informed Neural Network.

# 7
# Trajectory Reconstruction

In this chapter, we explore pMAGI's trajectory reconstruction performance, as compared against Physics-Informed Neural Networks (PINNs). Specifically, the question we are interested in is as follows: given noisy and sparse observations from an ODE-based dynamical system, can we reconstruct what the underlying ground-truth, unnoised trajectory would have looked like on our interval of observation? Put another way, in this chapter, we aim to explore pMAGI's in-sample interpolation abilities. Overall, we find that pMAGI performs very comparably to PINN, significantly outperforming its overparameterized competitor in many settings.

## 7.1 Experimental Setup

For generating all results in this chapter, we carry over our experimental setups for pMAGI and PINN from Section 5.1, with noise level multiplier $\alpha = 0.15$ and ten randomly-seeded datasets shared between both methods per setting. In all tables and figures, we will show the performances achieved by the best pMAGI and PINN hyperparameter variants. Following our practices in Section 4.3, to quantify trajectory reconstruction performance, let $\mathcal{X}_{true} = (\mathbf{X}_{true}, \mathbf{Y}_{true}, \mathbf{Z}_{true})$ refer to the true, unnoised trajectories of the relevant regime, where $\mathbf{X}_{true} = \{X(t) \mid t \in \tau_{inf}\}$ and $\mathbf{Y}_{true}, \mathbf{Z}_{true}$ are defined accordingly. Recall that $\tau_{inf}$ is the set of times where we are interested in our reconstructed trajectories.

Analogously, let $\hat{\mathcal{X}}_{inf} = (\hat{\mathbf{X}}_{inf}, \hat{\mathbf{Y}}_{inf}, \hat{\mathbf{Z}}_{inf})$ represent our reconstructed trajectory values for a given model variant, where $\hat{\mathbf{X}}_{inf} = \{\hat{X}_{inf}(t) \mid t \in \tau_{inf}\}$ and $\hat{\mathbf{Y}}_{inf}, \hat{\mathbf{Z}}_{inf}$ are defined accordingly. For PINNs, for each time $t \in \tau_{inf}$, $\hat{X}_{inf}(t)$ represents the point estimate outputted by the neural network at time $t$ for component $X$. For pMAGI, we let $\hat{X}_{inf}(t)$ be the posterior mean computed over all HMC samples. For an apples-to-apples comparison, we will only compare the point estimates of pMAGI versus PINN, as PINN does not have the capacity for generating uncertainty-quantified estimates. Like in Section 4.3, we quantify trajectory reconstruction performance using *scaled mean absolute error* (sMAE), defined for component $X$ as

$$L_X(\hat{\mathbf{X}}_{inf}; \mathbf{X}_{true}) = \frac{1}{|\tau_{inf}|} \sum_{t \in \tau_{inf}} \frac{|\hat{X}_{inf}(t) - X(t)|}{|X(t)|},$$

with errors on components $Y$ and $Z$ defined accordingly.

We begin by introducing some representative examples to emphasize the salient differences between pMAGI and PINN's performances. Then, we proceed to analyze pMAGI and PINN's performances in aggregate.

## 7.2 Selected Representative Examples

In this section, we provide a few illustrative examples of how pMAGI and PINN's outputed reconstructed trajectories differ *visually*, emphasizing the general performance advantages and disadvantages of the two methods. To improve readability and not dilute from the main message, we will show truncated figures in the main text (i.e., only showing one or two components and/or certain settings). For full transparency, because many of the full figures (and animations) are too large for even the appendices, we invite the reader to explore the full set of figures, stitched together into GIFs, at [bit.ly/PINN-vs-pMAGI-TrajRecons](bit.ly/PINN-vs-pMAGI-TrajRecons). For each truncated figure in the main text, we will also provide an immediate link to the relevant GIF. For clarity, the legends of the truncated and full figures and/or GIFs are identical.

Starting in order with the Stable (Canonical) regime in Figure 7.1, we observe that even though the Stable (Canonical) regime is the easiest to reconstruct, for many hyperparameter settings, PINNs have very high variance across randomly-seeded datasets. Specifically, at $\lambda = 100$ and $\lambda = 1000$, we see many volatile green lines, each corresponding to the reconstructed trajectory for a particular trial. This is likely an artifact of PINNs being overparameterized and not learning smooth, interpretable functional representations. In contrast, on all hyperparameter settings shown, our pMAGI variants have virtually no visible green lines — they are all overlapping with the lavender mean of the reconstructed trajectories, suggesting very minimal variance across trials. Thus, we argue that **pMAGI is less variable than PINN across trials.**

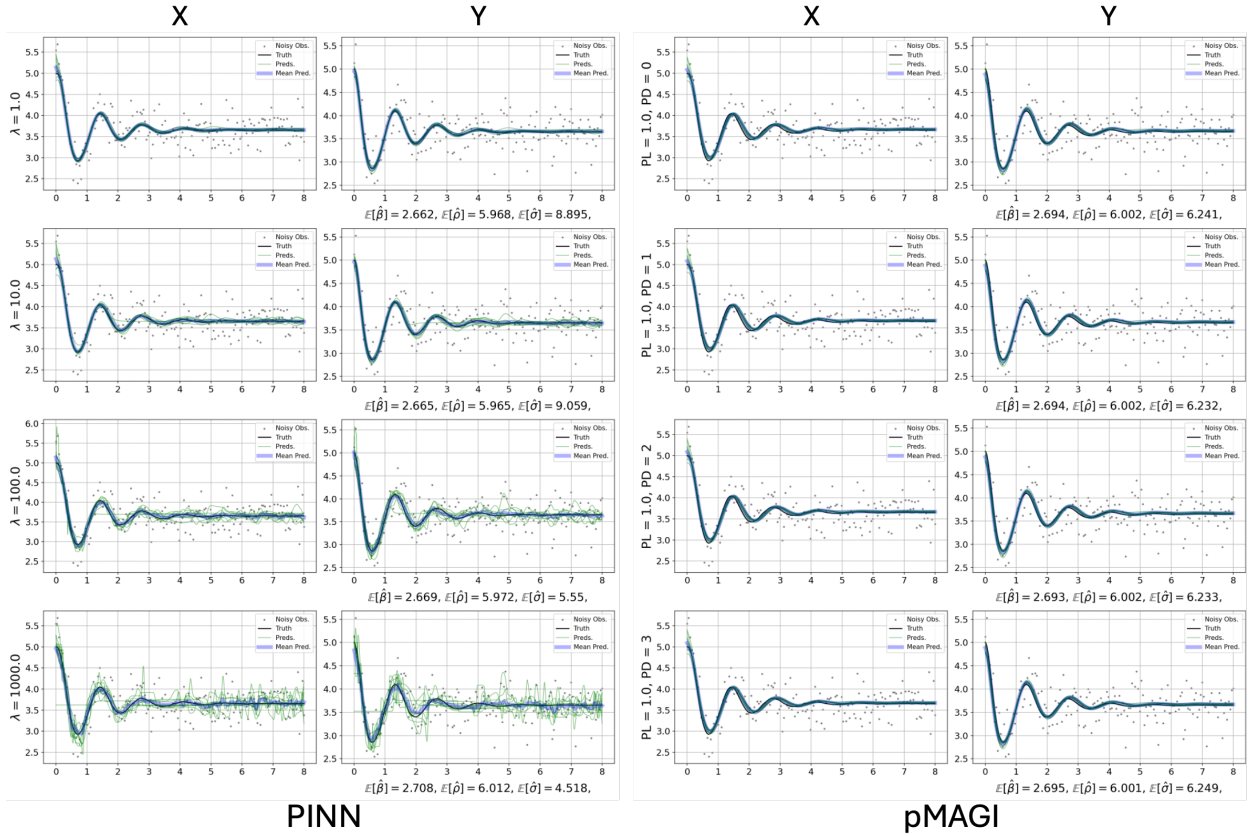

**Figure 7.1:** Selected PINN vs. pMAGI reconstructed trajectories on Stable (Canonical) with $T_{max} = 8.0, d_{obs} = 20$, and $T_{max,\text{pilot}} = 1.0$. The left two columns show the PINN's $X$ and $Y$ reconstructed trajectories, while the right two columns show pMAGI's $X$ and $Y$ reconstructed trajectories. Each row of our subplot grid corresponds to a different hyperparameter setting for PINN (the loss-type weighting $\lambda$) and pMAGI (the pilot discretization $D_{pilot}$). For each subplot, the x-axis is time $t$ from $t = 0$ to $t = T_{max}$. In each figure, the lavender curve represents the mean of the reconstructed trajectories across the ten randomly-seeded datasets, common to both models. The green lines represent individual reconstructed trajectories corresponding to each of the ten randomly-seeded datasets. The grey dots show the noisy and sparse observations, while the black line represents the ground truth. See the full GIF here.

Moving on to the Stable (Transient Chaos) regime, we see from Figure 7.2 that at a moderate observation density of $d_{obs} = 20$, all shown pMAGI variants match or outperform their PINN counterparts at $T_{max} = 4.0$. Yes, the PINN at $\lambda = 10.0$ looks accurate and robust, but the fact that only one of out of the four PINN hyperparameter variants shown succeeded, while all four of the pMAGI hyperparameter variants shown succeeded emphasizes another salient point: **pMAGI is much more robust to changes in hyperparameter tuning than**

**PINN, and is thus more practically-deployable for application settings.** Yes, the PINN at $\lambda = 100.0$ also seems to have a fairly accurate mean reconstructed trajectory, but the visible green lines suggest that there is still significant variance across trials — a problem that pMAGI does not have on any of the displayed settings.

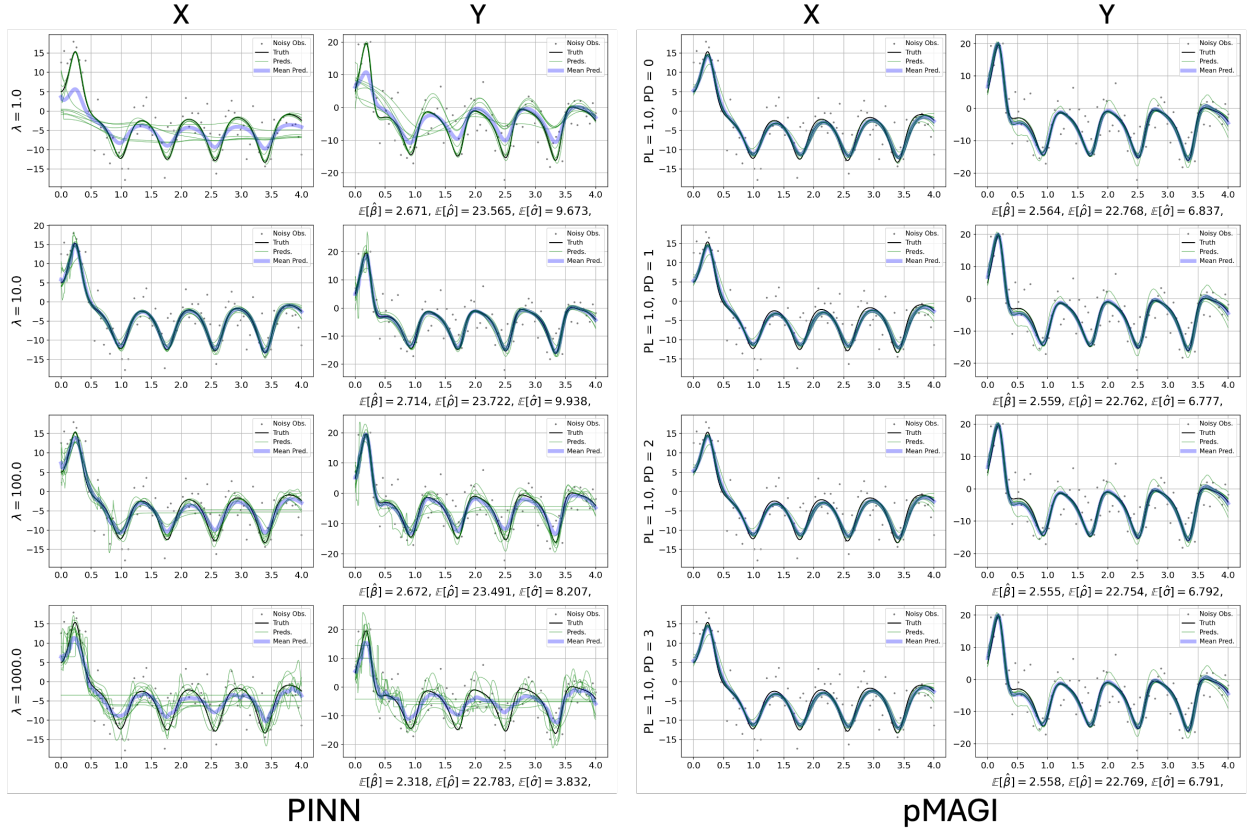

**Figure 7.2:** Selected PINN vs. pMAGI reconstructed trajectories on Stable (Transient Chaos) with $T_{max} = 4.0, d_{obs} = 20$, and $T_{max,\text{pilot}} = 1.0$. The left two columns show the PINN's $X$ and $Y$ reconstructed trajectories, while the right two columns show pMAGI's $X$ and $Y$ reconstructed trajectories. Each row of our subplot grid corresponds to a different hyperparameter setting for PINN (the loss-type weighting $\lambda$) and pMAGI (the pilot discretization $D_{pilot}$). For each subplot, the x-axis is time $t$ from $t = 0$ to $t = T_{max}$. In each figure, the lavender curve represents the mean of the reconstructed trajectories across the ten randomly-seeded datasets, common to both models. The green lines represent individual reconstructed trajectories corresponding to each of the ten randomly-seeded datasets. The grey dots show the noisy and sparse observations, while the black line represents the ground truth. See the full GIF here.

Remaining in the Stable (Transient Chaos) regime, but doubling $T_{max}$ to 8.0 and halving $d_{obs}$ to 10, the story is still relatively unchanged. Specifically, we see in Figure 7.3 that all four pMAGI variants shown clearly match or outperform all PINN variants shown, with respect to both accuracy (mean trajectory vs. ground truth) and variance (visibility of the green lines). Yes, the $\lambda = 10.0$ PINN variant performed well, but the mere fact that only one out of the four PINN variants shown succeeded, while all four pMAGI variants shown succeeded, again emphasizes the fact that **pMAGI is much less sensitive to hyperparameter choices than PINN**.

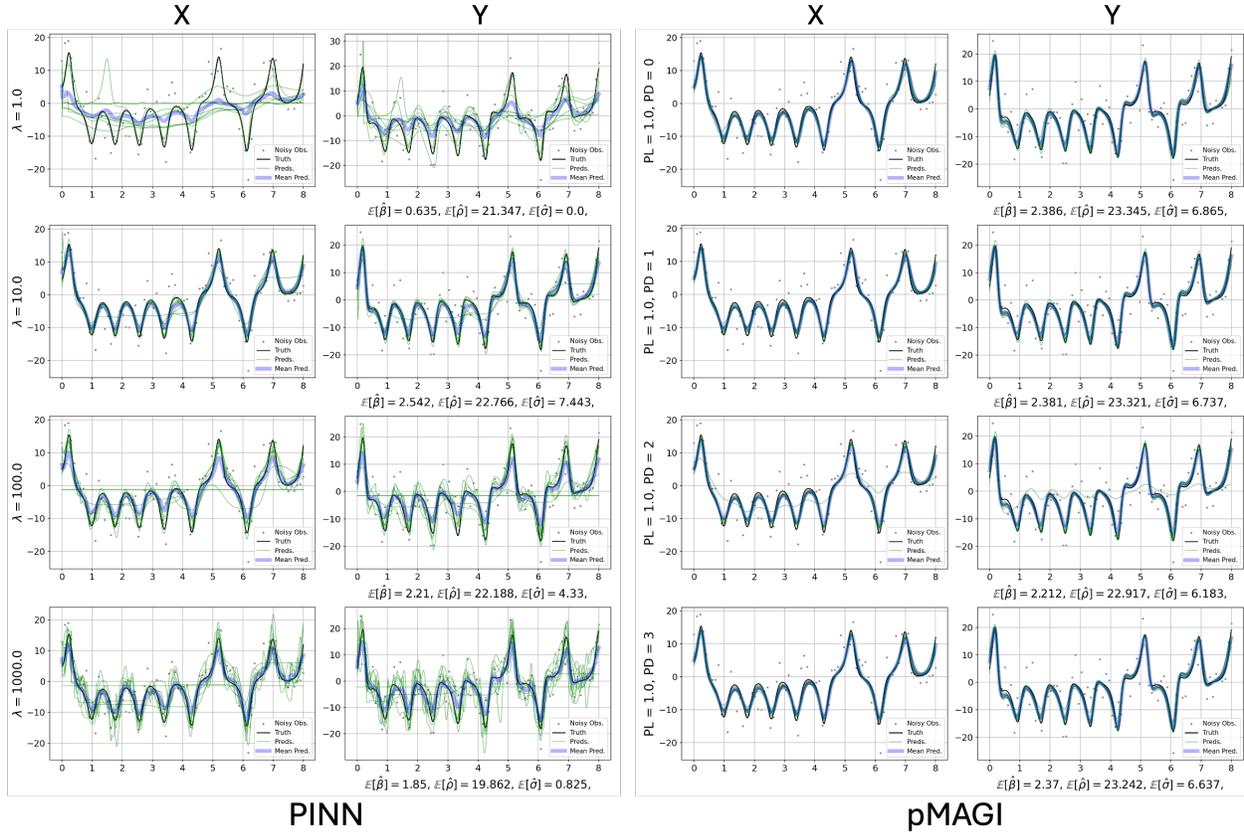

**Figure 7.3:** Selected PINN vs. pMAGI reconstructed trajectories on Stable (Transient Chaos) with $T_{max} = 8.0, d_{obs} = 10$, and $T_{max,\text{pilot}} = 1.0$. The left two columns show the PINN's $X$ and $Y$ reconstructed trajectories, while the right two columns show pMAGI's $X$ and $Y$ reconstructed trajectories. Each row of our subplot grid corresponds to a different hyperparameter setting for PINN (the loss-type weighting $\lambda$) and pMAGI (the pilot discretization $D_{pilot}$). For each subplot, the x-axis is time $t$ from $t = 0$ to $t = T_{max}$. In each figure, the lavender curve represents the mean of the reconstructed trajectories across the ten randomly-seeded datasets, common to both models. The green lines represent individual reconstructed trajectories corresponding to each of the ten randomly-seeded datasets. The grey dots show the noisy and sparse observations, while the black line represents the ground truth. See the full GIF here.

In the spirit of trustworthy science, we must tell the full story, including edge cases. Figure 7.4 shows PINN

and pMAGI's resultant reconstructed trajectories if we keep all settings from Figure 7.3, but halve $d_{obs}$ again down to just 5 noisy observations per unit time. In these scenarios, pMAGI fails to outperform PINN. Indeed, every hyperparameter setting of pMAGI shown fails to capture the ground truth even when computing mean trajectories across trials, while three out of the four PINN variants shown are able to capture the black ground truth curve with their lavender mean trajectory curves. Looking at pMAGI's $Y$ component trajectories, we realize that under low observation density settings like $d_{obs} = 5$, pMAGI may have a tendency to flat-line in its reconstructed trajectories. This is likely because the constant solution does satisfy the Lorenz equations, albeit trivially.

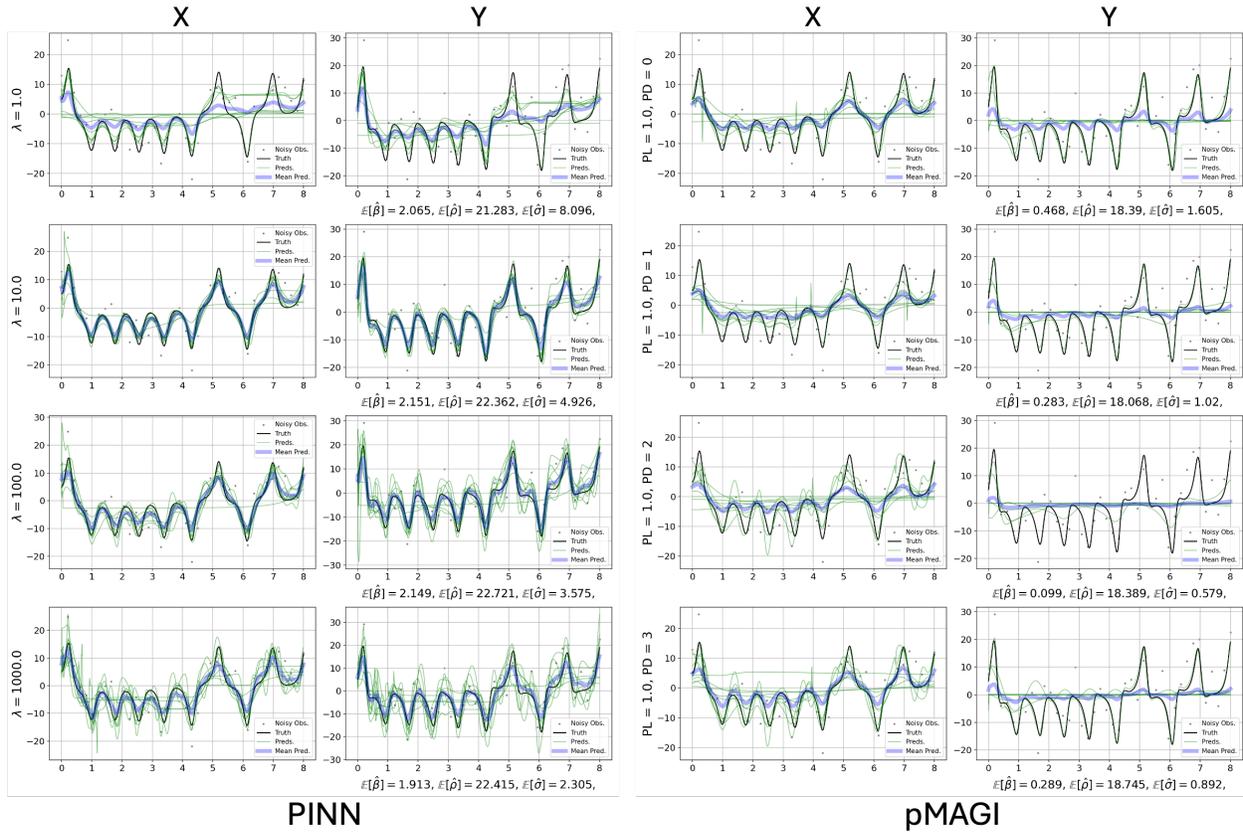

**Figure 7.4:** Selected PINN vs. pMAGI reconstructed trajectories on Stable (Transient Chaos) with $T_{max} = 8.0$, $d_{obs} = 5$, and $T_{max,\text{pilot}} = 1.0$. The left two columns show the PINN's $X$ and $Y$ reconstructed trajectories, while the right two columns show pMAGI's $X$ and $Y$ reconstructed trajectories. Each row of our subplot grid corresponds to a different hyperparameter setting for PINN (the loss-type weighting $\lambda$) and pMAGI (the pilot discretization $D_{pilot}$). For each subplot, the x-axis is time $t$ from $t = 0$ to $t = T_{max}$. In each figure, the lavender curve represents the mean of the reconstructed trajectories across the ten randomly-seeded datasets, common to both models. The green lines represent individual reconstructed trajectories corresponding to each of the ten randomly-seeded datasets. The grey dots show the noisy and sparse observations, while the black line represents the ground truth. See the full GIF here.

Nonetheless, taking both Figures 7.3 and 7.4 into account, one can extrapolate and conclude that the critical observation density $d_{obs}$ for successful pMAGI performance is somewhere between 5 and 10 noisy observations per unit time. By any standard, pMAGI is still a very data-efficient model. And, from the above figures, once pMAGI reaches that minimal $d_{obs}$ threshold, it quickly begins to meet, if not exceed, the PINN's performance.

Proceeding to the Chaotic (Butterfly) regime, we see in Figure 7.5 a scenario where pMAGI decisively outper-forms PINN. Specifically, at moderate observation density $d_{obs} = 20$ and long observation interval $T_{max} = 8.0$, all four of the pMAGI variants shown perform excellently on trajectory reconstruction, while none of the PINN variants are able to match this performance. This figure emphasizes our observation that **pMAGI will outperform PINN at moderate-to-high observation densities and long observation intervals.**

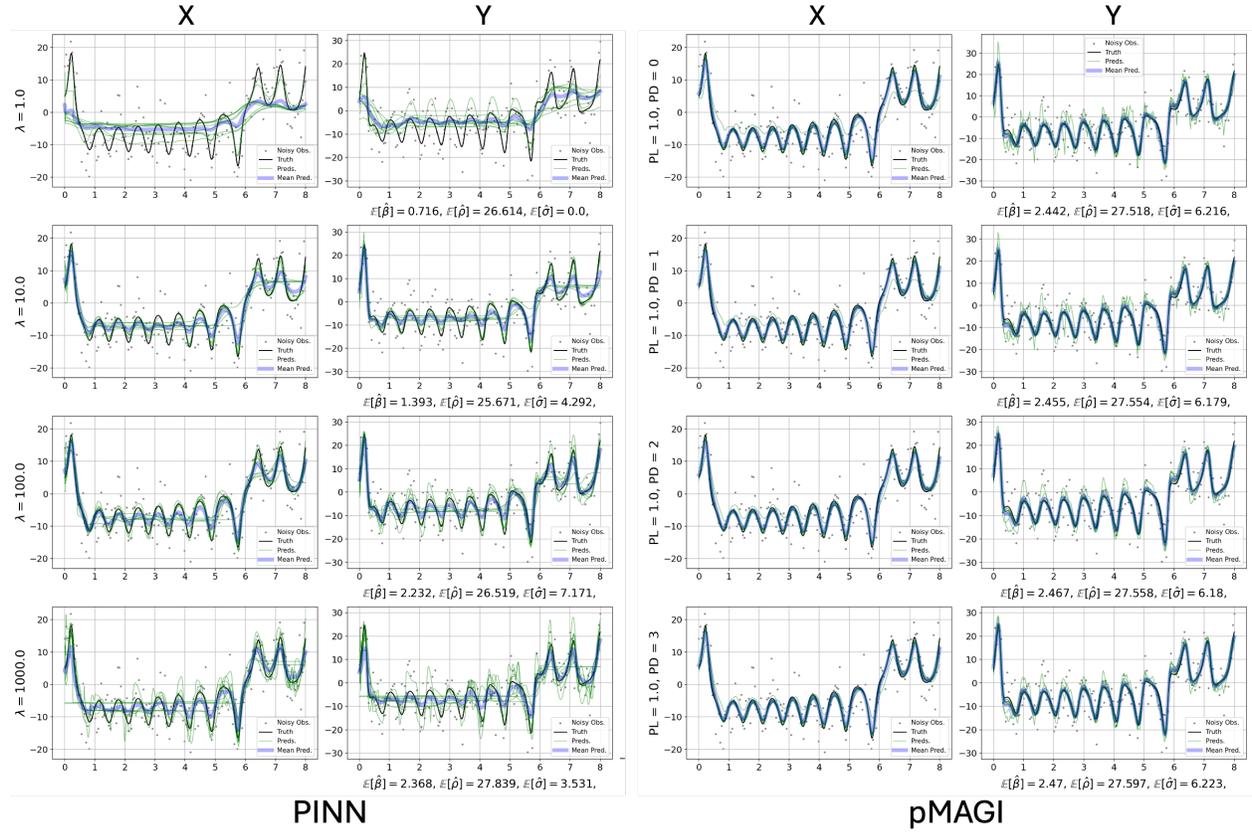

**Figure 7.5:** Selected PINN vs. pMAGI reconstructed trajectories on Chaotic (Butterfly) with $T_{max} = 8.0$, $d_{obs} = 20$, and $T_{max,\text{pilot}} = 1.0$. The left two columns show the PINN's $X$ and $Y$ reconstructed trajectories, while the right two columns show pMAGI's $X$ and $Y$ reconstructed trajectories. Each row of our subplot grid corresponds to a different hyperparameter setting for PINN (the loss-type weighting $\lambda$) and pMAGI (the pilot discretization $D_{pilot}$). For each subplot, the x-axis is time $t$ from $t = 0$ to $t = T_{max}$. In each figure, the lavender curve represents the mean of the reconstructed trajectories across the ten randomly-seeded datasets, common to both models. The green lines represent individual reconstructed trajectories corresponding to each of the ten randomly-seeded datasets. The grey dots show the noisy and sparse observations, while the black line represents the ground truth. See the full GIF here.

Finally, we move to the Chaotic (No Butterfly) regime. In Figure 7.6, we provide both PINN and pMAGI

with the highest signal possible: $d_{obs} = 40$ noisy observations per unit time, albeit on the smallest time interval possible with $T_{max} = 2.0$. Visually, we confirm that all pMAGI variants shown are able to outperform all PINN variants shown. The presence of all the green lines in the PINN figures suggests that even though the mean trajectories of PINN and pMAGI may be similar for some hyperparameter settings, PINN still has much higher variance than pMAGI. In contrast, pMAGI's green lines are barely visible in comparison.

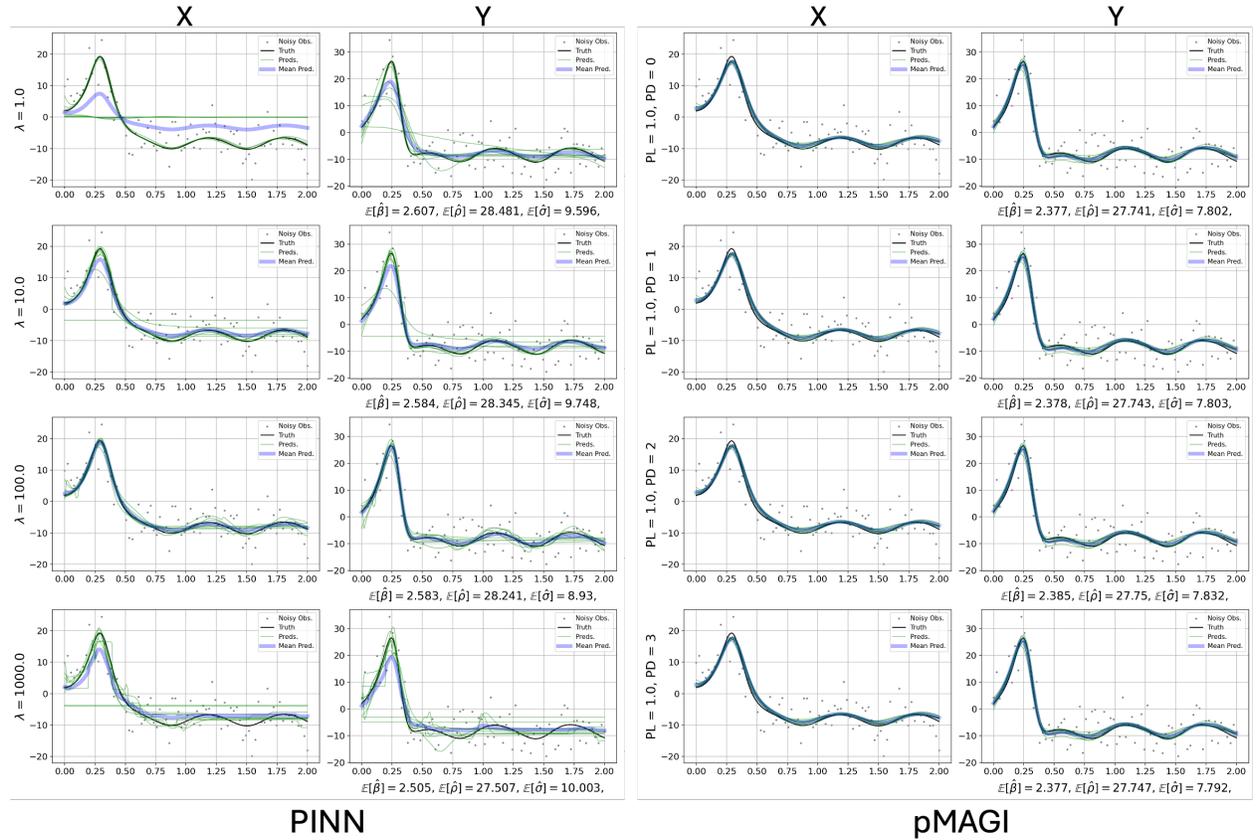

**Figure 7.6:** Selected PINN vs. pMAGI reconstructed trajectories on Chaotic (No Butterfly) with $T_{max} = 2.0$, $d_{obs} = 40$, and $T_{max, \text{pilot}} = 1.0$. The left two columns show the PINN's $X$ and $Y$ reconstructed trajectories, while the right two columns show pMAGI's $X$ and $Y$ reconstructed trajectories. Each row of our subplot grid corresponds to a different hyperparameter setting for PINN (the loss-type weighting $\lambda$) and pMAGI (the pilot discretization $D_{pilot}$). For each subplot, the x-axis is time $t$ from $t = 0$ to $t = T_{max}$. In each figure, the lavender curve represents the mean of the reconstructed trajectories across the ten randomly-seeded datasets, common to both models. The green lines represent individual reconstructed trajectories corresponding to each of the ten randomly-seeded datasets. The grey dots show the noisy and sparse observations, while the black line represents the ground truth. See the full GIF here.

## 7.3 Aggregate Benchmarks

Beginning with the Stable (Canonical) regime, we observe from Figure 7.7 that, overall, across all tested settings, pMAGI and PINN are both very successful at trajectory reconstruction on this regime, as indicated by the small y-axis values. We also observe that holding regime fixed, both models' trajectory reconstruction errors appear to decrease with increasing observation density $d_{obs}$. Crucially, we observe that for nearly all settings, with only a few exceptions, the box-and-whisker plots for pMAGI and PINN's errors overlap significantly. However, we do note that, upon visual inspection, PINN seems to incur on average slightly lower errors (by a few percentage points or even less than a percentage point) on most settings on this regime. In short, these performance differences are very minimal. At the same time, we must remember that PINN has hundreds if not thousands more parameters than pMAGI, which only requires parameterization of its mean and covariance functions. The PINN results shown below were outputed after $60K$ epochs of training, following van Herten et al.'s[49] example, while pMAGI variants were only run with $20K$ HMC steps total (accounting for the $4K$ HMC steps in the pilot module). **In terms of performance efficiency per parameter, it is undeniable that pMAGI emerges as the superior method.**

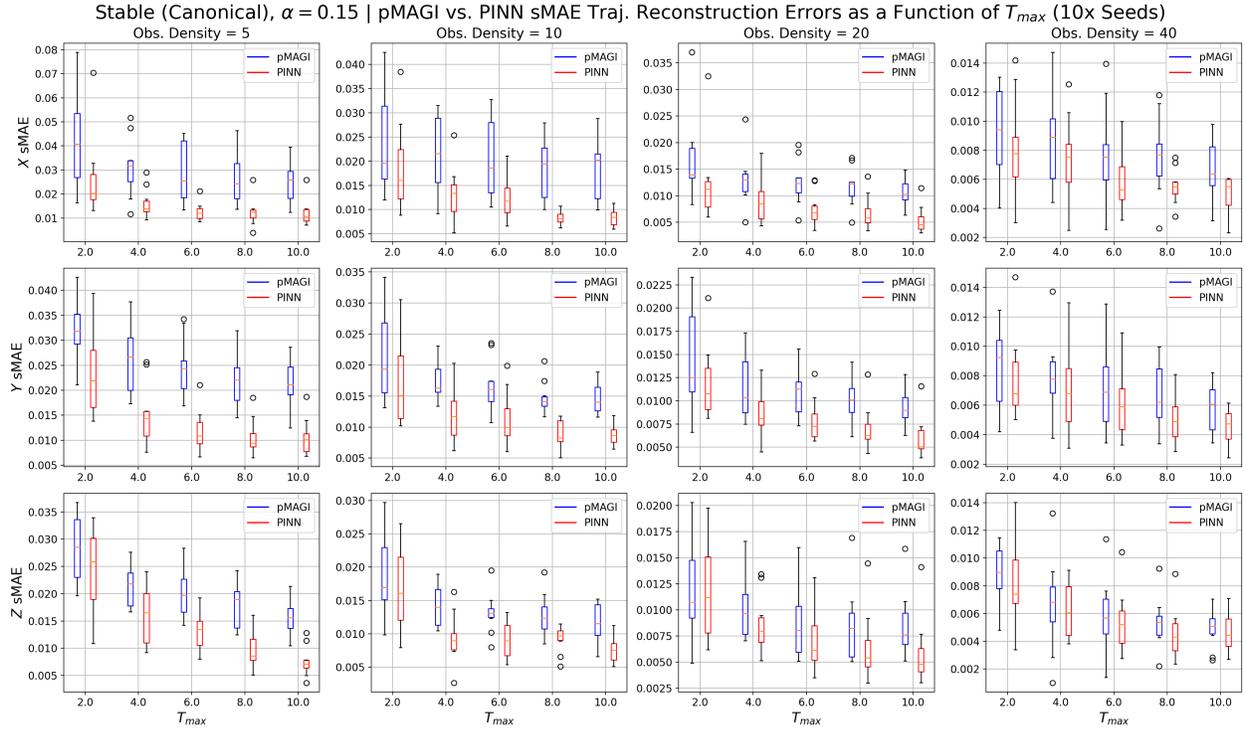

**Figure 7.7:** pMAGI vs. MAGI trajectory reconstruction on the Stable (Canonical) regime. Blue represents pMAGI, while red represents PINN. The x-axis of each subplot is $T_{max}$, the length of our noisy observation interval $t \in [0, T_{max}]$. Each row displays the sMAEs for a given component $X$, $Y$, or $Z$. Each column shows the models' performances holding the density of noisy observations per unit time $d_{obs}$ fixed. The box-and-whisker plots capture the error distributions across the ten randomly-seeded datasets.

Moving on to the Stable (Transient Chaos) regime, we observe from Figure 7.8 that this regime is objectively more difficult to perform trajectory reconstruction on than the Stable (Canonical) regime, as evidenced by the much larger values on the y-axes. Unpacking our results, we observe that, on the $X$ component, PINN appears to outperform pMAGI at $d_{obs} = 5$. By $d_{obs} = 10$ and $d_{obs} = 20$, the error distributions of the two models are effectively side-by-side. By $d_{obs} = 40$, however, while pMAGI experiences a major numerical instability issue at $T_{max} = 4.0$, PINN experiences significantly larger errors at $T_{max} = 8.0$ and $T_{max} = 10.0$.

On the $Y$ component, we observe that at almost all observation densities $d_{obs}$ and interval lengths $T_{max}$, pMAGI incurs lower, if not significantly lower, errors than PINN. Most notably, if we focus on the $d_{obs} = 40$ subplot, we observe that while the pMAGI box-and-whisker plots are so concentrated near 0 (except for $T_{max} = 4.0$), the corresponding PINN box-and-whisker plots are significantly higher, not even accounting for PINN's outlier errors on certain trials. On the $Z$ component, we recognize that PINN outperforms pMAGI at $d_{obs} = 5$, but this performance delta quickly dissipates by $d_{obs} = 10$ and $d_{obs} = 20$, with pMAGI significantly outperforming PINN on the longer time intervals of $T_{max} = 8.0$ and $10.0$, on $d_{obs} = 20$ and $40$.

Overall, the main takeaway on the Stable (Transient Chaos) regime is that while pMAGI may have a disadvantage against PINNs on low observation densities (i.e., $d_{obs} = 5$), pMAGI will perform on par with if not better than PINN with sufficient observation density, especially when reconstructing longer time intervals like $T_{max} = 8.0$ and $10.0$, echoing our conclusions in the previous section. Again taking into consideration that pMAGI is much more parameter-efficient than PINN, pMAGI more than presents itself in a favorable light.

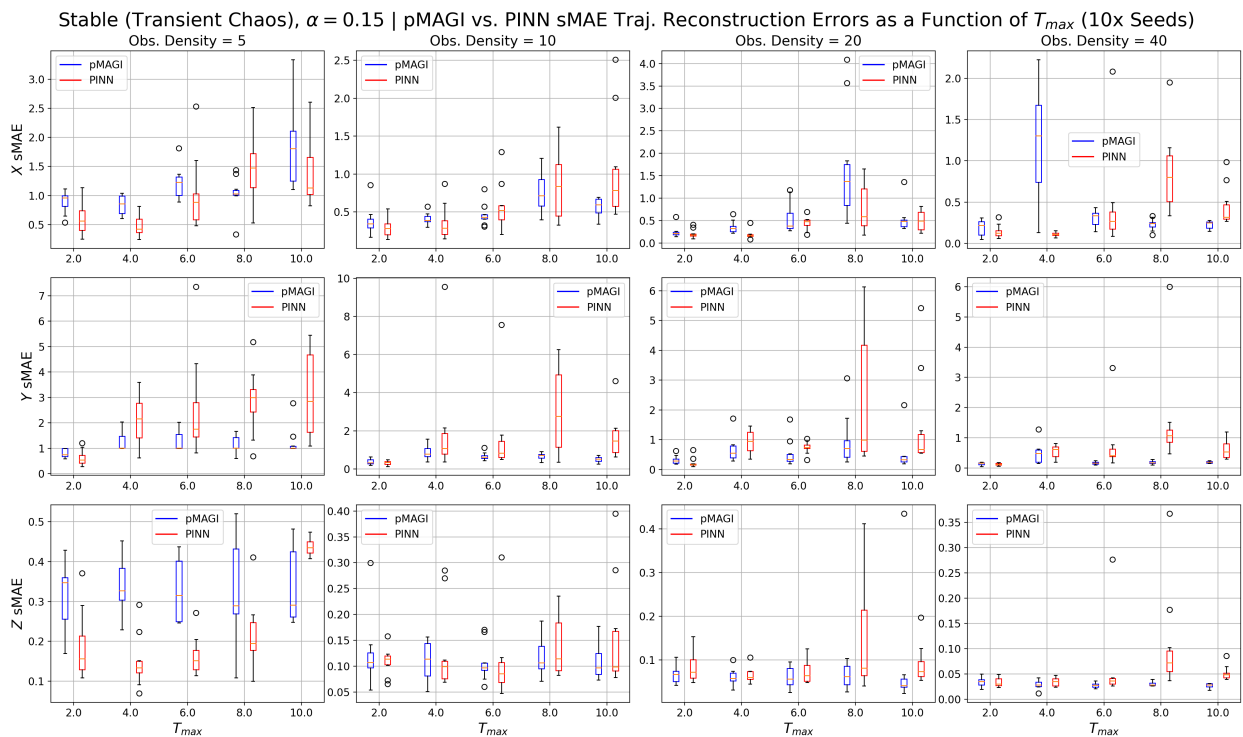

**Figure 7.8:** pMAGI vs. PINN trajectory reconstruction on the Stable (Transient Chaos) regime. Blue represents pMAGI, while red represents PINN. The x-axis of each subplot is $T_{max}$, the length of our noisy observation interval $t \in [0, T_{max}]$. Each row displays the sMAEs for a given component $X$, $Y$, or $Z$. Each column shows the models' performances holding the density of noisy observations per unit time $d_{obs}$ fixed. The box-and-whisker plots capture the error distributions across the ten randomly-seeded datasets.

Looking at the Chaotic (Butterfly) regime in Figure 7.9, we observe a somewhat similar story as the Stable (Transient Chaos) regime. Specifically, at $d_{obs} = 5$, pMAGI overall loses out to PINN, but does incur lower errors on $Y$ at $T_{max} = 6.0, 8.0,$ and $10.0$. By the time we reach $d_{obs} = 10$ and $d_{obs} = 20$, we observe that the performance gap has shrunken considerably, with pMAGI pretty consistently outperforming PINN at $T_{max} = 8.0$ and $10.0$, and even sometimes at smaller $T_{max}$ values. At $d_{obs} = 40$, pMAGI seems to outperform PINN on all components and interval lengths, with the sole exception of $T_{max} = 2.0$ on the $X$ component. Indeed, our overall story becomes more apparent: **pMAGI will match or outperform PINN at moderate to high observation densities $d_{obs}$, with pMAGI's largest performance advantages on long observation intervals.**

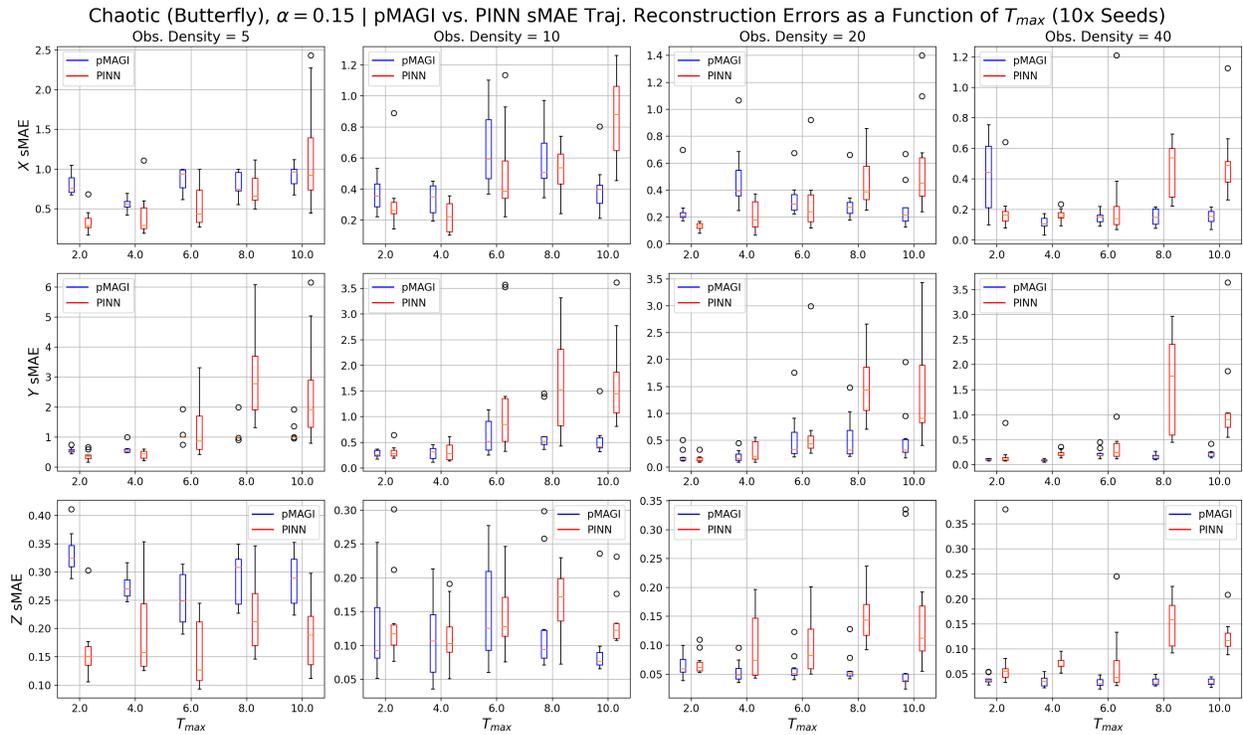

**Figure 7.9:** pMAGI vs. PINN trajectory reconstruction on the Chaotic (Butterfly) regime. Blue represents pMAGI, while red represents PINN. The x-axis of each subplot is $T_{max}$, the length of our noisy observation interval $t \in [0, T_{max}]$. Each row displays the sMAEs for a given component $X$, $Y$, or $Z$. Each column shows the models' performances holding the density of noisy observations per unit time $d_{obs}$ fixed. The box-and-whisker plots capture the error distributions across the ten randomly-seeded datasets.

Finally, on the Chaotic (No Butterfly) regime as shown in Figure 7.10, we first begin by acknowledging that pMAGI experienced significant numerical instability issues on $T_{max} = 8.0, 4.0$, and $2.0$ on observation densities $d_{obs} = 5, 10$, and $20$, respectively. As such, the resultant distortion of the subplots makes it very difficult to analyze the remainder of the results in these subplots — that being said, the alternative of rescaling these plots and/or omitting the numerical instabilities did not seem to be the right thing to do in terms of reporting the full truth. However, looking at the $X$ component at $d_{obs} = 40$, we note that pMAGI decisively outperforms PINN at $T_{max} \geq 6.0$, with rough parity at $T_{max} = 2.0$ and $4.0$. Looking at the $Y$ and $Z$ components, we once again observe that pMAGI loses out to PINN at $d_{obs} = 5$. But, this performance gap shrinks considerably by $d_{obs} = 10$, with pMAGI meeting or exceeding PINN performance on all settings at $d_{obs} = 20$ and consistently outperforming PINN by $d_{obs} = 40$.

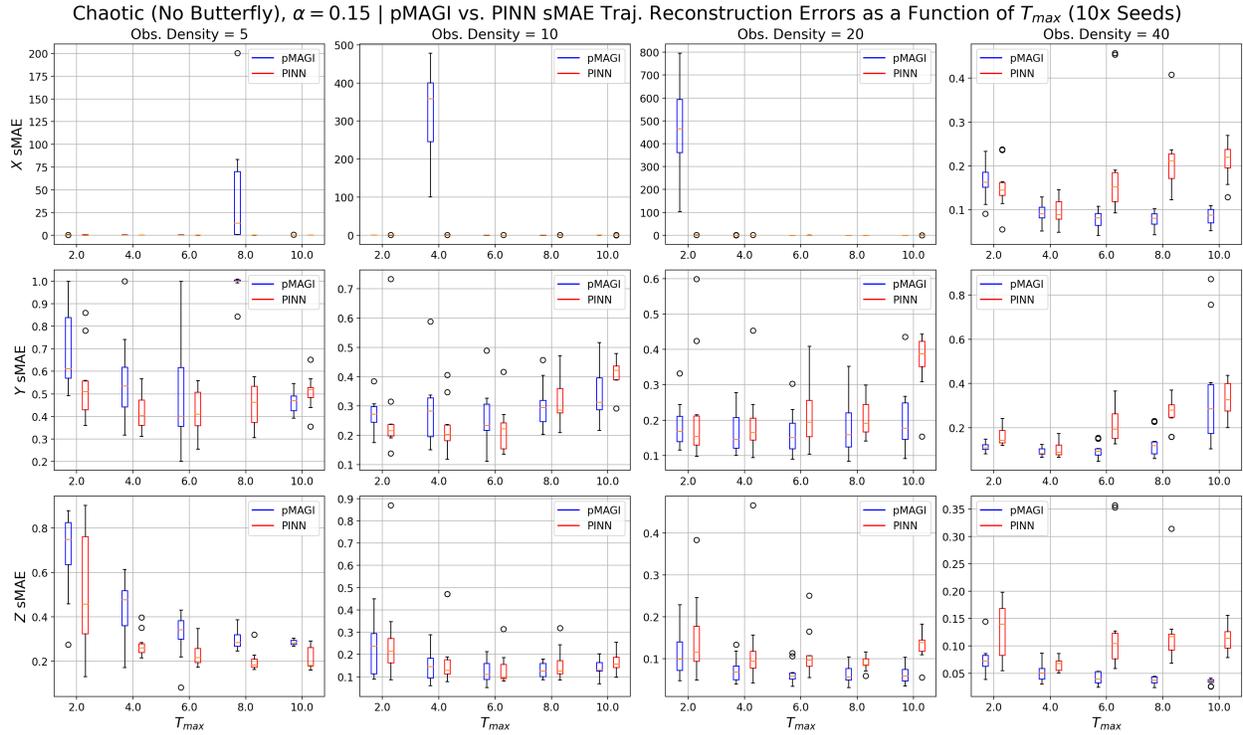

**Figure 7.10:** pMAGI vs. PINN trajectory reconstruction on the Chaotic (No Butterfly) regime. Blue represents pMAGI, while red represents PINN. The x-axis of each subplot is $T_{max}$, the length of our noisy observation interval $t \in [0, T_{max}]$. Each row displays the sMAEs for a given component $X$, $Y$, or $Z$. Each column shows the models' performances holding the density of noisy observations per unit time $d_{obs}$ fixed. The box-and-whisker plots capture the error distributions across the ten randomly-seeded datasets.

## 7.4 Main Takeaways

In this chapter, we have extensively demonstrated that **pMAGI holds six decisive advantages over PINN.** First, pMAGI is much more parameter-efficient than PINN. Second, at moderate to high observation density settings, pMAGI will consistently outperform PINN. Third, at moderate to long observation intervals, pMAGI will also consistently outperform PINN. Fourth, pMAGI tends to have lower variance across trials than PINN. Fifth, pMAGI is much more robust to hyperparameter values than PINN. Sixth, pMAGI is naturally capable of uncertainty quantification, while PINN is not. To tell the full story, we do acknowledge that, in some cases, PINN may outperform pMAGI at the lowest observation density settings. **However, all in all, it is clear to see that pMAGI presents itself as an overall stronger model for trajectory reconstruction than the heavily-overparameterized PINN.**

In the next chapter, we will introduce Pilot MAGI Sequential Prediction (PMSP), a novel method for predicting the trajectories of ODE-based dynamical systems multiple timesteps into the *future*.

# 8
# Prediction

In this chapter, we introduce Pilot MAGI Sequential Prediction (PMSP), a method for predicting the trajectories of ODE-based dynamical systems multiple time steps into the future, given only noisy and sparse observations $\mathcal{X}_{obs}$ at a set of timepoints $t \in \tau_{obs}$, with $0 \leq t \leq T_{max}$. We will then compare PMSP against Physics-Informed Neural Networks (PINNs) on the task of predicting the trajectories of our Lorenz system test beds $\Delta T_{\text{pred}} = 0.5, 1.0, 2.0$, and $4.0$ into the future from either a starting $T_{max} = 2.0$ or a starting $T_{max} = 6.0$, under noise level multiplier settings of $\alpha = 1.5 \times 10^{-5}$ and $\alpha = 0.15$, coupled with our other standard test bed settings from the previous chapters. We demonstrate that our PMSP method outperforms or, at the very least, is a strong competitor, to PINNs on the task of future trajectory prediction in many settings.

Before we start on the technical details, we provide Figure 8.1 to visually emphasize the extent to which PMSP can outperform PINNs on the future prediction task. Specifically, the colored lines in the first row of Figure 8.1 represent three PMSP variants' future predictions on each component of the Chaotic (Butterfly) regime, up to $\Delta T_{pred} = 4.0$ timesteps into the future. The colored lines in the second row represent three PINN variants' future predictions on the exact same setting. In each subplot, the unnoised ground truth is represented with a black line. Visually, we see clearly that while **all three PMSP variants shown can output extremely accurate predictions for the entire $\Delta T_{pred} = 4.0$ timesteps into the future, all three PINN variants quickly flatline in their predictions, incurring massive errors.** In the coming sections, we will explain in full detail what each of the PMSP and PINN variants mean (like "PMSP(ROP)"), and the statistical motivations of our methodology.

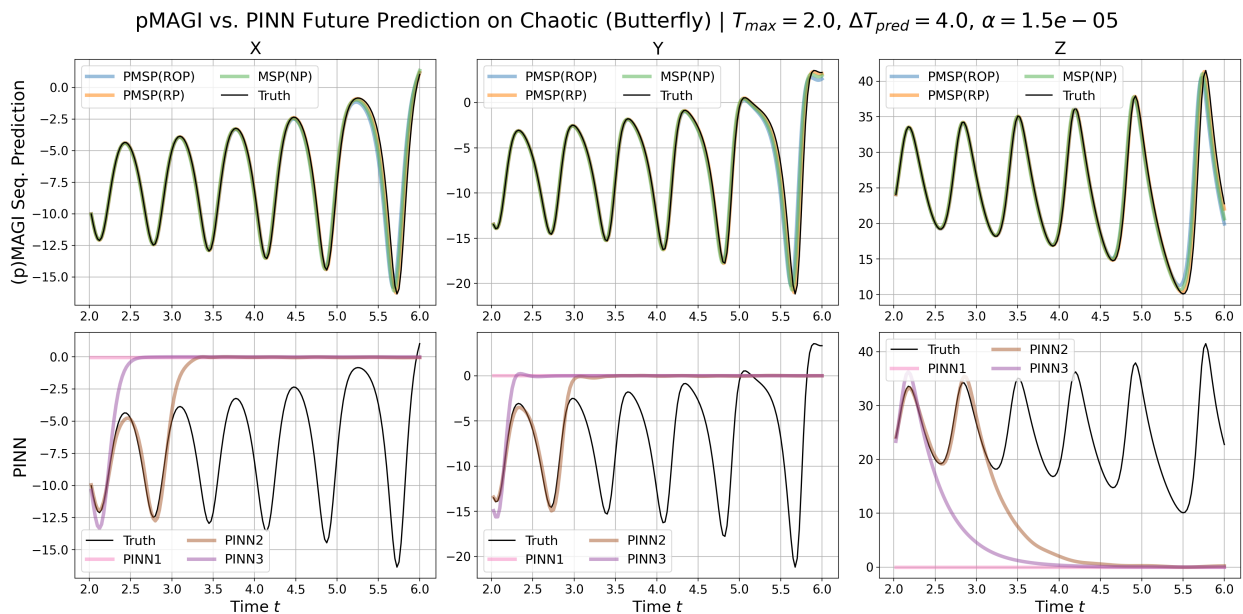

**Figure 8.1:** Selected Pilot MAGI Sequential Prediction (PMSP) vs. PINN future prediction trajectories on the Chaotic (Butterfly) regime at $\alpha = 1.5 \times 10^{-5}$, $T_{max} = 2.0$, and $\Delta T_{pred} = 4.0$. PMSP model variants are shown in the first row, while PINN models are shown in the second row, with all prediction curves color-coded by model variant. The x-axis for each subplot is time $t$ on the prediction interval $t \in [2.0, 2.0 + \Delta T_{pred}]$. Visually, we can estimate model performance by examining how closely the predicted trajectories match the ground truth (in black).

## 8.1 Motivation for Novel Pilot MAGI Sequential Prediction (PMSP) Method

Given their Bayesian sampling structures, MAGI and pMAGI can already, *in theory*, perform prediction out-of-the-box. For a concrete example, suppose that we have noisy and sparse observations from a Lorenz system trajectory $\mathcal{X}_{obs}$ at times $t \in \{0, 0.1, 0.2, \ldots, 10.0\} = \tau_{obs}$, and we would like to predict the future (unnoised) trajectory of this system $\Delta T_{\text{pred}} = 5$ time steps into the future at times $t \in \{10.1, 10.2, \ldots, 15.0\} = \tau_{add}$. Let $\mathcal{X}_{add}$ be an $|\tau_{add}| \times 3$ matrix containing all NaN values. By augmenting our $\mathcal{X}_{obs}$ and $\tau_{obs}$ with $\mathcal{X}_{add}$ and $\tau_{add}$, respectively, and passing in our modified $\mathcal{X}_{obs}$ and $\tau_{obs}$ into MAGISolver, we will obtain sampled trajectory component values at times $t \in \tau_{obs}$, which includes both our in-sample and prediction timesteps. From these sampled trajectory component values, we can then directly extract MAGI or pMAGI's predictions at our desired future time steps. For example, we can obtain point-predictions by taking the means of our samples for each component at each time step, and obtain predictive intervals by taking the relevant quantiles. Of course, like pMAGI and MAGI, we can also specify additional discretization to evaluate our error variable $W$ at a higher resolution.

However, in practice, the method described above is almost always unsuccessful. By expanding the number of components we must sample from using HMC by $|\tau_{add}| \times 3$ values, we have rendered it extremely difficult for our HMC sampler to converge to the ideal neighborhood of the sampling space. One solution for overcoming this obstacle is to make predictions in smaller sequential steps. Instead of directly appending so many values to $\mathcal{X}_{obs}$ and $\tau_{obs}$, what if we first try to predict $\Delta T_{step} = 0.5$ timesteps into the future? Having generated predictions from $T = 10$ to $T = 10.5$ (our first sequential prediction step), we can try to warm-start the MAGISolver and try to predict from $T = 10$ to $T = 11$ (our second prediction step), and so on and so forth, until we have finally generated predictions on our full desired interval from $T = 10$ to $T = 15$. What do we mean by warm-start? Recall that MAGISolver allows us to exogenously specify initial values/guesses for $\hat{\theta}$ and $\hat{\mathcal{X}}_{inf}$ to help the HMC sampler more quickly converge to high probability density sampling regions. For this chapter, let EWSI stand for this crucial concept of *exogenous warm-start initialization*.

During our second prediction step, we can construct a warm-start guess for $\hat{\mathcal{X}}_{inf}$ by using the last HMC sam-ple of the trajectory components from our first prediction step as our initialization for times $t = 0$ to $t = 10.5$ ("in-sample"), which we call $\hat{\mathcal{X}}_{step,init,IS}$. For the time interval $t = 10.5$ to $t = 11$, we can start by extracting the components' values at the last timestep ($t = 10.5$) obtained from the final HMC sample of the first prediction step. Then, we can numerically integrate forward by $\Delta T_{step} = 0.5$ timesteps, employing the last HMC sample of our parameters $\hat{\theta}_{inf}$ from the same first prediction step as governing parameters. This process allows us to construct an initial guess $\hat{\mathcal{X}}_{step,init,OS}$ for our system's trajectory spanning the period from $t = 10.5$ to $t = 11$, which we will refer to as "out-of-sample" for this second prediction step. Stitching together our initializations for $t \in [0, 10.5]$ and $t \in [10.5, 11]$, we now have a full-warm start initialization $\hat{\mathcal{X}}_{step,init}$ that we can pass into our MAGISolver. In a similar idea, we can construct a warm-start guess for $\hat{\theta}$ by directly taking the last HMC sample of our parameters from our first prediction step. With warm-start initialization values $\hat{\mathcal{X}}_{step,init}$ and $\hat{\theta}$ fully defined, we can run our MAGISolver at this second prediction step, and repeat this process until we have generated predictions for our full desired interval. Hopefully, because we are warm-starting our algorithm, our HMC samplers can converge more quickly and efficiently, ultimately yielding more accurate predictions. Given this sequential setup, we name our algorithm Pilot MAGI Sequential Prediction (PMSP).

Until now, we have avoided discussion of the "Pilot" portion of PMSP. There are a few reasonable methods to incorporate the pilot modules for estimation of $(\phi_1, \phi_2)$ into this sequential algorithm. Let the variable PM, short for *pilot mode*, describe how we intend to incorporate the pilot module idea. One idea is to simply not use a pilot module, which we will encode as PM $= NP$ ("No Pilot"). Another idea is to simply estimate $(\phi_1, \phi_2)$ once from $t = 0$ to $t = 1$ and use these values for the entire sequential prediction process, which we will encode as PM $= LP$ ("Left Pilot"). Similarly, one could estimate $(\phi_1, \phi_2)$ once from $t = \max(0, T_{max} - 1)$ to $t = T_{max}$ and use these values for the entire sequential prediction process, if one believes that estimated $(\phi_1, \phi_2)$ values closer to the prediction time interval are likely to be more useful — we encode this approach as PM $= RP$ ("Right Pilot"). Now, given that we are generating predictions for a potentially-chaotic system, whose dynamics may change drastically over time, one might want to re-estimate $(\phi_1, \phi_2)$ after each sequential prediction step in an "online manner." We encode these approaches as PM $= LOP$ ("Left Online Pilot") and PM $= ROP$ ("Right

Online Pilot") based on which interval we used to generate our initial estimates of $(\phi_1, \phi_2)$ on the first sequential prediction step.

As a quick proof-of-concept, in the figure below, we demonstrate out-of-the-box prediction (i.e., just appending NaNs and expanding $\tau_{inf}$ in one step) using pMAGI versus sequential prediction (PMSP) on the chaotic (butterfly) regime, with noisy and sparse observations provided from $t = 0.0$ to $t = 2.0$ and predictions generated for $t = 2.0$ to $t = 7.0$. We see that using pMAGI right out-of-the-box, we are only able to generate useful predictions for around $\Delta T_{\text{pred}} = 0.5$ timesteps, before the predicted trajectories flatline. This behavior is not unexpected because $(X, Y, Z)$ adopting a constant solution over time *does* satisfy the Lorenz equations, but this is clearly not reflective of our chaotic (butterfly) system. In contrast, our PMSP sequential prediction method is able to generate accurate predictions at least $\Delta T_{\text{pred}} = 4.0$ timesteps into the future. Our method seems to work!

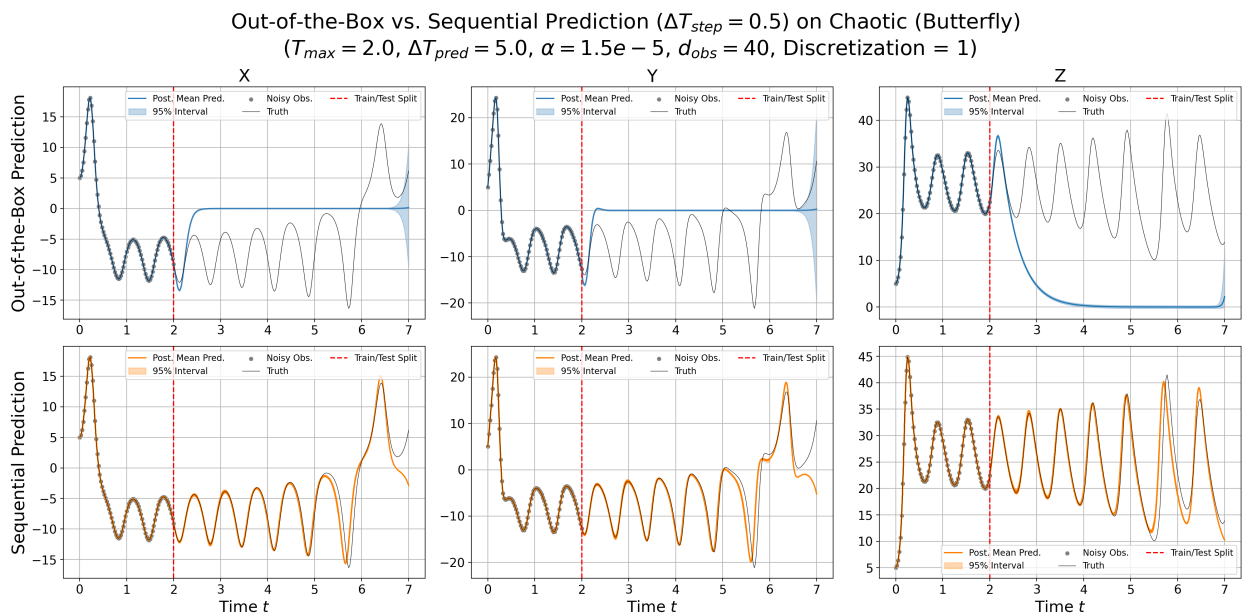

**Figure 8.2:** Out-of-the-Box vs. Sequential Prediction future prediction trajectories on the Chaotic (Butterfly) regime. The interval band corresponds to a $95\%$ prediction interval, while the solid colored line corresponds to the point-prediction obtained via taking the posterior mean of our samples after the last prediction timestep. The end of our in-sample "training" observations is marked with the dotted vertical line, the unnoised ground truth is marked as a black line, and the sparse and noisy observations are marked as grey dots.

## 8.2 Algorithmic Description of PMSP

Now, we provide a more computational description of the PMSP in Algorithm 4. Specifically, $\mathcal{X}_{\text{obs}}$ and $\tau_{\text{obs}}$ represents our in-sample noisy and sparse observations (i.e., our training data) and the set of time steps that we have these observations, respectively. $T_{max}$ represents the last time point of our training data and $D$ represents the discretization that we will apply on our inputs to MAGISolver to obtain a finer-grained computation of our error variable $W$. $\Delta T_{\text{pred}}$ represents how many time steps we wish to generate predictions into the future (e.g., $\Delta T_{\text{pred}} = 5$ in our example in the previous section). $\Delta T_{\text{step}}$ represents our stepsize for each sequential prediction step (e.g., $\Delta T_{\text{step}} = 0.5$ in our example in the previous section). PM represents our *pilot mode*, as discussed earlier. Practically, let $N_{HMC,\text{init}}$ be the number of HMC steps we will run for our first sequential prediction step. Because we are increasing the dimensionality of our sampling space at each sequential prediction step, we will increase the number of HMC steps desired at each call of MAGISolver at each sequential prediction step to better facilitate convergence. Let $N_{HMC,\text{peak}}$ denote the number of HMC steps we will use at our *last* sequential prediction step. Heuristically, after the first sequential prediction step, we will make the number of HMC steps desired at each sequential prediction step proportional to the length of the time interval we are generating prediction samples for.

Practically, one can also save the predictions generated at each sequential prediction step as checkpoints, as it is possible that, at later prediction timesteps, extending our prediction horizon for too long may harm overall convergence, and negatively affect our predictions at earlier timesteps, too. Please see the full details of the PMSP algorithm in the Algorithm 4 display below, with ample comments. In the algorithm below, the subscript *IS* refers to "in-sample" and *OS* to "out-of-sample." If a pilot module is required, we run our initial pilot modules using 4001 HMC steps and any subsequent pilot modules using only 101 HMC steps, as $\sigma$ no longer needs to nor should be inferred from predicted time intervals, and $(\phi_1, \phi_2)$ estimation happens before the HMC sampler begins. As alluded to earlier, the notation $[-1]$ means to query the last sample from the previous collection of HMC samples. Similarly, the notation $\hat{\mathcal{X}}_{prev}[-1][-1]$ means to query the components' values at the last time step contained in the last HMC sample from the previous sequential prediction step. When reading the below algorithm, if PM $= NP$ ("No Pilot"), one can safely ignore all lines involving a pilot module.

## Algorithm 4 Pilot MAGI Sequential Prediction (PMSP)

**Input:** $\mathcal{X}_{\text{obs}}$ - noisy in-sample observations, $\tau_{\text{obs}}$ - in-sample observed time steps, $T_{\text{max}}$ - end of in-sample observations, $D$ - discretization level, $\Delta T_{\text{pred}}$ - full prediction horizon, $\Delta T_{\text{step}}$ - sequential prediction stepsize, PM - pilot mode, $N_{HMC,\text{init}}$ - number of HMC steps for first sequential prediction step, $N_{HMC,\text{peak}}$ - number of HMC steps for last sequential prediction step

**Output:** Samples of predicted trajectories $\hat{\mathcal{X}}$pred,step and samples of parameters $\hat{\theta}$pred,step after last prediction step.

1: **function** PMSP($\mathcal{X}_{\text{obs}}, \tau_{\text{obs}}, T_{\text{max}}, D, \Delta T_{\text{fc}}, \Delta T_{\text{step}}$, PM, $N_{HMC,\text{init}}, N_{HMC,\text{peak}}$)
2:
3:      # Initializing settings and data structures
4:      $T_{\text{pred}} \leftarrow T_{\text{max}} + \Delta T_{\text{pred}}$; $N_{\text{steps}} \leftarrow \lceil \Delta T_{\text{pred}} / \Delta T_{\text{step}} \rceil$
5:      $\tau_{\text{add}} \leftarrow (40 \cdot \Delta T_{\text{pred}})$ equally-spaced points between $t = T_{\text{max}}$ (excl.) and $t = T_{\text{pred}}$ (incl.)
6:      $\tau_{\text{pred}} \leftarrow$ concatenate $(\tau_{\text{obs}}, \tau_{\text{add}})$; $\mathcal{X}_{\text{pred}} \leftarrow$ append NaNs to the end of $\mathcal{X}_{\text{obs}}$, corresponding to $\tau_{\text{pred}}$
7:      $\tau_{\text{pred}} \leftarrow$ discretize $\tau_{\text{pred}}$ with level $D$; $\mathcal{X}_{\text{pred}} \leftarrow$ discretize $\mathcal{X}_{\text{pred}}$ with level $D$ to match $\tau_{\text{pred}}$
8:
9:      # Running our initial pilot module for estimating $(\hat{\phi}_1, \hat{\phi}_2), \hat{\sigma}$
10:      **if** PM $\in$ {LP, LOP} **then**
11:          $\tau_{\text{pilot}} \leftarrow \{t \in \tau_{\text{pred}} \mid 0 \le t \le \min(1, T_{\text{max}})\}$; $\mathcal{X}_{\text{pilot}} \leftarrow$ corresponding subset from $\mathcal{X}_{\text{pred}}$
12:      **else if** PM $\in$ {RP, ROP} **then**
13:          $\tau_{\text{pilot}} \leftarrow \{t \in \tau_{\text{pred}} \mid \max(0, T_{\text{max}} - 1) \le t \le T_{\text{max}}\}$; $\mathcal{X}_{\text{pilot}} \leftarrow$ corresponding subset from $\mathcal{X}_{\text{pred}}$
14:      **end if**
15:      $(\hat{\phi}_1, \hat{\phi}_2), \hat{\sigma}, \hat{\theta}_{\text{pilot}} \leftarrow$ MAGISolver($\mathcal{X}_{\text{pilot}}, \tau_{\text{pilot}}$; $(\mathbf{f}, \nabla_\theta f, \nabla_\mathbf{x} f), \mathcal{S}$)
16:
17:      # Creating cache variables to facilitate warm-starts between time steps.
18:      $\hat{\mathcal{X}}_{\text{prev}}, \hat{\theta}_{\text{prev}}, \tau_{\text{prev}} \leftarrow$ None, None, None
19:
20:      **for** $n_{\text{step}} \leftarrow 1$ **to** $N_{\text{step}}$ **do**      ▷ Proceed through each step of our sequential prediction routine
21:          $T_{\text{pred,step}} \leftarrow T_{\text{max}} + (n_{\text{step}} * \Delta T_{\text{step}})$; $\mathcal{S} \leftarrow \{\}$      ▷ how far to predict on this step? Also, reset MAGISolver exogenous settings.
22:          $\tau_{\text{step}} \leftarrow \{t \in \tau_{\text{pred}} \mid t \le T_{\text{pred,step}}\}$; $\mathcal{X}_{\text{step}} \leftarrow$ corresponding subset of $\mathcal{X}_{\text{pred}}$
23:
24:          # Check if we need to re-estimate $(\hat{\phi}_1, \hat{\phi}_2)$
25:          **if** $n_{\text{step}} \ne 1$ and PM $\in$ {LOP, ROP} **then**
26:              $T_{\text{pilot,end}} \leftarrow \max(\tau_{\text{prev}})$; $T_{\text{pilot,start}} \leftarrow T_{\text{pilot,end}} - \Delta T_{\text{step}}$
27:              $\tau_{\text{pilot}} \leftarrow \{t \in \tau_{\text{prev}} \mid T_{\text{pilot,start}} \le t \le T_{\text{pilot,end}}\}$; $\mathcal{X}_{\text{pilot}} \leftarrow$ corresponding subset of $\hat{\mathcal{X}}_{prev}$
28:              $(\hat{\phi}_1, \hat{\phi}_2) \leftarrow$ re-estimate using pilot module with $\tau_{\text{pilot}}, \mathcal{X}_{\text{pilot}}$. Store in $\mathcal{S}$.
29:          **end if**
30:
31:          # Special settings if we are working on our first prediction step
32:          **if** $n_{\text{step}} == 1$ **then**
33:              Set $N_{HMC,\text{step}} \leftarrow N_{HMC,\text{init}}$ with burn-in ratio 0.5. Store in $\mathcal{S}$.
34:              $\hat{\theta}_{\text{step,init}} \leftarrow$ colMeans($\hat{\theta}_{\text{pilot}}$). Store in $\mathcal{S}$.      ▷ specify EWSI for parameters $\theta$ using parameter-wise means from pilot samples
35:          **else**
36:              Set $N_{HMC,\text{step}} \leftarrow \frac{T_{\text{pred,step}}}{T_{\text{pred}}} \cdot N_{HMC,\text{peak}}$ with burn-in ratio 0.2. Store in $\mathcal{S}$.      ▷ The more components to sample, the more HMC steps.
37:              $\hat{\theta}$step,init $\leftarrow \hat{\theta}$prev$[-1]$. Store in $\mathcal{S}$.      ▷ specify EWSI for parameters $\theta$ using last sample from previous prediction step
38:              $\hat{\mathcal{X}}$step,init,IS $\leftarrow \hat{\mathcal{X}}$prev$[-1]$      ▷ specify EWSI for "in-sample" trajectory for $0 \le t \le T_{\text{pred,step}} - \Delta T_{\text{step}}$
39:
40:              # Specify EWSI for "out-of-sample" trajectory for $T$pred,step $- \Delta T_{\text{step}} < t \le T_{\text{pred,step}}$.
41:              $\hat{\mathcal{X}}$step,init,OS $\leftarrow$ numerically-integrate from $\hat{\mathcal{X}}$prev$[-1][-1]$ with parameters $\hat{\theta}$prev$[-1]$
42:
43:              # Construct our full EWSI for $\mathcal{X}$ for this prediction step.
44:              $\hat{\mathcal{X}}$step,init $\leftarrow$ concatenate ($\hat{\mathcal{X}}$step,init,IS, $\hat{\mathcal{X}}$step,init,OS). Store in $\mathcal{S}$.
45:          **end if**
46:
47:          # Perform our MAGISolver call at this step to output prediction (samples) for this step
48:          $\hat{\mathcal{X}}$pred,step, $\hat{\theta}$pred,step $\leftarrow$ MAGISolver($\mathcal{X}$step, $\tau_{\text{step}}$; $(\mathbf{f}, \nabla_\theta f, \nabla_\mathbf{x} f), \mathcal{S}$)      ▷ generate new step predictions
49:          $\hat{\mathcal{X}}$prev, $\hat{\theta}$prev, $\tau_{\text{prev}} \leftarrow \hat{\mathcal{X}}$pred,step, $\hat{\theta}$pred,step, $\tau_{\text{step}}$      ▷ update cache variables
50:      **end for**
51:
52:      **Return** $\hat{\mathcal{X}}$pred,step, $\hat{\theta}$pred,step, after last step.
53: **end function**

## 8.3 PINN Competitor Methods and Experimental Setup

Having introduced the PMSP method, we now proceed to extensively benchmark it against our main competitor method — the Physics-Informed Neural Network (PINN). For all methods, we will use all four Lorenz test bed regimes with a fixed in-sample observation set from $t = 0$ to $t = T_{max}$ for $T_{max} \in \{2.0, 6.0\}$, with in-sample observation densities of $d_{obs} = 5, 10, 20, 40$ for PINNs and only $d_{obs} = 10, 40$ for PMSP due to time and computation restraints. We will focus on predicting the systems' trajectories $\Delta T_{\text{pred}} \in \{0.5, 1.0, 2.0, 4.0\}$ time steps into the future, with our prediction test set fixed at 40 evenly-spaced values per unit time regardless of in-sample $d_{obs}$. In practice, we originally intended to predict up to $\Delta T_{\text{pred}} = 5.0$ steps into the future, but selected the above $\Delta T_{\text{pred}}$ values because they were successive powers of 2, giving us potentially nicer scaling intuition.

To address the disparity in $d_{obs}$ settings for PINNS vs. PMSP and to keep tables and figures within reasonable limits, when computing summary metrics like prediction error, we will take the lowest error produced over any setting of in-sample $d_{obs}$, though, almost always, as one would guess, the lowest errors are achieved when there is as much training data as possible (i.e., $d_{obs} = 40$). Sometimes, a PINN might achieve a lower prediction error using $d_{obs} = 20$ versus $d_{obs} = 40$, so we will give competitor methods the best light possible and take whatever lowest prediction error was achieved across all $d_{obs}$ values. Because of computation and time constraints, we restrict our focus to noise level multipliers $\alpha = 1.5 \times 10^{-5}$ and $\alpha = 0.15$, and only perform one trial per model and data setting, using the same common datasets for both PMSP and PINN for consistency. It should be noted that due to various (non-preventable) technical issues with the cluster, not all PMSP experiment settings were able to be completed, even among this more restricted set, but we will still show that even with just a portion of the hyperparameter settings tested, PMSP still decisively outperforms PINN (which did have all experimental settings finish running) in many, many situations. It is very possible that demonstrated PMSP performance would have been even higher had all hyperparameter variants finished running.

Specifically, to present PINNs in the best light possible, we implement three different PINN methods for prediction: PINN1, PINN2, and PINN3. For all three methods, we will use the codebase written by van Herten et al.[49] with 3 dense-layers and 32 nodes in each layer. We will also try $\lambda \in \{0.1, 1.0, 10.0, 100.0, 1000.0\}$ for all three methods, with an Adam optimizer using learning rate $\eta = 0.01$, following van Herten et al.[49]. We will only show the <u>best</u> hyperparameter value's results in our figures and tables for each PINN method on a given setting. For all three methods, we will first train the PINN for 60000 epochs on both reconstruction loss and physics-based loss on our in-sample noisy observations, weighted by $\lambda$. Then, we will freeze the parameters containing the inferred values of $\theta$, before starting our next phase of training. Let $\hat{\mathbf{X}}_{\text{freeze}}(T_{max})$ represent a PINN's prediction at time $t = T_{max}$ immediately after the initial 60000 epochs of training and before the next phase of training (which we presume will be close to the true $\mathbf{X}(T_{max})$, given the results in the previous chapters).

For PINN1, we will continue training our neural network purely on the physics-based loss on the time points of our desired prediction test set specified above for a second, final round of 60000 epochs.

For PINN2, let $\text{NN}(t) \in \mathbb{R}^3$ represent the forward-pass function of our neural network immediately after the first 60000 epochs of training. To hard-enforce our boundary constraint of $\hat{\mathbf{X}}_{\textit{freeze}}(T_{max})$ at our last observed timestep, for the next 60000 epochs of training, we will output our forward-pass as follows:

$$\text{NN2}(t) = \frac{t - T_{max}}{\Delta T_{\text{pred}}} \text{NN}(t) + \frac{T_{max} + \Delta T_{\text{pred}} - t}{\Delta T_{\text{pred}}} \hat{\mathbf{X}}_{\text{freeze}}(T_{max}).$$

The reason for this NN2 modification is to ensure that regardless of what training occurs during the second, final set of 60000 epochs, our PINN will output $\hat{\mathbf{X}}_{\text{freeze}}(T_{max})$ at the boundary point of $t = T_{max}$, a bit analogous to standard numerical solvers. With this setup, on the second set of 60000 epochs, we will only need to train PINN2 on the physics-based loss on the time points of our desired prediction set, just like with PINN1.

As an intermediate option between the no-boundary-constraint setup of PINN1 and the hard-boundary constraint setup of PINN2, we will let PINN3 represent a soft-boundary option. As such, for its second, final set of 60000 epochs of training, PINN3 will have a loss function composed of physics-based loss on the time points of our desired prediction set and an one-observation "reconstruction" loss of $||\text{NN}(T_{max}) - \hat{\mathbf{X}}_{\text{freeze}}(T_{max})||_2^2$, weighted by the hyperparameter $\lambda$.

For PMSP, we will try all possible pilot modes PM $\in \{NP, LP, RP, LOP, ROP\}$, discretizations $D \in \{1, 2\}$, and prediction stepsizes $\Delta T_{\text{step}} \in \{0.5, 1.0, 5.0\}$ (with the last option corresponding to out-of-the-box prediction). For $\Delta T_{step} = 0.5$, we set $N_{HMC,\text{init}} = N_{HMC,\text{peak}} = 16001$. For $\Delta T_{\text{step}} = 1.0$, we set $N_{HMC,\text{init}} = N_{HMC,\text{peak}} = 32001$. Finally, for $\Delta T_{\text{step}} = 5.0$, we set $N_{HMC,\text{init}} = N_{HMC,\text{peak}} = 128001$. We chose these values to roughly balance out the total computation performed as a function of the total HMC steps and the average number of components sampled per sequential prediction step.

For all methods and variants, we will quantify prediction error using scaled Mean Absolute Error (sMAE). Specifically, let $\tau_{pred}$ be the set of prediction time points, $\mathbf{X}(t)$ represent the true, unnoised values of our system at time $t$, and $\hat{\mathcal{X}}_{pred} = \{\hat{\mathbf{X}}(t) \mid t \in \tau_{pred}\}$ represent a given set of predictions from a model. Then, sMAE for each component $i$ of our system is defined as follows:

$$\text{sMAE}_i\left(\hat{\mathcal{X}}_{pred}\right) = \frac{1}{|\tau_{pred}|} \sum_{t \in \tau_{pred}} \frac{|X_i(t) - \hat{X}_i(t)|}{|X_i(t)|}.$$

In practice, we generate the ground truth targets for our prediction experiments using numerical integration via SciPy's solve_ivp module [50] with absolute and relative tolerance both set to $1 \times 10^{-10}$. While numerical errors will accrue with any numerical solver when integrating chaotic systems, we can be reasonably confident that our generated ground truth trajectories are reflective of the theoretical truth on our intervals of interest because at such high precision tolerances, all solvers in the solve_ivp module outputed visually-instinguishable trajectories.

## 8.4 Aggregate Results

From Tables 8.1 and 8.2, we see that at noise level multiplier $\alpha = 1.5 \times 10^{-5}$, for both $T_{max} = 2$ and $T_{max} = 6$, across all test bed regimes and prediction horizons $\Delta T_{\text{pred}}$, PMSP variants decisively outperform all three PINN variants in virtually all tested settings. The difference in sMAE between PMSP and PINN variants is oftentimes a matter of multiple orders of magnitude. In general, we observe for all model variants that prediction error increases as $\Delta T_{\text{pred}}$ increases, which intuitively makes sense because we are straying farther away from the signal contained in our in-sample training data. Looking at the aforementioned two tables, it is also clear that PMSP(LOP) and PMSP(ROP), our two PMSP variants that update $(\phi_1, \phi_2)$ in an online manner after each prediction step, perform the best, matching our intuitions and hypotheses.

| Regime | $\Delta T_{pred}$ | MSP(NP) | PMSP(LP) | PMSP(RP) | PMSP(LOP) | PMSP(ROP) | PINN1 | PINN2 | PINN3 |
|---|---|---|---|---|---|---|---|---|---|
| Stable (Canonical) | 0.5 | **0.0000** | **0.0000** | **0.0000** | **0.0000** | **0.0000** | 0.0385 | 0.0002 | 0.0004 |
| Stable (Canonical) | 1.0 | **0.0000** | **0.0000** | **0.0000** | **0.0000** | **0.0000** | 0.0314 | 0.0004 | 0.0009 |
| Stable (Canonical) | 2.0 | **0.0000** | **0.0000** | **0.0000** | **0.0000** | **0.0000** | 0.0216 | 0.0004 | 0.0012 |
| Stable (Canonical) | 4.0 | **0.0000** | 0.0001 | 0.0001 | 0.0001 | 0.0001 | 0.0128 | 0.0006 | 0.0016 |
| Stable (Transient Chaos) | 0.5 | 0.0020 | 0.0009 | 0.0006 | 0.0007 | **0.0004** | 0.8889 | 0.1701 | 0.1708 |
| Stable (Transient Chaos) | 1.0 | 0.0151 | 0.0095 | 0.0043 | 0.0012 | **0.0002** | 0.7856 | 0.4679 | 0.5003 |
| Stable (Transient Chaos) | 2.0 | 0.1292 | 0.0301 | **0.0239** | 0.0555 | 0.0431 | 0.9981 | 0.8449 | 0.8814 |
| Stable (Transient Chaos) | 4.0 | 0.1616 | 0.0993 | 0.1051 | **0.0113** | 0.0588 | 0.9983 | 0.9191 | 0.9403 |
| Chaotic (Butterfly) | 0.5 | **0.0005** | 0.0007 | 0.0008 | 0.0006 | **0.0005** | 0.4605 | 0.0436 | 0.0961 |
| Chaotic (Butterfly) | 1.0 | 0.0016 | 0.0023 | 0.0010 | 0.0016 | **0.0005** | 0.4239 | 0.0669 | 0.1346 |
| Chaotic (Butterfly) | 2.0 | 0.0047 | 0.0091 | **0.0008** | 0.0014 | 0.0015 | 0.5463 | 0.4476 | 0.3298 |
| Chaotic (Butterfly) | 4.0 | 0.2037 | 0.0640 | 0.0687 | **0.0289** | 0.0669 | 0.9956 | 0.7289 | 0.8933 |
| Chaotic (No Butterfly) | 0.5 | 0.0004 | **0.0002** | **0.0002** | 0.0005 | **0.0002** | 0.1766 | 0.0642 | 0.0319 |
| Chaotic (No Butterfly) | 1.0 | 0.0009 | 0.0004 | 0.0002 | **0.0001** | 0.0002 | 0.1681 | 0.0967 | 0.0493 |
| Chaotic (No Butterfly) | 2.0 | 0.0030 | 0.0019 | 0.0004 | **0.0002** | **0.0002** | 0.1792 | 0.1433 | 0.0881 |
| Chaotic (No Butterfly) | 4.0 | 0.0177 | 0.0069 | **0.0005** | 0.0082 | 0.0064 | 0.2076 | 0.1901 | 0.1578 |

**Table 8.1:** Scaled mean absolute errors (sMAE) of models towards predicting the future with training interval $T_{max} = 2$ and noise multiplier $\alpha = 1.5 \times 10^{-5}$. MSP(NP) refers to *MAGI Sequential Prediction with no pilot*. PMSP refers to *Pilot MAGI Sequential Prediction*, and the codes in parentheses correspond to the *pilot mode* PM. The model(s) that incurred the lowest error is bolded in each row. **Lower values indicate better performance.**

It is also interesting how the majority of all PMSP variants incur virtually *no error* on the Stable (Canonical) regime, up to four decimal places. This makes sense intuitively because the stable (canonical) regime effectively converges to a fixed point by $t = 7.0$, which should be relatively straightforward. In contrast, while the PINN variants do also incur the lowest errors on Stable (Canonical) compared to the other regimes, the fact that their errors are much higher than that of PMSP's on the Stable (Canonical) regime suggests that PINNs may have difficulty learning generalizable solutions for even relatively-simple systems. The complexity of the overparameterized PINN architectures may result in various unpredictable artifacts, non-smoothness, and other complexities in their learned representations.

| Regime | $\Delta T_{pred}$ | MSP(NP) | PMSP(LP) | PMSP(RP) | PMSP(LOP) | PMSP(ROP) | PINN1 | PINN2 | PINN3 |
|---|---|---|---|---|---|---|---|---|---|
| Stable (Canonical) | 0.5 | **0.0000** | **0.0000** | **0.0000** | **0.0000** | **0.0000** | 0.0011 | 0.0009 | 0.0007 |
| Stable (Canonical) | 1.0 | **0.0000** | **0.0000** | **0.0000** | **0.0000** | **0.0000** | 0.0010 | 0.0011 | 0.0010 |
| Stable (Canonical) | 2.0 | **0.0000** | **0.0000** | **0.0000** | **0.0000** | **0.0000** | 0.0007 | 0.0008 | 0.0007 |
| Stable (Canonical) | 4.0 | 0.0133 | **0.0000** | **0.0000** | **0.0000** | **0.0000** | 0.0004 | 0.0005 | 0.0005 |
| Stable (Transient Chaos) | 0.5 | 0.0220 | 0.0255 | **0.0187** | 0.0240 | 0.0216 | 0.9963 | 1.0597 | 1.3368 |
| Stable (Transient Chaos) | 1.0 | 0.0663 | **0.0092** | 0.0154 | 0.0070 | 0.0605 | 0.9981 | 1.0668 | 1.0486 |
| Stable (Transient Chaos) | 2.0 | 0.0632 | 0.0341 | 0.0579 | 0.0259 | **0.0254** | 0.9999 | 0.9267 | 1.0161 |
| Stable (Transient Chaos) | 4.0 | **0.2620** | 0.7334 | 0.6680 | 0.7053 | 0.6609 | 1.0000 | 0.9586 | 1.0130 |
| Chaotic (Butterfly) | 0.5 | 0.0023 | 0.0022 | **0.0014** | 0.0022 | 0.0016 | 0.6238 | 0.0742 | 0.0814 |
| Chaotic (Butterfly) | 1.0 | 0.0031 | 0.0039 | 0.0056 | **0.0021** | **0.0021** | 0.9933 | 0.6830 | 0.7245 |
| Chaotic (Butterfly) | 2.0 | 0.0435 | 0.1790 | **0.0123** | 0.0312 | 0.0345 | 1.0000 | 0.8705 | 0.8316 |
| Chaotic (Butterfly) | 4.0 | 0.7768 | 1.0395 | 0.7237 | **0.0563** | 0.3043 | 0.9988 | 0.9451 | 0.9224 |
| Chaotic (No Butterfly) | 0.5 | 0.0003 | 0.0007 | **0.0001** | 0.0007 | **0.0001** | 0.2397 | 0.0636 | 0.0682 |
| Chaotic (No Butterfly) | 1.0 | 0.0027 | 0.0012 | 0.0007 | **0.0003** | 0.0007 | 0.2597 | 0.1357 | 0.1375 |
| Chaotic (No Butterfly) | 2.0 | 0.0055 | 0.0059 | 0.0037 | **0.0009** | 0.0020 | 0.2871 | 0.2174 | 0.2181 |
| Chaotic (No Butterfly) | 4.0 | **0.0094** | 0.0417 | 0.0116 | 0.0245 | 0.0258 | 0.4243 | 0.3906 | 0.4194 |

**Table 8.2:** Scaled mean absolute errors (sMAE) of models towards predicting the future with training interval $T_{max} = 6$ and noise multiplier $\alpha = 1.5 \times 10^{-5}$. MSP(NP) refers to *MAGI Sequential Prediction with no pilot*. PMSP refers to *Pilot MAGI Sequential Prediction*, and the codes in parentheses correspond to the *pilot mode* PM. The model(s) that incurred the lowest error is bolded in each row. **Lower values indicate better performance.**

However, at the much higher noise level multiplier of $\alpha = 0.15$ with $T_{max} = 2.0$, we observe that PMSP variants seem to lose some of their decisive performance advantages. From Table 8.3, we see that PMSP variants still win on the Stable (Canonical) and Chaotic (Butterfly) regimes. However, the PINN variants outperform PMSP on the Chaotic (No Butterfly) variants, though the differences in error are overall much less decisive than in the $\alpha = 1.5 \times 10^{-5}$ settings. On the Stable (Transient Chaos) setting, PMSP variants win for predicting 0.5 and 1.0 time steps into the future, but PINN1 outperforms the PMSP variants for predicting 2.0 and 4.0 time steps ahead. However, this is at most a limited, pyrrhic victory for PINN because the magnitudes of the sMAE errors for all models at 2.0 and 4.0 time steps ahead imply that these predictions are all effectively unusable. We will explore in the next section what these high-error predictions actually look like.

| Regime | $\Delta T_{pred}$ | MSP(NP) | PMSP(LP) | PMSP(RP) | PMSP(LOP) | PMSP(ROP) | PINN1 | PINN2 | PINN3 |
|---|---|---|---|---|---|---|---|---|---|
| Stable (Canonical) | 0.5 | 0.0109 | 0.0074 | 0.0050 | 0.0051 | **0.0033** | 0.0311 | 0.0099 | 0.0209 |
| Stable (Canonical) | 1.0 | **0.0068** | 0.0071 | 0.0091 | 0.0090 | 0.0096 | 0.0297 | 0.0071 | 0.0212 |
| Stable (Canonical) | 2.0 | 0.0067 | 0.0082 | 0.0065 | 0.0056 | **0.0050** | 0.0214 | 0.0074 | 0.0145 |
| Stable (Canonical) | 4.0 | 0.0071 | **0.0060** | 0.0071 | 0.0075 | 0.0072 | 0.0148 | 0.0071 | 0.0109 |
| Stable (Transient Chaos) | 0.5 | 0.0616 | 0.0574 | 0.0857 | **0.0488** | 0.0748 | 0.6074 | 0.3271 | 0.6131 |
| Stable (Transient Chaos) | 1.0 | 0.1931 | 0.1834 | **0.1514** | 0.3538 | 0.2211 | 0.7730 | 0.4380 | 0.7289 |
| Stable (Transient Chaos) | 2.0 | 2.0649 | 2.3658 | 1.7159 | 1.5347 | 1.3956 | **0.7658** | 0.7975 | 0.7977 |
| Stable (Transient Chaos) | 4.0 | 1.0402 | 1.0732 | 1.9446 | 0.7855 | 0.9358 | **0.7750** | 0.8785 | 0.8794 |
| Chaotic (Butterfly) | 0.5 | **0.0656** | 0.0808 | 0.3008 | 0.0729 | 0.1191 | 0.3465 | 0.1536 | 0.4234 |
| Chaotic (Butterfly) | 1.0 | 0.2127 | 0.0904 | 0.3021 | **0.0272** | 0.1613 | 0.3757 | 0.2477 | 0.3970 |
| Chaotic (Butterfly) | 2.0 | 0.2531 | **0.1933** | 0.3438 | 0.1958 | 0.3503 | 0.3834 | 0.4196 | 0.5072 |
| Chaotic (Butterfly) | 4.0 | 1.7706 | 1.2043 | 0.8590 | **0.7819** | 0.7892 | 0.8201 | 1.1952 | 0.9189 |
| Chaotic (No Butterfly) | 0.5 | 0.1376 | 0.0929 | 0.2461 | 0.0974 | 0.2831 | 0.1577 | **0.0734** | 0.1391 |
| Chaotic (No Butterfly) | 1.0 | 0.1809 | 0.1391 | 0.2381 | 0.1567 | 0.2779 | 0.1608 | **0.0964** | 0.1178 |
| Chaotic (No Butterfly) | 2.0 | 0.1963 | 0.1900 | 0.2573 | 0.2417 | 0.4382 | 0.1692 | 0.1304 | **0.0925** |
| Chaotic (No Butterfly) | 4.0 | 0.2083 | 0.1976 | 0.2561 | 0.2866 | 0.5363 | 0.1921 | 0.1795 | **0.1551** |

**Table 8.3:** Scaled mean absolute errors (sMAE) of models towards predicting the future with training interval $T_{max} = 2$ and noise multiplier $\alpha = 0.15$. MSP(NP) refers to *MAGI Sequential Prediction with no pilot*. PMSP refers to *Pilot MAGI Sequential Prediction*, and the codes in parentheses correspond to the *pilot mode* PM. The model(s) that incurred the lowest error is bolded in each row. **Lower values indicate better performance.**

Finally, at high noise $\alpha = 0.15$ with $T_{max} = 6.0$, from Table 8.4, we observe that while PINN2 and PINN3 outperform the PMSP variants on Stable (Canonical), the actual sMAE error values are very, very close to each other and not indicative of a decisive victory. On the remaining three regimes, PMSP variants consistently outperform the PINN variants, oftentimes with decisively smaller errors.

| Regime | $\Delta T_{pred}$ | MSP(NP) | PMSP(LP) | PMSP(RP) | PMSP(LOP) | PMSP(ROP) | PINN1 | PINN2 | PINN3 |
|---|---|---|---|---|---|---|---|---|---|
| Stable (Canonical) | 0.5 | 0.0035 | 0.0040 | 0.0034 | 0.0043 | 0.0035 | 0.0035 | **0.0020** | 0.0022 |
| Stable (Canonical) | 1.0 | 0.0038 | 0.0038 | 0.0032 | 0.0039 | 0.0039 | 0.0037 | **0.0030** | 0.0031 |
| Stable (Canonical) | 2.0 | 0.0066 | 0.0035 | 0.0035 | 0.0039 | 0.0080 | 0.0036 | **0.0032** | **0.0032** |
| Stable (Canonical) | 4.0 | 0.0284 | 0.0047 | 0.0054 | 0.0074 | 0.0108 | 0.0035 | 0.0036 | **0.0033** |
| Stable (Transient Chaos) | 0.5 | 0.1500 | 0.2535 | 0.2733 | **0.1246** | 0.2407 | 0.9942 | 1.4885 | 1.1612 |
| Stable (Transient Chaos) | 1.0 | 0.6113 | 0.4907 | **0.1713** | 0.3647 | 0.4032 | 0.9887 | 1.1576 | 1.0339 |
| Stable (Transient Chaos) | 2.0 | 0.7341 | 0.7018 | **0.2508** | 0.4461 | 0.5896 | 0.9923 | 1.0104 | 1.0100 |
| Stable (Transient Chaos) | 4.0 | **0.7997** | 2.4401 | 1.3265 | 1.9178 | 1.2932 | 0.9989 | 0.9767 | 1.0041 |
| Chaotic (Butterfly) | 0.5 | 0.1528 | 0.1707 | 0.1132 | 0.1290 | **0.0976** | 0.5770 | 0.4113 | 0.5594 |
| Chaotic (Butterfly) | 1.0 | 0.3291 | **0.1606** | 0.2438 | 0.4607 | 0.2824 | 1.0000 | 0.5647 | 0.8645 |
| Chaotic (Butterfly) | 2.0 | **0.3054** | 1.1322 | 0.3756 | 0.9622 | 0.8246 | 1.0000 | 0.8689 | 0.9246 |
| Chaotic (Butterfly) | 4.0 | 2.3444 | 1.2732 | 2.5630 | 0.9448 | **0.8940** | 0.9981 | 1.0687 | 0.9614 |
| Chaotic (No Butterfly) | 0.5 | 0.0958 | 0.0865 | 0.0699 | **0.0531** | 0.0557 | 0.2496 | 0.2232 | 0.2483 |
| Chaotic (No Butterfly) | 1.0 | 0.1114 | 0.1245 | 0.0873 | 0.1240 | **0.0868** | 0.2755 | 0.2653 | 0.2714 |
| Chaotic (No Butterfly) | 2.0 | **0.1114** | 0.1185 | 0.1204 | 0.1427 | 0.1205 | 0.2962 | 0.2953 | 0.2904 |
| Chaotic (No Butterfly) | 4.0 | **0.1830** | 0.3701 | 0.2596 | 0.3633 | 0.3620 | 0.4421 | 0.4076 | 0.4337 |

**Table 8.4:** Scaled mean absolute errors (sMAE) of models towards predicting the future with training interval $T_{max} = 6$ and noise multiplier $\alpha = 0.15$. MSP(NP) refers to *MAGI Sequential Prediction with no pilot*. PMSP refers to *Pilot MAGI Sequential Prediction*, and the codes in parentheses correspond to the *pilot mode* PM. The model(s) that incurred the lowest error is bolded in each row. **Lower values indicate better performance.**

Taking all tables into account, we can confidently argue that, generally, **PMSP outperforms PINN on the ODE-based dynamical systems prediction task**, as evaluated using our four Lorenz test beds. Note that we have not even considered PMSP's inherent Bayesian uncertainty quantification, relative interpretability, and parameter efficiency yet — all of which are certainly advantages in favor of PMSP.

## 8.5 Selected Settings for Detailed Analysis

Referring back to Figure 8.1 from the start of this chapter, at $\alpha = 1.5 \times 10^{-5}$, recall that for the Chaotic (Butterfly) regime under the specified settings, all three PINN variants eventually flat-line in their predictions, while all PMSP variants shown accurately output predictions right on top of the ground truth.

Just to provide another visual example of PMSP's performance dominance, Figure 8.3 juxtaposes the future predictions of some PMSP variants versus that of the PINN variants on the Stable (Transient Chaos) regime at $\alpha = 1.5 \times 10^{-5}$. Indeed, this regime is much more difficult to output predictions for, given the change in dynam-ics around $t = 5.0$. Yet, all three PMSP variants shown are still able to output accurate predictions for at least $\Delta T_{pred} = 3.0$ timesteps into the future, while all PINN variants quickly devolve towards outputting a flat line as their prediction. At $\alpha = 1.5 \times 10^{-5}$, the prediction curves of PINN and PMSP on the Stable (Canonical) and Chaotic (No Butterfly) regimes tell very similar stories, so we relegate these corresponding figures to Appendix A.8 to avoid redundancy.

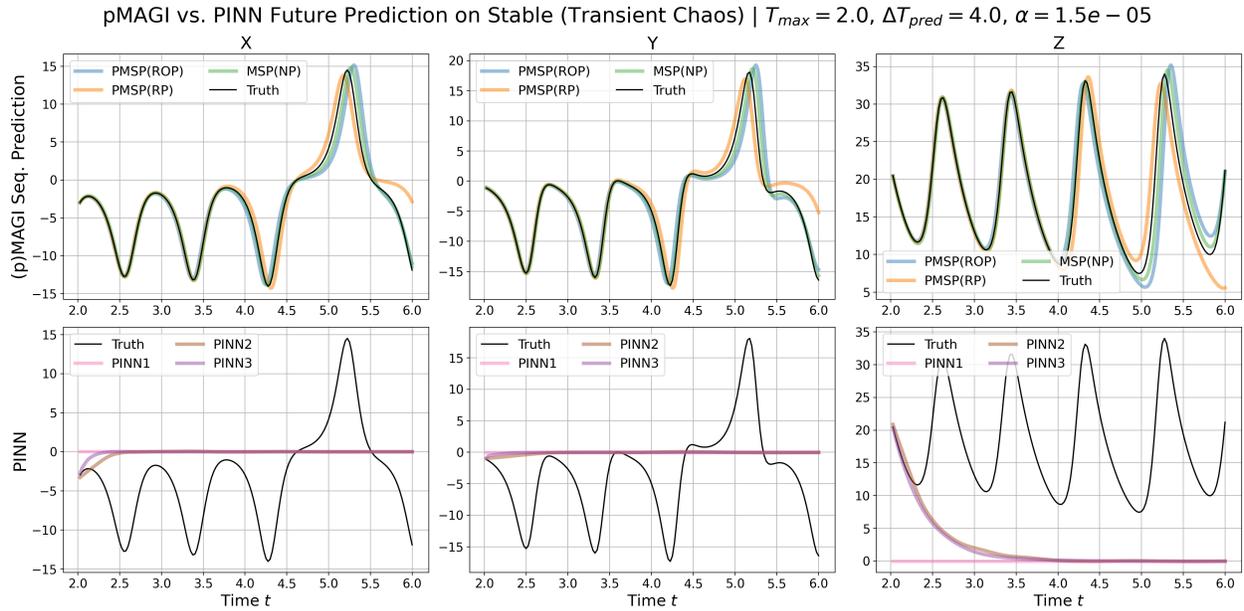

**Figure 8.3:** Selected Pilot MAGI Sequential Prediction (PMSP) vs. PINN future prediction trajectories on the Stable (Transient Chaos) regime at $\alpha = 1.5 \times 10^{-5}$, $T_{max} = 2.0$, and $\Delta T_{fc} = 4.0$. PMSP model variants are shown in the first row, while PINN models are shown in the second row, with all prediction curves color-coded by model variant. The x-axis for each subplot is time $t$ on the prediction interval $t \in [2.0, 2.0 + \Delta T_{\text{pred}}]$. Visually, we can estimate model performance by examining how closely the predicted trajectories match the ground truth (in black).

In Figure 8.4, we examine PMSP vs. PINN prediction curves on the Chaotic (No Butterfly) regime at $\alpha = 0.15$ — a set of settings where PINN2 and PINN3 incurred lower prediction errors than our PMSP variants at $T_{max} = 2.0$, and at all prediction horizons $\Delta T_{\text{pred}} = \{0.5, 1.0, 2.0, 4.0\}$. Yet, looking at the prediction curves, it is clear that the predictions generated by PINN1, PINN2, and PINN3 are all simply unusable for any practical purposes past $\Delta T_{\text{pred}} = 1.0$ at best. Both PINN and PMSP do not generate remotely useful prediction curves, and it is clear from the subplots that PINN only appears to accrue lower prediction errors because PMSP's predictions seemed to be shifted a few units upwards and away from the mean of the ground truth trajectories. Because the other three regimes' figures convey similar messages, we relegate them to Appendix A.8 as well.

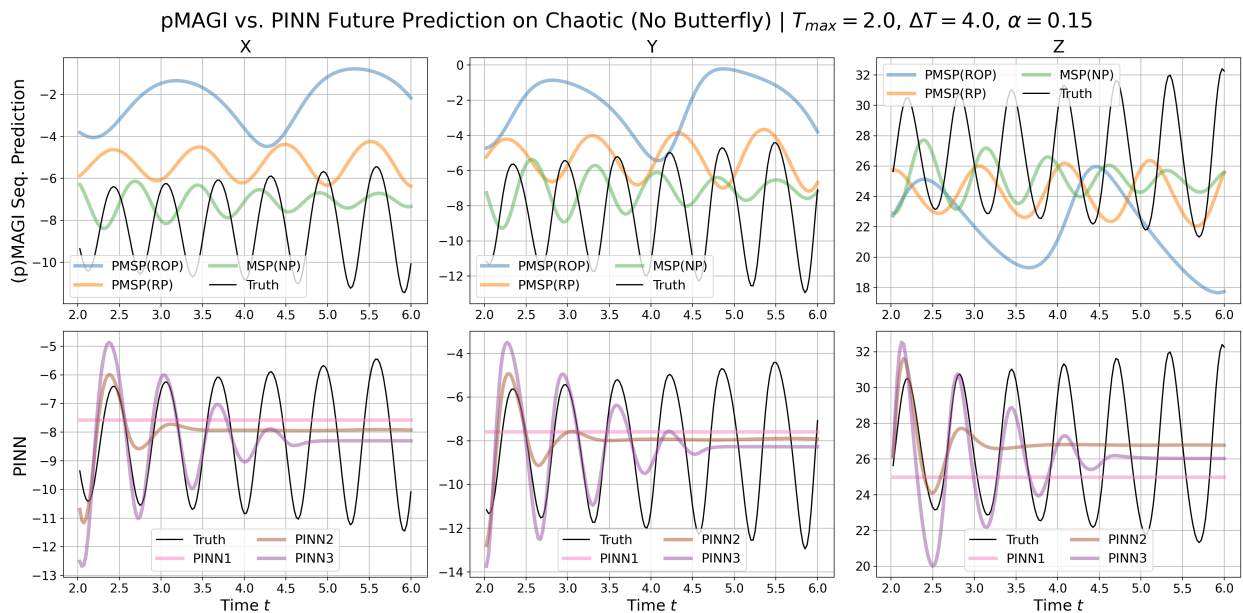

**Figure 8.4:** Selected Pilot MAGI Sequential Prediction (PMSP) vs. PINN future prediction trajectories on the Chaotic (No Butterfly) regime at $\alpha = 0.15$, $T_{max} = 2.0$, and $\Delta T_{\text{pred}} = 4.0$. PMSP model variants are shown in the first row, while PINN models are shown in the second row, with all prediction curves color-coded by model variant. The x-axis for each subplot is time $t$ on the prediction interval $t \in [2.0, 2.0 + \Delta T_{\text{pred}}]$. Visually, we can estimate model performance by examining how closely the predicted trajectories match the ground truth (in black).

## 8.6 Main Takeaways

In sum, it is very clear that, conditional on ability to output practically-usable future predictions, **PMSP is a decisively stronger future prediction method than PINN**, especially at low noise levels. Of course, PMSP also has the critical advantage of having natural uncertainty quantification, which none of the PINN-based methods we tested can attest to. Indeed, PSMP is a powerful new tool for ODE future prediction.

# 9
# Discussion, Future Work, and Conclusion

## 9.1 Discussion

In this thesis, we introduced two novel methods building off of the MAGI method[55]. First, we introduced Pilot MAGI (pMAGI), a numerically-stabler and higher-performance upgrade of the base MAGI method. Second, we introduced Pilot MAGI Sequential Prediction (PMSP), an exciting new tool that allows one to predict the trajectory of ODE-based dynamical systems multiple timesteps into the future, given only sparse and noisy ob-servations as input. Through extensive benchmarking on our four Lorenz-based test beds, we demonstrate that compared to competitor methods, especially the Physics-Informed Neural Network (PINN), pMAGI and PSMP are not only significantly more computationally- and parameter-efficient, but also consistently higher-performing in many settings on the tasks of parameter inference, trajectory reconstruction, and future prediction. Of course, in contrast to the state-of-the-art competitor methods, pMAGI and PMSP also come with natural uncertainty quantification, given their Bayesian model foundations. **By any definition, pMAGI and PMSP are powerful, Bayesian, uncertainty-quantified competitors to the PINN.**

During our investigations of pMAGI's inference abilities, we also encountered some potential new insights regarding the inherent identifiability of certain Lorenz-based dynamical systems — posing fascinating questions at the intersection of computational methods and dynamical systems theory. We also explored applying pMAGI towards performing probabilistic binary classification on whether an observed system was stable or chaotic, given only sparse and noisy observations. In the process, we proposed an MCMC-based estimator that proved to be quite well-calibrated and intuitive for most noise levels tested.

## 9.2 Future Work

### 9.2.1 Missing Component Situations

One natural line of questions left unanswered by this work is: how would pMAGI, Particle Swarm Optimization (PSO), Differential Evolution (DE), and PINN fare under missing component situations? Specifically, what if one only observed the *X* and *Y* components of the Lorenz system? Can we still infer the parameters governing the system? Can we perhaps reconstruct the missing component *Z*'s trajectories? How would our answers change between stable and chaotic regimes? Are the parameters $\theta = (\beta, \rho, \sigma)$ even uniquely-identifiable if we only observe some components of the system? Do we at least need to specify some initial condition on *Z*, following the paradigm of numerical-integration? We have already begun preliminary explorations of these questions, though there is still much work to be done. For one, initial testing suggests that $(\phi_1, \phi_2)$ estimation — a crucial compo-nent of pMAGI — becomes very numerically unstable under missing component situations. Can we construct a more robust $(\phi_1, \phi_2)$ hyperparameter estimation procedure?

### 9.2.2 APPLICATIONS TO INFECTIOUS DISEASE FORECASTING

Given the success of PMSP, one immediate direction of future work is to apply our methodology towards real-world forecasting problems. In the near future, we hope to explore the combination of PMSP and compartmental ODE models, such as the basic SIR model [20], towards characterizing and forecasting the spread of infectious diseases. Compared to deep-learning, autoregressive, and various black-box models, PMSP-powered disease forecasting models come equipped with the natural advantages of interpretability (we can directly examine the posterior distributions of the parameters governing the spread of disease) and uncertainty quantification without the need for additional tools like conformal prediction. Given that many infectious diseases have dynamics which change over time, another line of exciting work would be to modify pMAGI and PMSP for online parameter inference, trajectory reconstruction, and future prediction.

### 9.2.3 FURTHER COMPUTATIONAL EFFICIENCY

While pMAGI yields significant numerical-stability improvements over MAGI, it is still by no means perfect. We also know that Hamiltonian Monte Carlo (HMC) will struggle in very high-dimensional sampling settings. As such, another direction of future work is to explore replacing the HMC sampling component in pMAGI with a potentially even more computationally-efficient and numerically-stabler Variational Bayes' module [48] that approximate the posterior distribution rather than directly samples from it.

## 9.3 CONCLUSION

We hope researchers and other end-users in any discipline that employs ODEs for modeling natural phenomena can benefit from our contributions of pMAGI and PMSP.

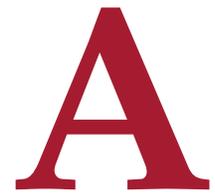

# A
# Additional Figures

These appendices contain figures that supplement main ideas and arguments from the main text. Oftentimes, the performance trends and/or differences of various models are very consistent over testbed regimes, so to avoid unnecessary repetition, we chose to showcase the most representative and salient examples in the main text and relegate the remaining results to the appendices. Other times, for larger figure panels with 20-30+ subplots, we chose to zoom in on only a subset of these subplots for readability and flow in the main text and relegate the full figure panels to the appendices.

## A.1 Additional Figures for Numerical Problems with MAGI (All Regimes)

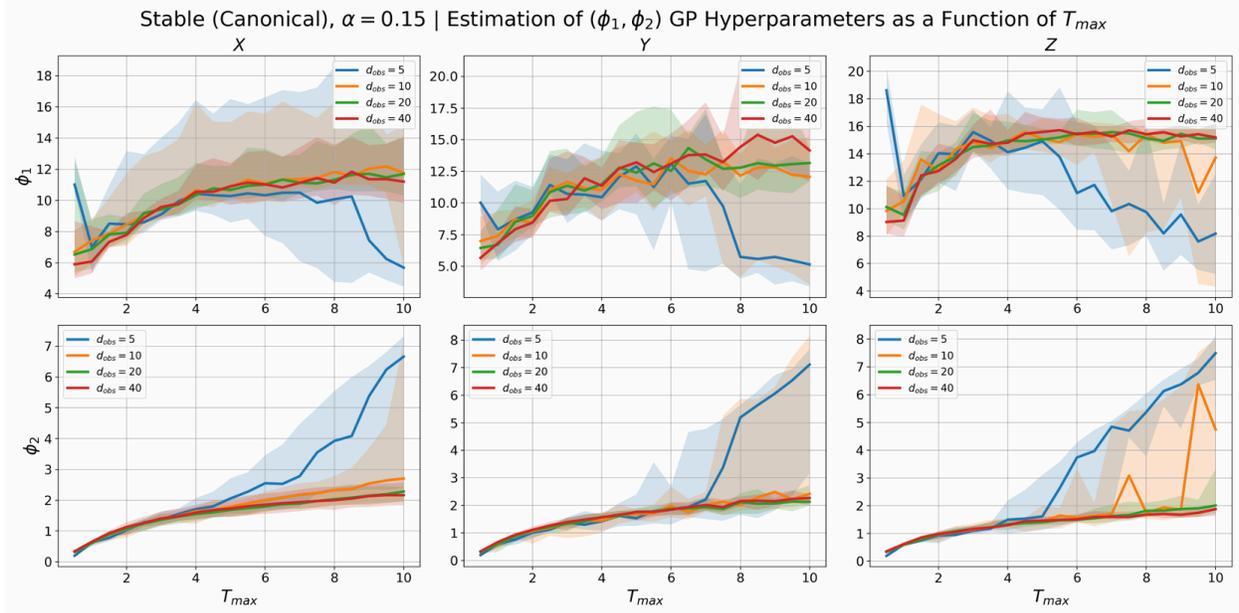

**Figure A.1:** Estimated $(\phi_1, \phi_2)$ values (y-axis) for each Lorenz component as a function of $T_{max}$ (x-axis) on the Stable (Canonical) regime with noise level multiplier $\alpha = 0.15$, averaged across ten randomly-seeded datasets. Specifically, we are showing estimated $(\phi_1, \phi_2)$ values on the interval $t \in [0, T_{max}]$ as points on the curve. The colors indicate the density of noisy observations $d_{obs}$ per unit time. The correspondingly-colored intervals demonstrate the minimum and maximum estimated hyperparameter values across the ten randomly-seeded datasets.

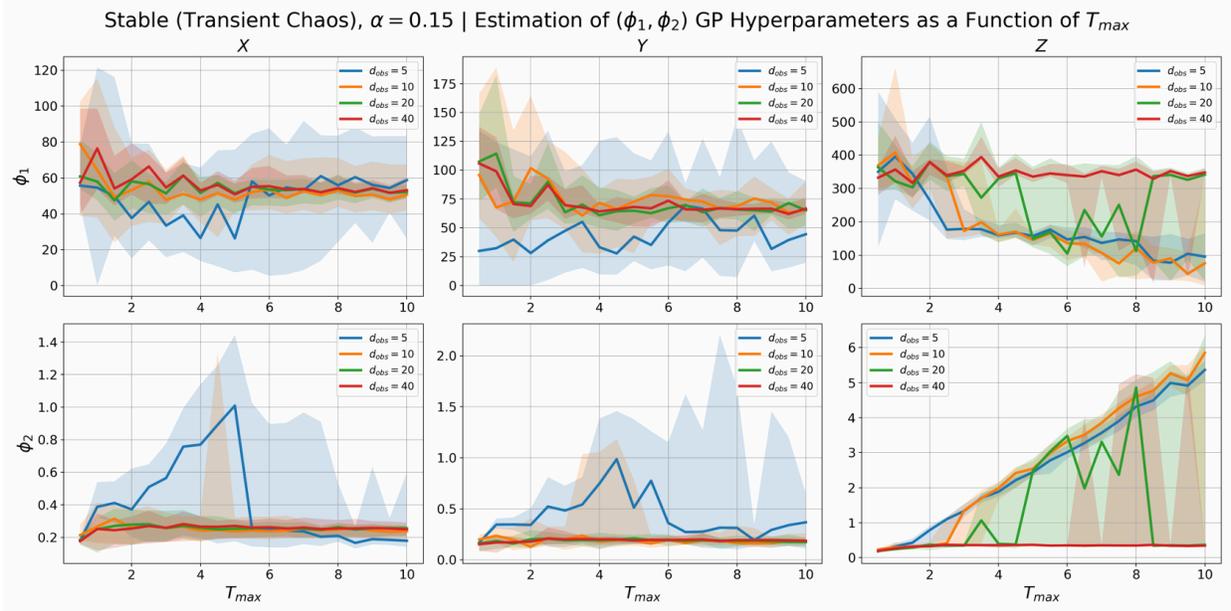

**Figure A.2:** Estimated $(\phi_1, \phi_2)$ values (y-axis) for each Lorenz component as a function of $T_{max}$ (x-axis) on the Stable (Transient Chaos) regime with noise level multiplier $\alpha = 0.15$, averaged across ten randomly-seeded datasets. Specifically, we are showing estimated $(\phi_1, \phi_2)$ values on the interval $t \in [0, T_{max}]$ as points on the curve. The colors indicate the density of noisy observations $d_{obs}$ per unit time. The correspondingly-colored intervals demonstrate the minimum and maximum estimated hyperparameter values across the ten randomly-seeded datasets.

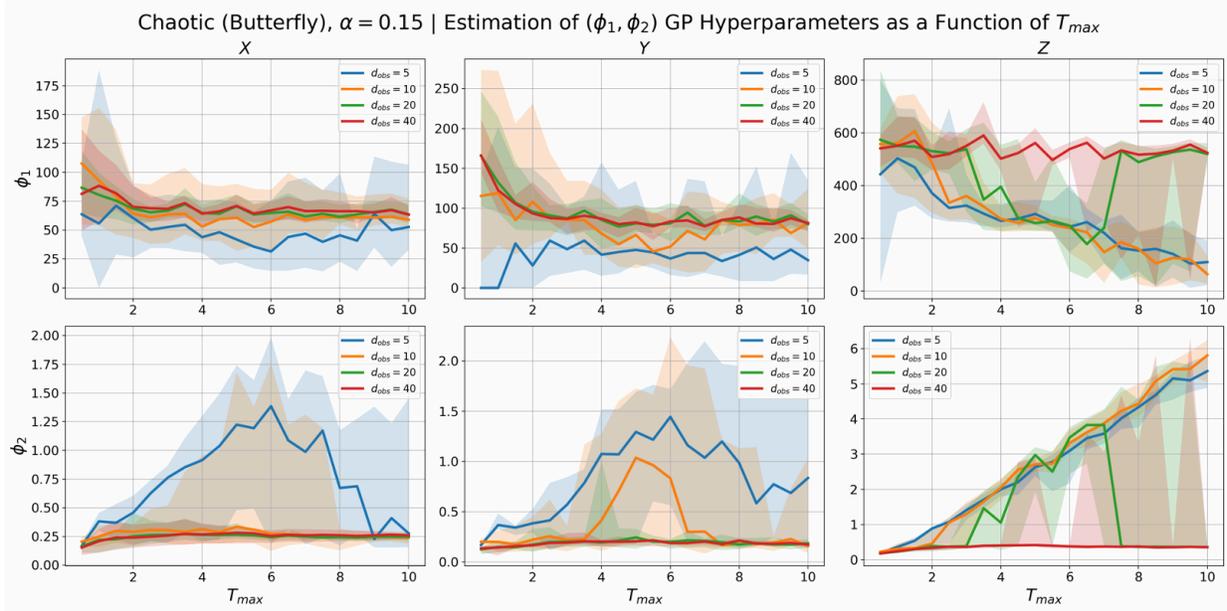

**Figure A.3:** Estimated $(\phi_1, \phi_2)$ values (y-axis) for each Lorenz component as a function of $T_{max}$ (x-axis) on the Chaotic (Butterfly) regime with noise level multiplier $\alpha = 0.15$, averaged across ten randomly-seeded datasets. Specifically, we are showing estimated $(\phi_1, \phi_2)$ values on the interval $t \in [0, T_{max}]$ as points on the curve. The colors indicate the density of noisy observations $d_{obs}$ per unit time. The correspondingly-colored intervals demonstrate the minimum and maximum estimated hyperparameter values across the ten randomly-seeded datasets.

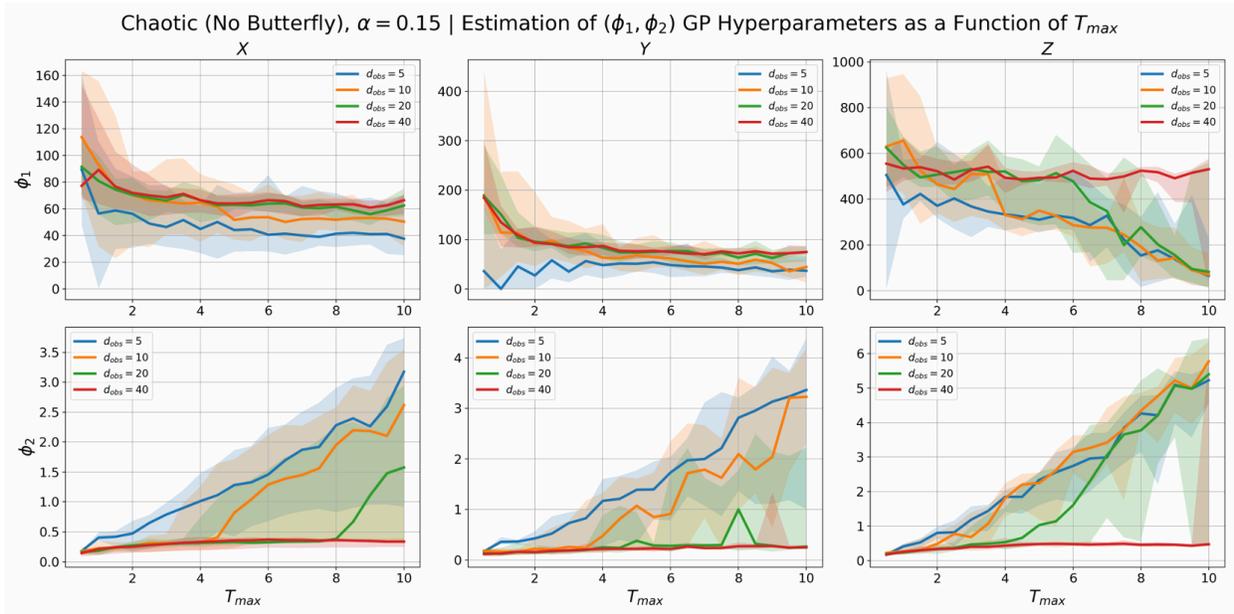

**Figure A.4:** Estimated $(\phi_1, \phi_2)$ values (y-axis) for each Lorenz component as a function of $T_{max}$ (x-axis) on Chaotic (No Butterfly) regime with noise level multiplier $\alpha = 0.15$, averaged across ten randomly-seeded datasets. Specifically, we are showing estimated $(\phi_1, \phi_2)$ values on the interval $t \in [0, T_{max}]$ as points on the curve. The colors indicate the density of noisy observations $d_{obs}$ per unit time. The correspondingly-colored intervals demonstrate the minimum and maximum estimated hyperparameter values across the ten randomly-seeded datasets.

## A.2 Additional Figures for pMAGI vs. MAGI Parameter Inference (All Regimes)

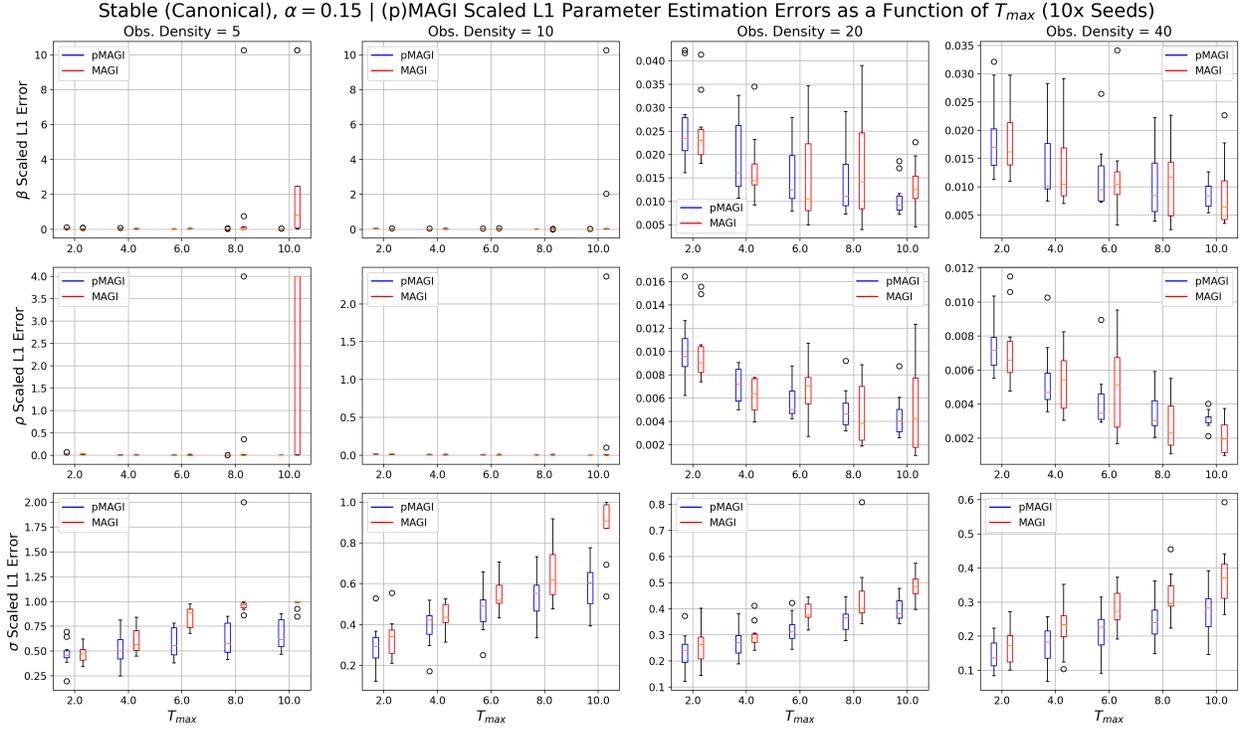

**Figure A.5:** pMAGI vs. MAGI Parameter Inference on the Stable (Canonical) regime. Each row corresponds to one parameter in $(\beta, \rho, \sigma)$ and each column corresponds to the density of noised observations per unit time $d_{obs} \in \{5, 10, 20, 40\}$. The $x$-axis for each subplot corresponds to the length of the observation interval, i.e. $t \in [0.0, T_{max}]$. The box-and-whisker plots show the distribution of the scaled L1 parameter estimation errors for the best pMAGI vs. MAGI variants across the 10 random-seeded datasets. Blue (always on the left) corresponds to pMAGI while red (always on the right) corresponds to MAGI. **Lower errors indicate better performance.**

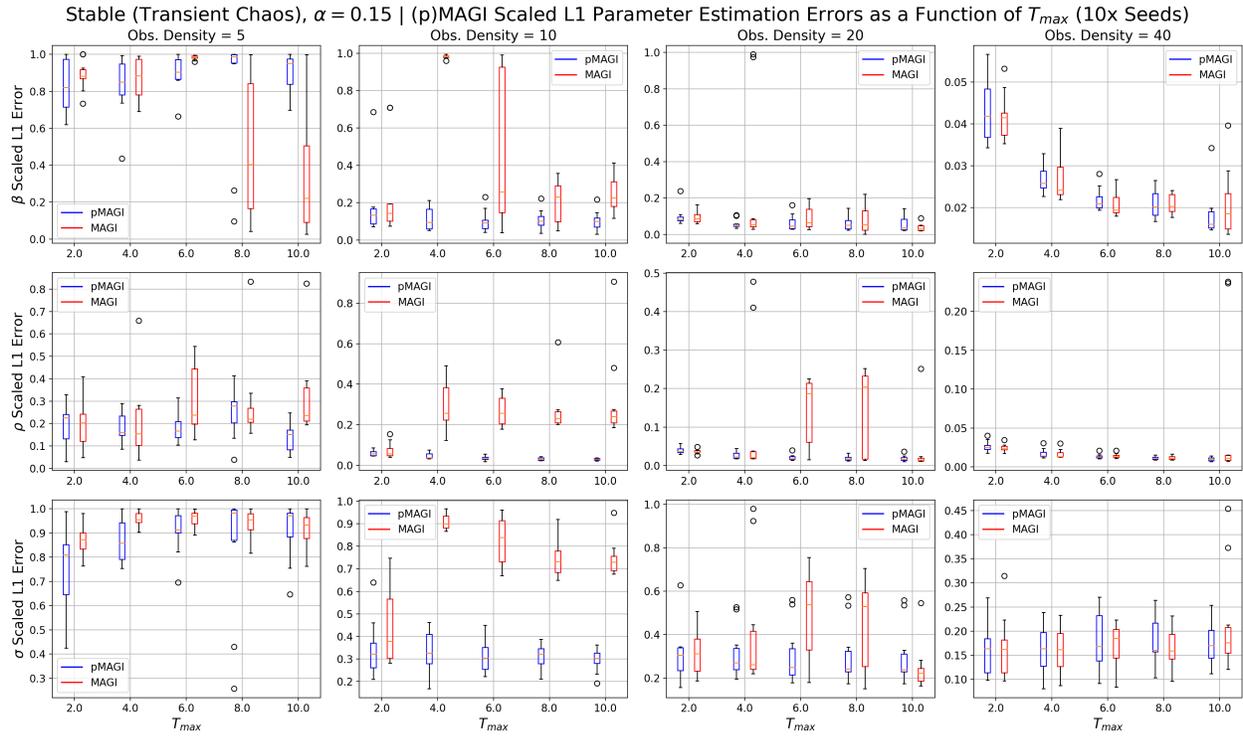

**Figure A.6:** pMAGI vs. MAGI Parameter Inference on the Stable (Transient Chaos) regime. Each row corresponds to one parameter in $(\beta, \rho, \sigma)$ and each column corresponds to the density of noised observations per unit time $d_{obs} \in \{5, 10, 20, 40\}$. The $x$-axis for each subplot corresponds to the length of the observation interval, i.e. $t \in [0.0, T_{max}]$. The box-and-whisker plots show the distribution of the scaled L1 parameter estimation errors for the best pMAGI vs. MAGI variants across the 10 random-seeded datasets. Blue (always on the left) corresponds to pMAGI while red (always on the right) corresponds to MAGI. **Lower errors indicate better performance.**

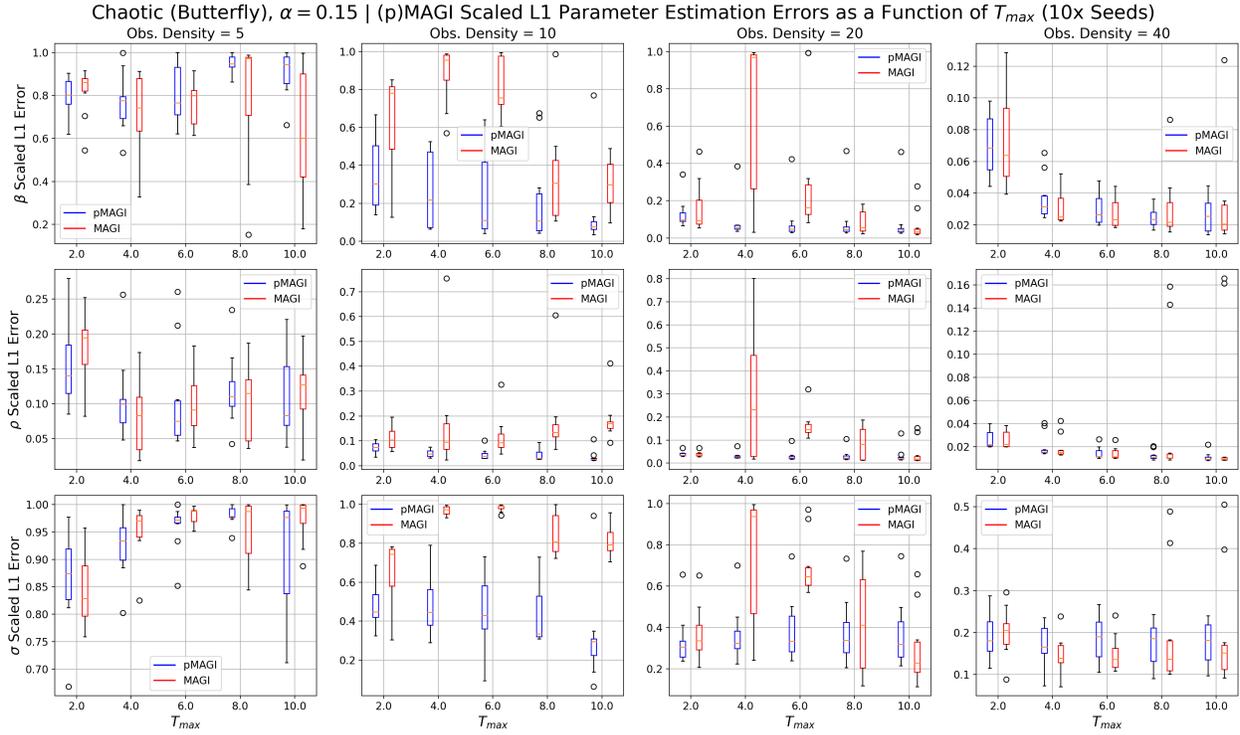

**Figure A.7:** pMAGI vs. MAGI Parameter Inference on the Chaotic (Butterfly) regime. Each row corresponds to one parameter in $(\beta, \rho, \sigma)$ and each column corresponds to the density of noised observations per unit time $d_{obs} \in \{5, 10, 20, 40\}$. The $x$-axis for each subplot corresponds to the length of the observation interval, i.e. $t \in [0.0, T_{max}]$. The box-and-whisker plots show the distribution of the scaled L1 parameter estimation errors for the best pMAGI vs. MAGI variants across the 10 random-seeded datasets. Blue (always on the left) corresponds to pMAGI while red (always on the right) corresponds to MAGI. **Lower errors indicate better performance.**

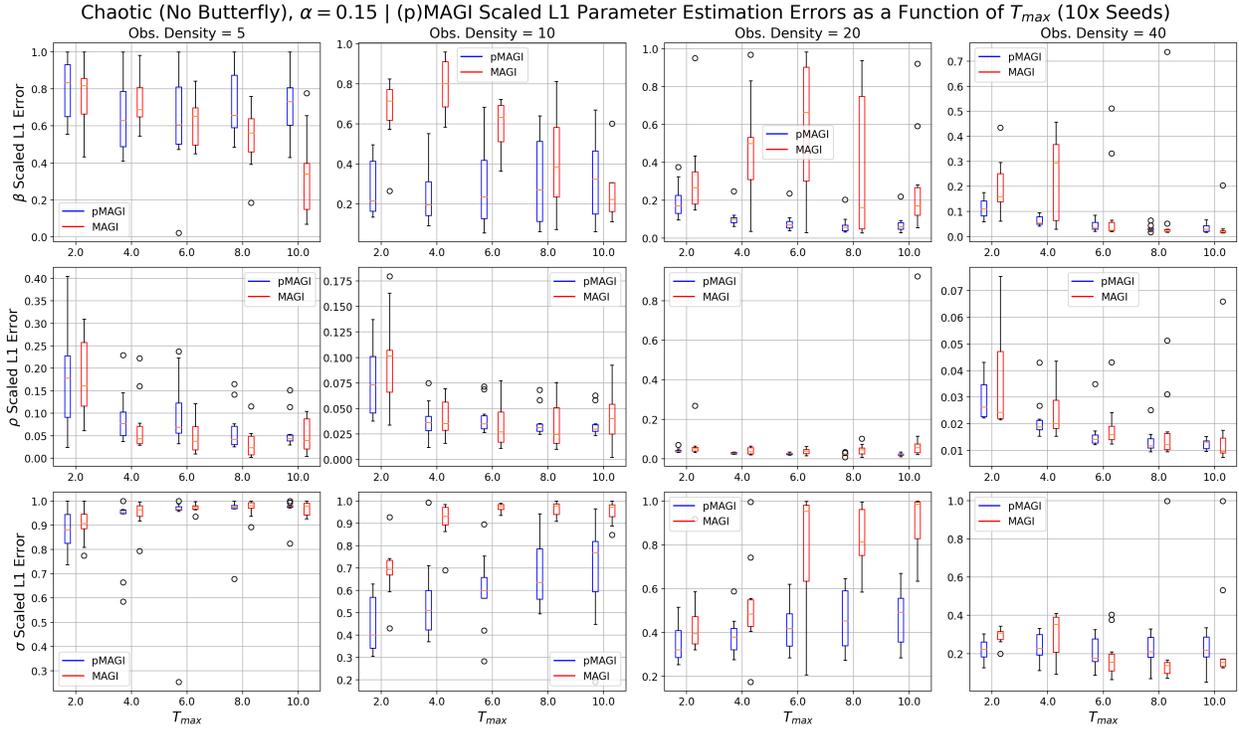

**Figure A.8:** pMAGI vs. MAGI Parameter Inference on the Chaotic (No Butterfly) regime. Each row corresponds to one parameter in $(\beta, \rho, \sigma)$ and each column corresponds to the density of noised observations per unit time $d_{obs} \in \{5, 10, 20, 40\}$. The $x$-axis for each subplot corresponds to the length of the observation interval, i.e. $t \in [0.0, T_{max}]$. The box-and-whisker plots show the distribution of the scaled L1 parameter estimation errors for the best pMAGI vs. MAGI variants across the 10 random-seeded datasets. Blue (always on the left) corresponds to pMAGI while red (always on the right) corresponds to MAGI. **Lower errors indicate better performance.**

## A.3 Additional Figures for pMAGI vs. MAGI Trajectory Reconstruction (All Regimes)

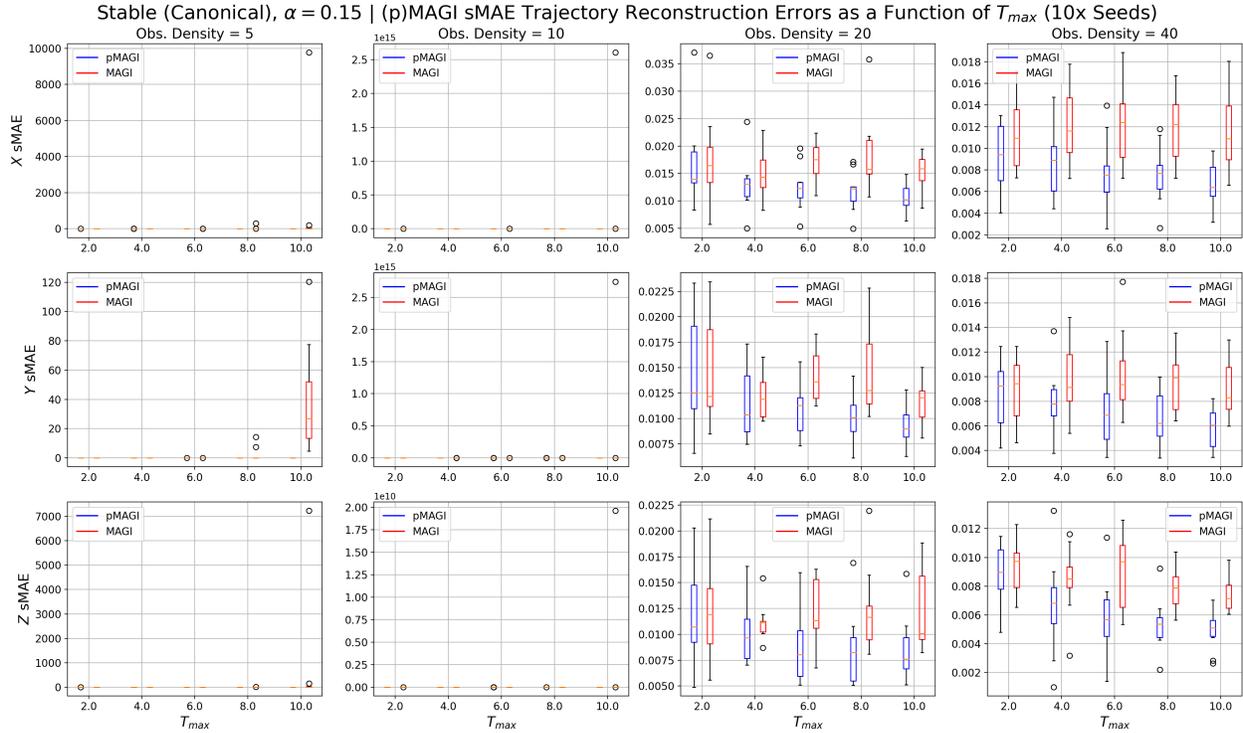

**Figure A.9:** pMAGI vs. MAGI Trajectory Reconstruction on the Stable (Canonical) regime. Each row corresponds to one component in $(X, Y, Z)$ and each column corresponds to the density of noised observations per unit time $d_{obs} \in \{5, 10, 20, 40\}$. The $x$-axis for each subplot corresponds to the length of the observation interval, i.e. $t \in [0.0, T_{max}]$. The box-and-whisker plots show the distribution of the sMAE errors for the best pMAGI vs. MAGI variants across the 10 random-seeded datasets. Blue (always on the left) corresponds to pMAGI while red (always on the right) corresponds to MAGI. **Lower errors indicate better performance.**

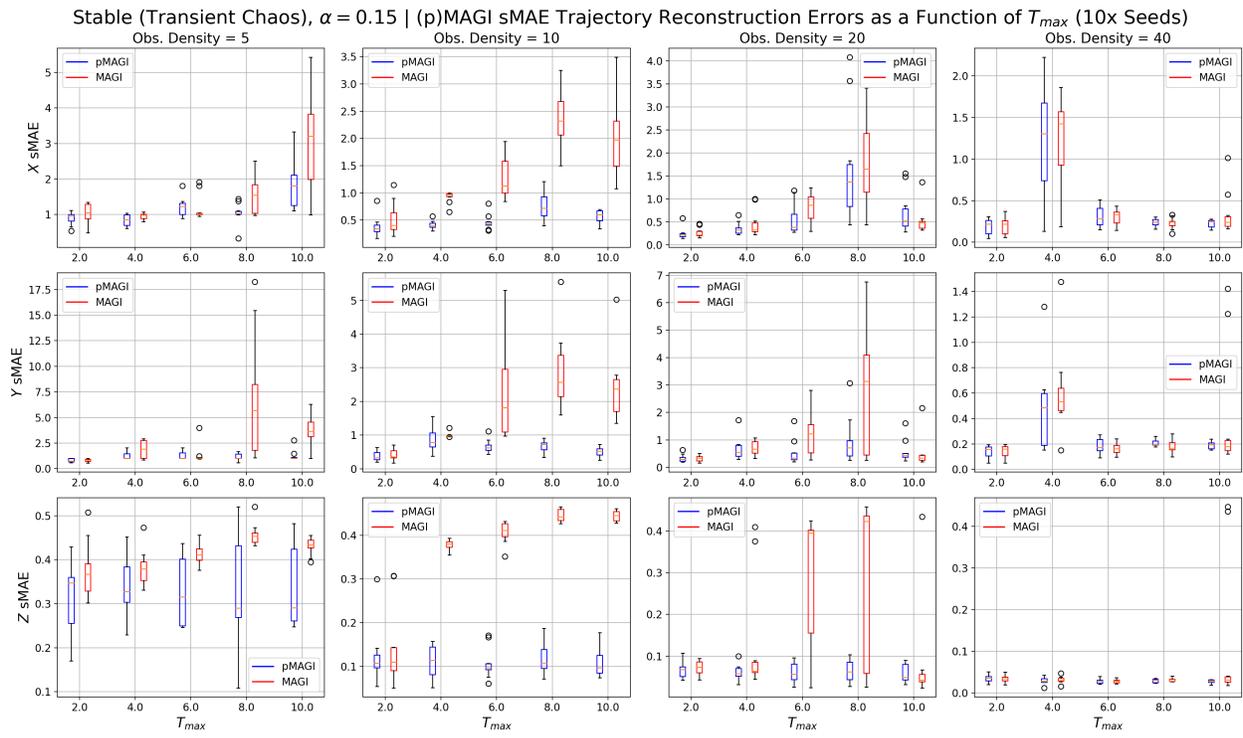

**Figure A.10:** pMAGI vs. MAGI Trajectory Reconstruction on the Stable (Transient Chaos) regime. Each row corresponds to one component in $(X, Y, Z)$ and each column corresponds to the density of noised observations per unit time $d_{obs} \in \{5, 10, 20, 40\}$. The $x$-axis for each subplot corresponds to the length of the observation interval, i.e. $t \in [0.0, T_{max}]$. The box-and-whisker plots show the distribution of the sMAE errors for the best pMAGI vs. MAGI variants across the 10 random-seeded datasets. Blue (always on the left) corresponds to pMAGI while red (always on the right) corresponds to MAGI. **Lower errors indicate better performance.**

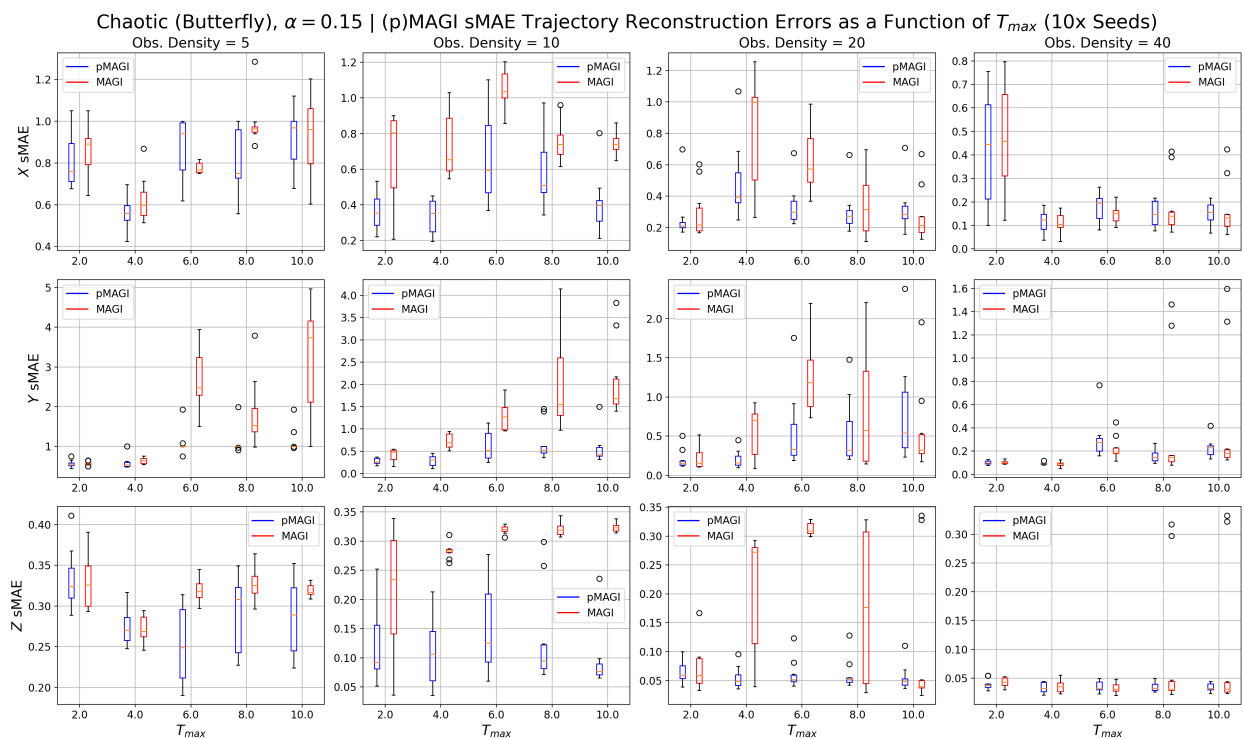

**Figure A.11:** pMAGI vs. MAGI Trajectory Reconstruction on the Chaotic (Butterfly) regime. Each row corresponds to one component in $(X, Y, Z)$ and each column corresponds to the density of noised observations per unit time $d_{obs} \in \{5, 10, 20, 40\}$. The $x$-axis for each subplot corresponds to the length of the observation interval, i.e. $t \in [0.0, T_{max}]$. The box-and-whisker plots show the distribution of the sMAE errors for the best pMAGI vs. MAGI variants across the 10 random-seeded datasets. Blue (always on the left) corresponds to pMAGI while red (always on the right) corresponds to MAGI. **Lower errors indicate better performance.**

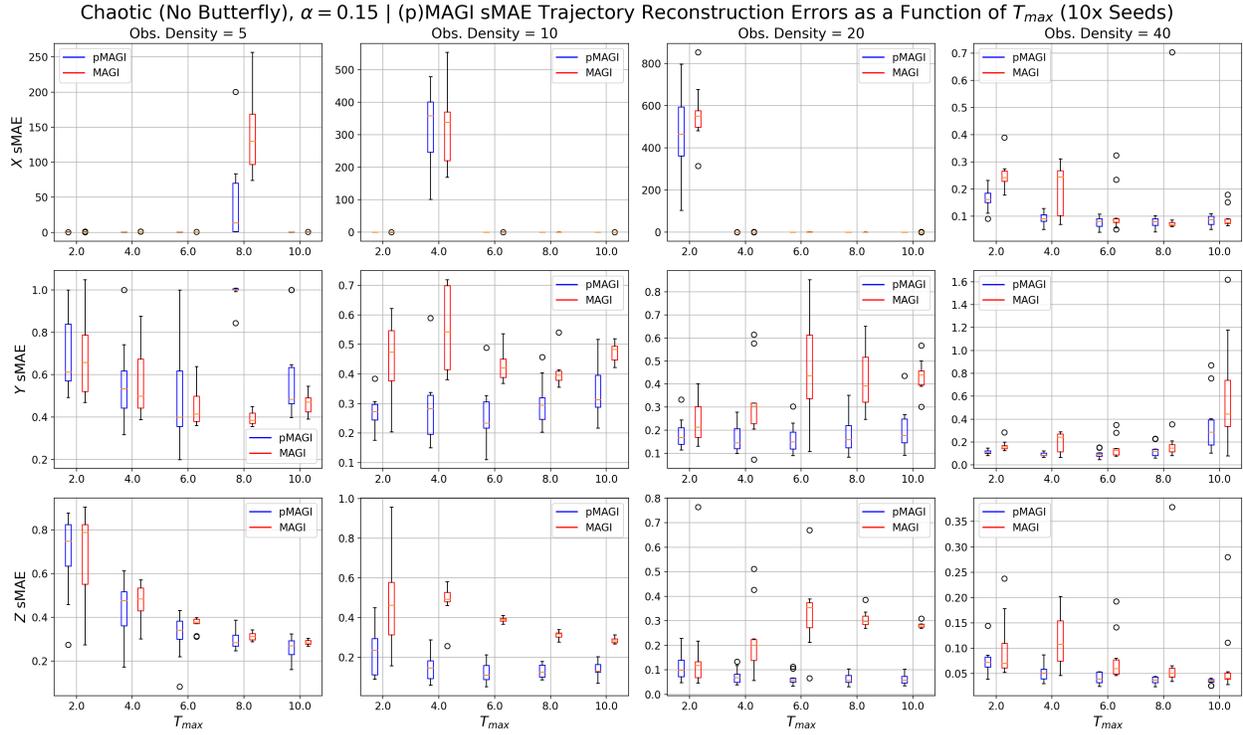

**Figure A.12:** pMAGI vs. MAGI Trajectory Reconstruction on the Chaotic (No Butterfly) regime. Each row corresponds to one component in $(X, Y, Z)$ and each column corresponds to the density of noised observations per unit time $d_{obs} \in \{5, 10, 20, 40\}$. The $x$-axis for each subplot corresponds to the length of the observation interval, i.e. $t \in [0.0, T_{max}]$. The box-and-whisker plots show the distribution of the sMAE errors for the best pMAGI vs. MAGI variants across the 10 random-seeded datasets. Blue (always on the left) corresponds to pMAGI while red (always on the right) corresponds to MAGI. **Lower errors indicate better performance.**

## A.4 Additional Figures for pMAGI vs. PSO and DE Parameter Inference (All Regimes)

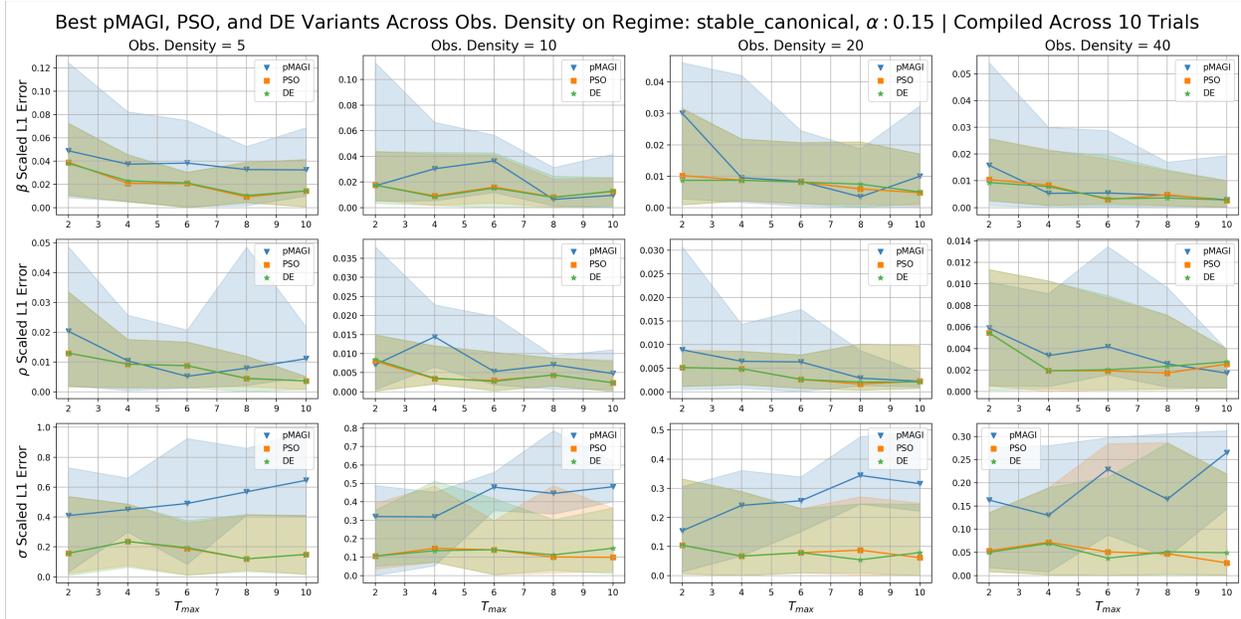

**Figure A.13:** pMAGI vs. PSO and DE Parameter Inference on the Stable (Canonical) regime. The x-axis of each subplot is $T_{max}$, the length of our noisy observation interval $t \in [0, T_{max}]$. Blue represents pMAGI, orange represents PSO, and green represents DE. The solid lines indicate the average errors across all ten randomly-seeded trials. The colored confidence bands represent the maximum and minimum errors accrued on the given setting across ten randomly-seeded trials. **Lower errors indicate better performance.**

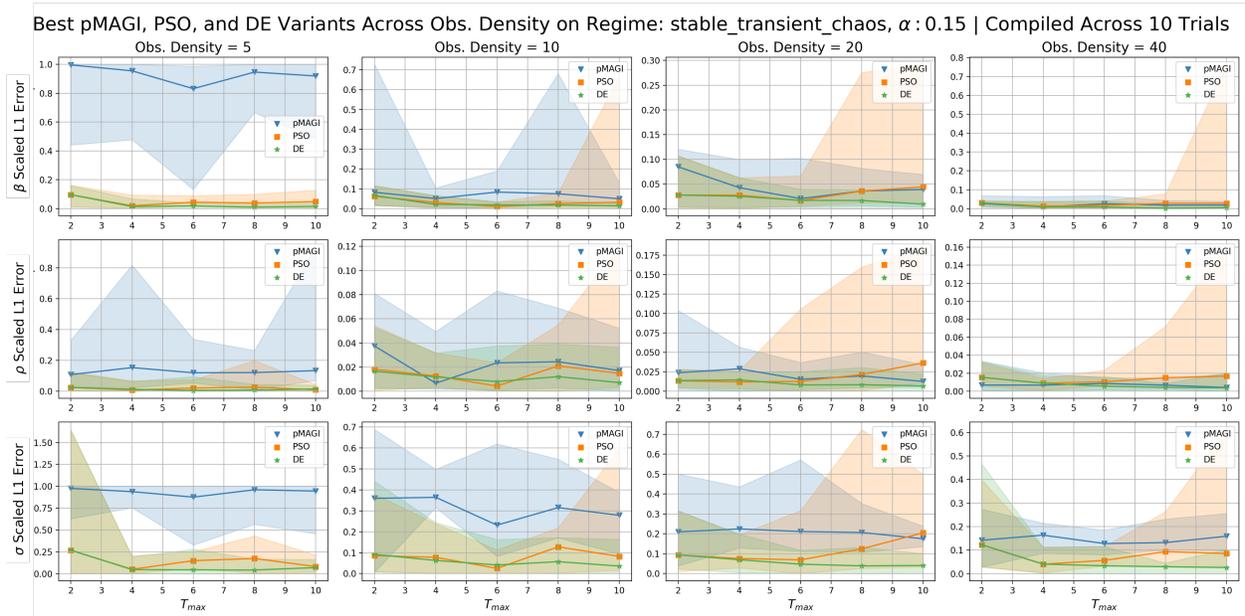

**Figure A.14:** pMAGI vs. PSO and DE Parameter Inference on the Stable (Transient Chaos) regime. The x-axis of each subplot is $T_{max}$, the length of our noisy observation interval $t \in [0, T_{max}]$. Blue represents pMAGI, orange represents PSO, and green represents DE. The solid lines indicate the average errors across all ten randomly-seeded trials. The colored confidence bands represent the maximum and minimum errors accrued on the given setting across ten randomly-seeded trials. **Lower errors indicate better performance.**

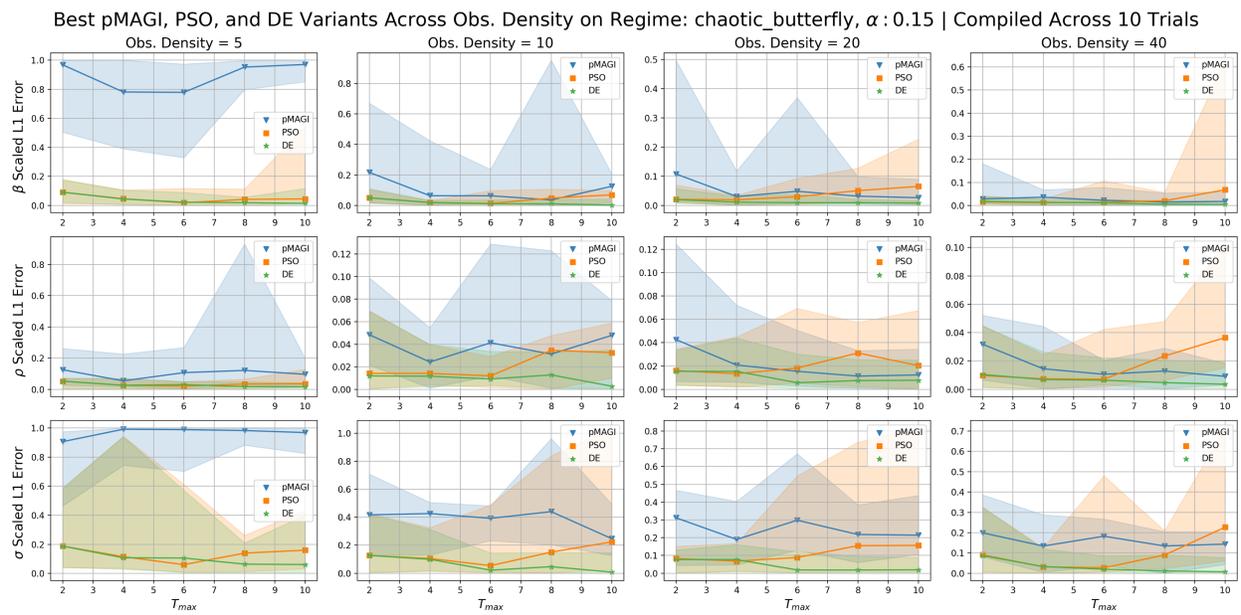

**Figure A.15:** pMAGI vs. PSO and DE Parameter Inference on the Chaotic (Butterfly) regime. The x-axis of each subplot is $T_{max}$, the length of our noisy observation interval $t \in [0, T_{max}]$. Blue represents pMAGI, orange represents PSO, and green represents DE. The solid lines indicate the average errors across all ten randomly-seeded trials. The colored confidence bands represent the maximum and minimum errors accrued on the given setting across ten randomly-seeded trials. **Lower errors indicate better performance.**

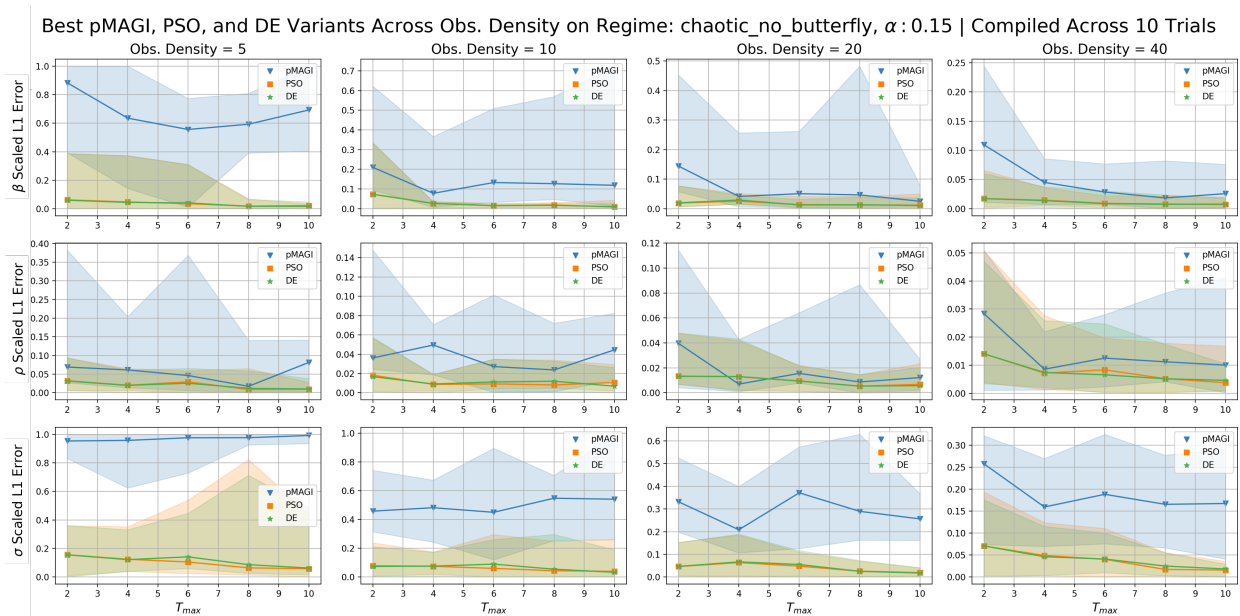

**Figure A.16:** pMAGI vs. PSO and DE Parameter Inference on the Chaotic (No Butterfly) regime. The x-axis of each subplot is $T_{max}$, the length of our noisy observation interval $t \in [0, T_{max}]$. Blue represents pMAGI, orange represents PSO, and green represents DE. The solid lines indicate the average errors across all ten randomly-seeded trials. The colored confidence bands represent the maximum and minimum errors accrued on the given setting across ten randomly-seeded trials. **Lower errors indicate better performance.**

## A.5 Additional Figures for pMAGI vs. PINN Parameter Inference (All Regimes)

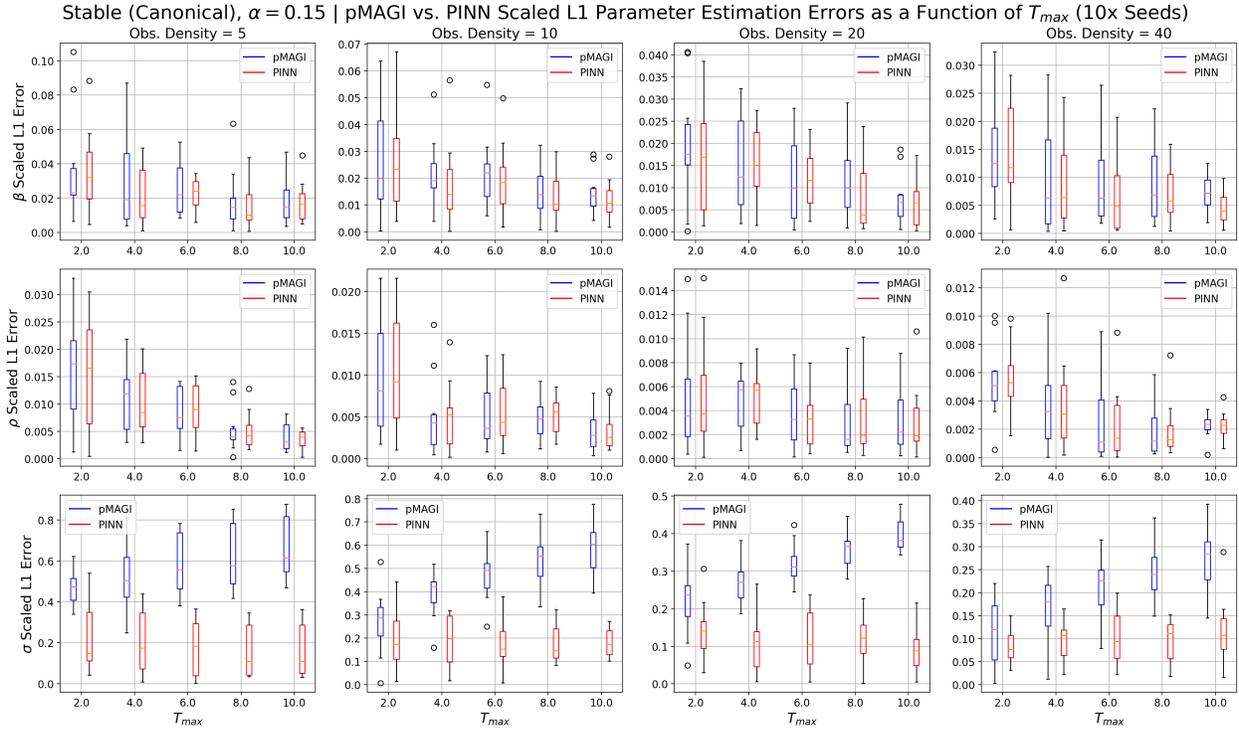

**Figure A.17:** pMAGI vs. PINN Parameter Inference on the Stable (Canonical) regime. Blue represents pMAGI, while red represents PINN. The x-axis of each subplot is $T_{max}$, the length of our noisy observation interval $t \in [0, T_{max}]$. **Lower error values indicate better performance.**

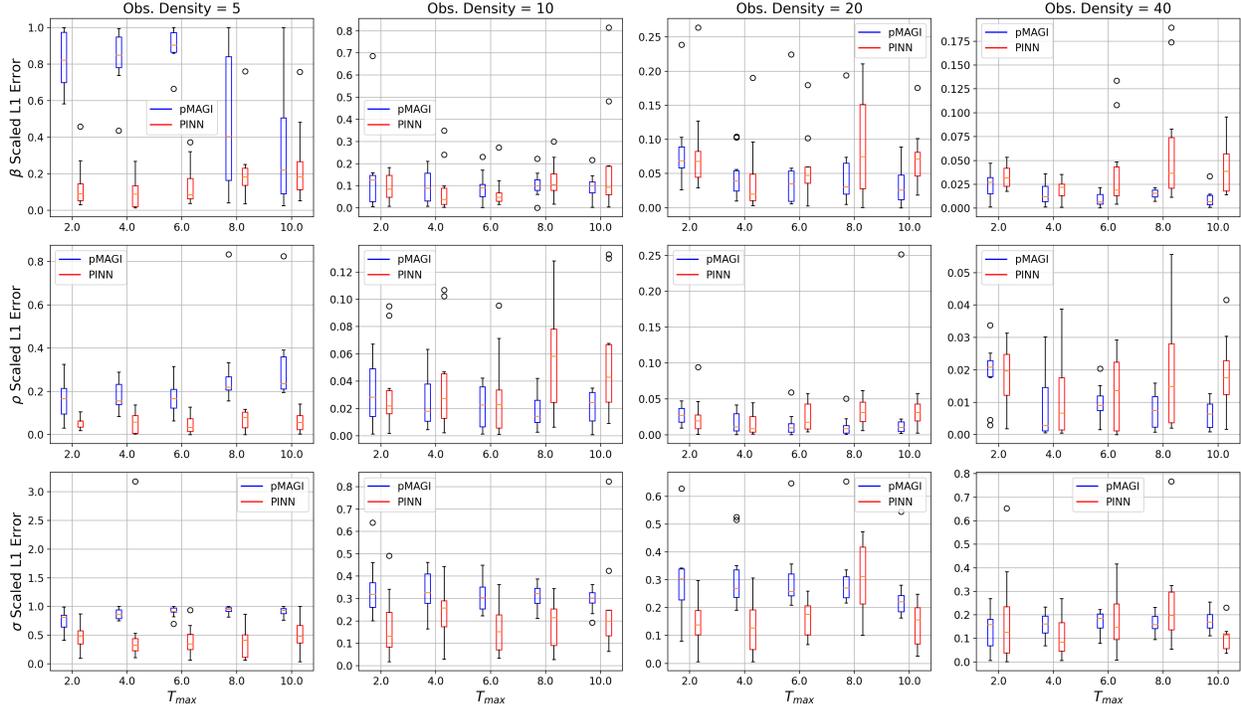

**Figure A.18:** pMAGI vs. PINN Parameter Inference on the Stable (Transient Chaos) regime. Blue represents pMAGI, while red represents PINN. The x-axis of each subplot is $T_{max}$, the length of our noisy observation interval $t \in [0, T_{max}]$. **Lower error values indicate better performance.**

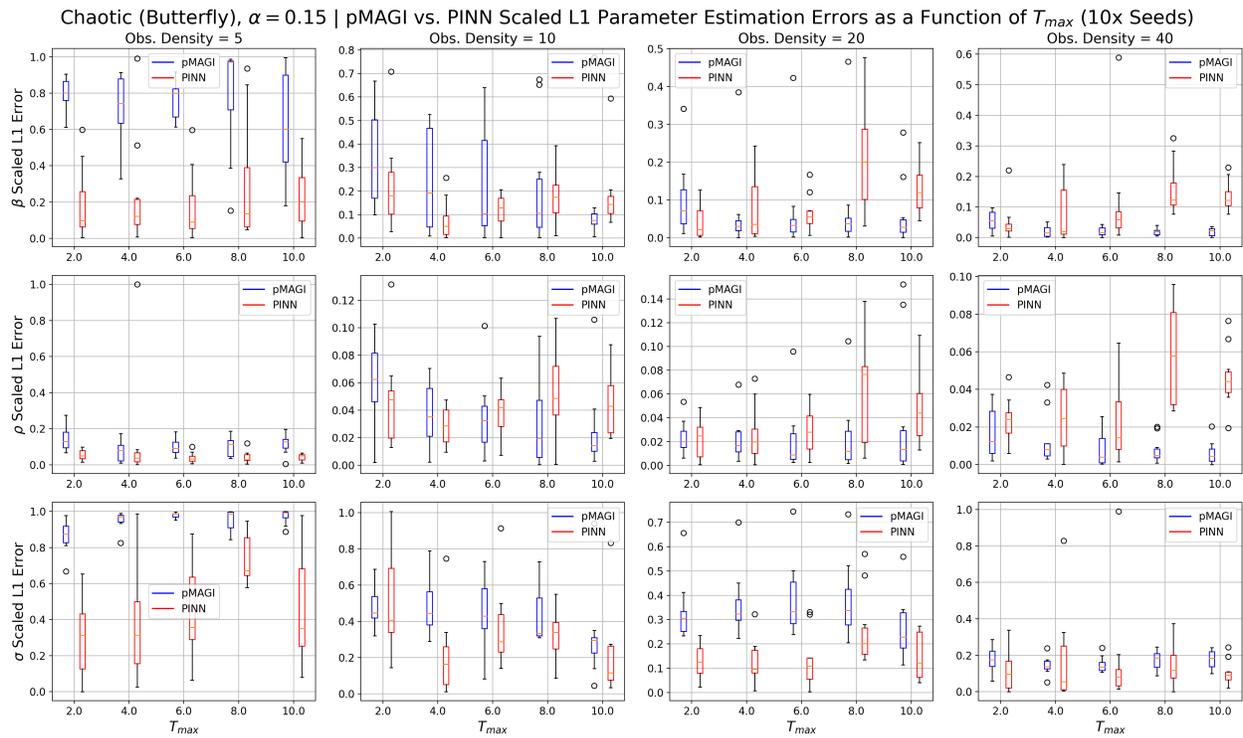

Figure A.19: pMAGI vs. PINN Parameter Inference on the Chaotic (Butterfly) regime. Blue represents pMAGI, while red represents PINN. The x-axis of each subplot is $T_{max}$, the length of our noisy observation interval $t \in [0, T_{max}]$. **Lower error values indicate better performance.**

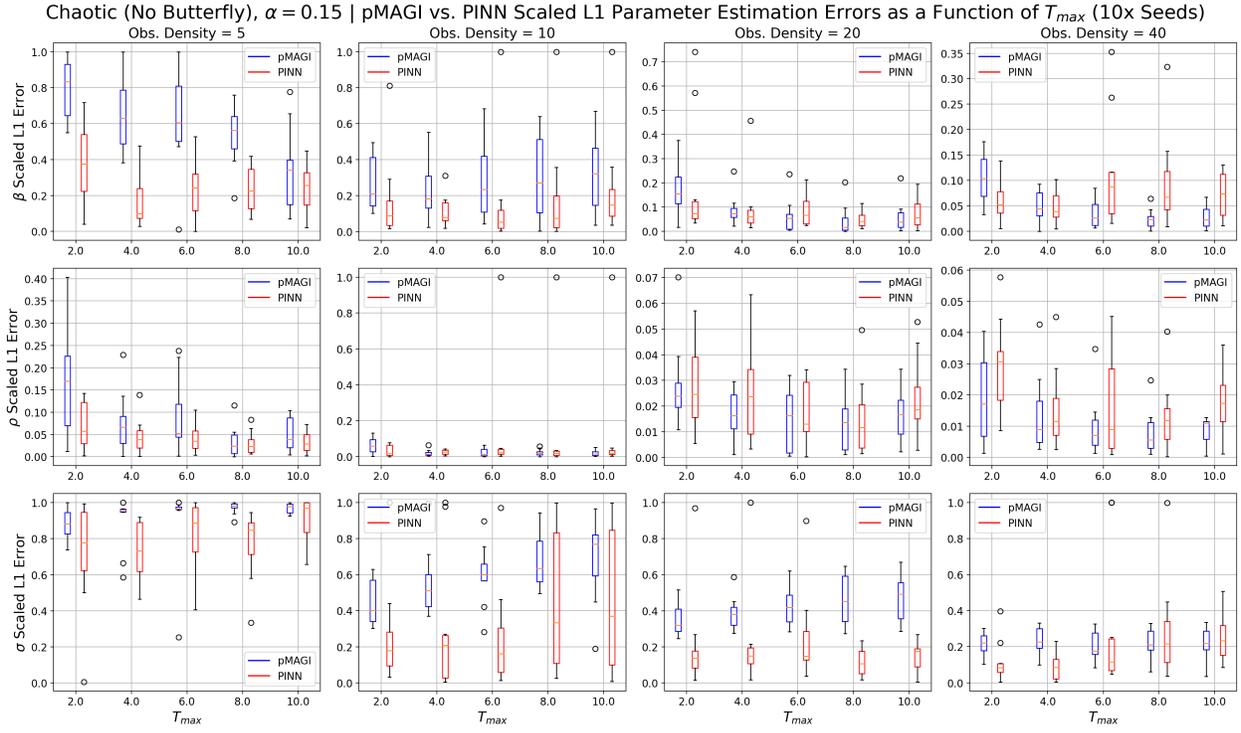

**Figure A.20:** pMAGI vs. PINN Parameter Inference on the Chaotic (No Butterfly) regime. Blue represents pMAGI, while red represents PINN. The x-axis of each subplot is $T_{max}$, the length of our noisy observation interval $t \in [0, T_{max}]$. **Lower error values indicate better performance.**

## A.6 Additional Figures for Lorenz System Identifiability

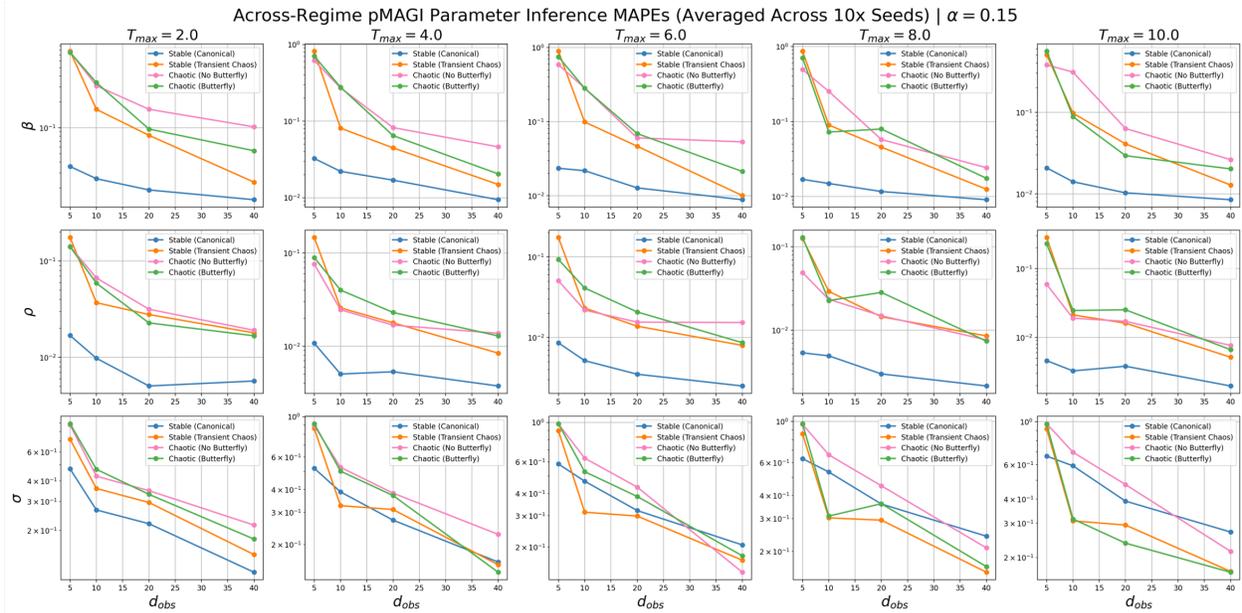

**Figure A.21:** Lorenz parameter identifiability at $\alpha = 0.15$ as quantified with MAPE. The x-axis of each subplot shows the observation density $d_{obs}$. The y-axis of each subplot, rendered on the log-scale for readability, shows the MAPE on the given parameter. The solid lines correspond to the mean errors averaged across ten randomly-seeded datasets. The error curves corresponding to the Stable (Canonical) regime are shown in blue, the Stable (Transient Chaos) in orange, the Chaotic (No Butterfly) in pink, and the Chaotic (Butterfly) in green. **Lower error values indicate better performance, which imply stronger identifiability.**

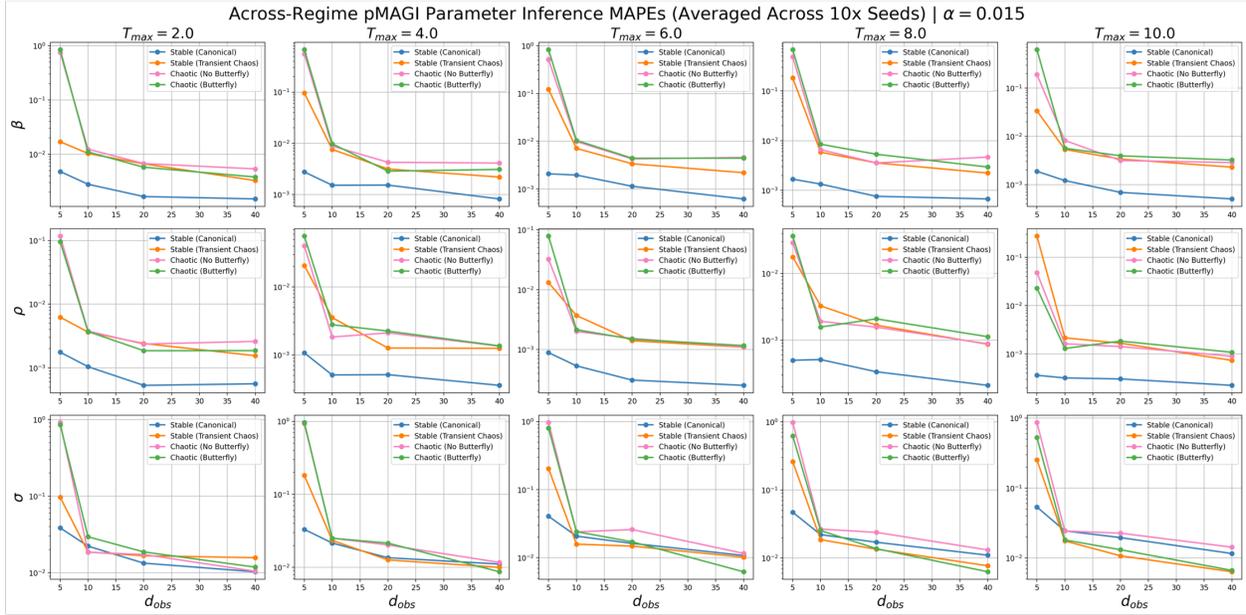

**Figure A.22:** Lorenz parameter identifiability at $\alpha = 0.015$ as quantified with MAPE. The x-axis of each subplot shows the observation density $d_{obs}$. The y-axis of each subplot, rendered on the log-scale for readability, shows the MAPE on the given parameter. The solid lines correspond to the mean errors averaged across ten randomly-seeded datasets. The error curves corresponding to the Stable (Canonical) regime are shown in blue, the Stable (Transient Chaos) in orange, the Chaotic (No Butterfly) in pink, and the Chaotic (Butterfly) in green. **Lower error values indicate better performance, which imply stronger identifiability.**

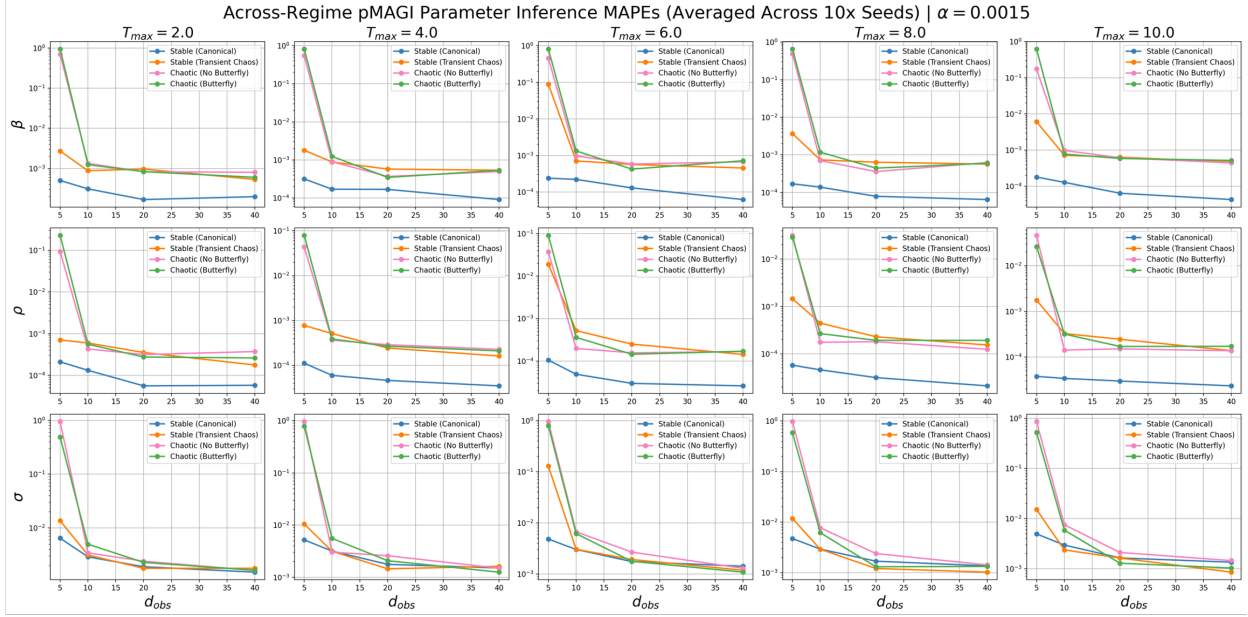

**Figure A.23:** Lorenz parameter identifiability at $\alpha = 1.5 \times 10^{-3}$ as quantified with MAPE. The x-axis of each subplot shows the observation density $d_{obs}$. The y-axis of each subplot, rendered on the log-scale for readability, shows the MAPE on the given parameter. The solid lines correspond to the mean errors averaged across ten randomly-seeded datasets. The error curves corresponding to the Stable (Canonical) regime are shown in blue, the Stable (Transient Chaos) in orange, the Chaotic (No Butterfly) in pink, and the Chaotic (Butterfly) in green. **Lower error values indicate better performance, which imply stronger identifiability.**

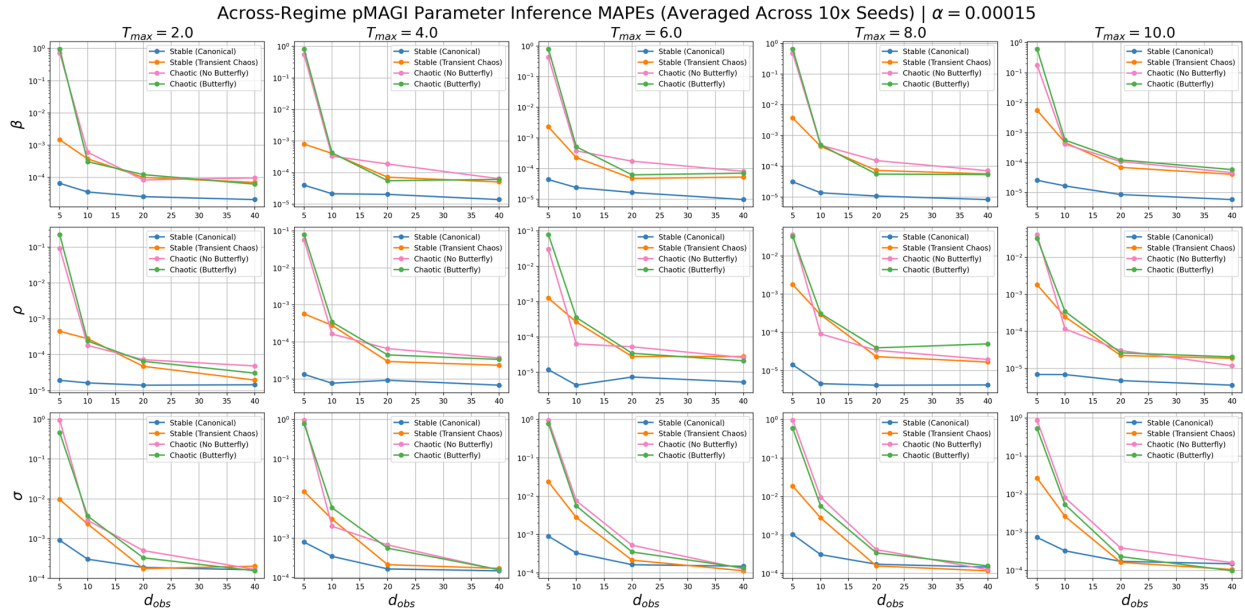

**Figure A.24:** Lorenz parameter identifiability at $\alpha = 1.5 \times 10^{-4}$ as quantified with MAPE. The x-axis of each subplot shows the observation density $d_{obs}$. The y-axis of each subplot, rendered on the log-scale for readability, shows the MAPE on the given parameter. The solid lines correspond to the mean errors averaged across ten randomly-seeded datasets. The error curves corresponding to the Stable (Canonical) regime are shown in blue, the Stable (Transient Chaos) in orange, the Chaotic (No Butterfly) in pink, and the Chaotic (Butterfly) in green. **Lower error values indicate better performance, which imply stronger identifiability.**

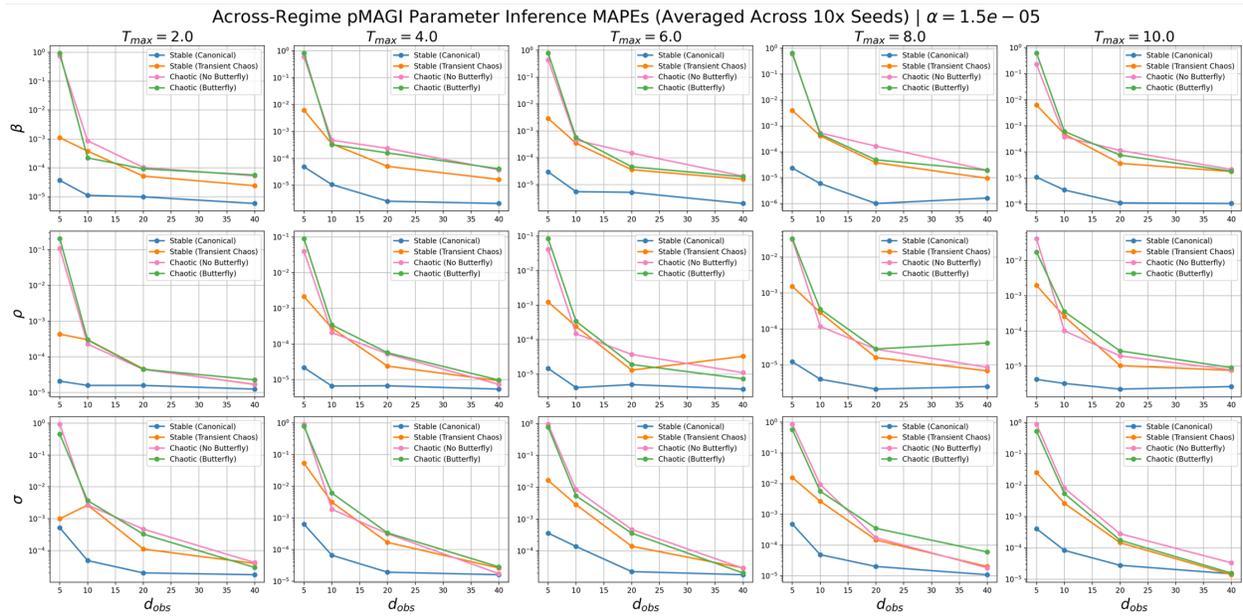

**Figure A.25:** Lorenz parameter identifiability at $\alpha = 1.5 \times 10^{-5}$ as quantified with MAPE. The x-axis of each subplot shows the observation density $d_{obs}$. The y-axis of each subplot, rendered on the log-scale for readability, shows the MAPE on the given parameter. The solid lines correspond to the mean errors averaged across ten randomly-seeded datasets. The error curves corresponding to the Stable (Canonical) regime are shown in blue, the Stable (Transient Chaos) in orange, the Chaotic (No Butterfly) in pink, and the Chaotic (Butterfly) in green. **Lower error values indicate better performance, which imply stronger identifiability.**

## A.7 Additional Figures for Probabilistic Binary Classification

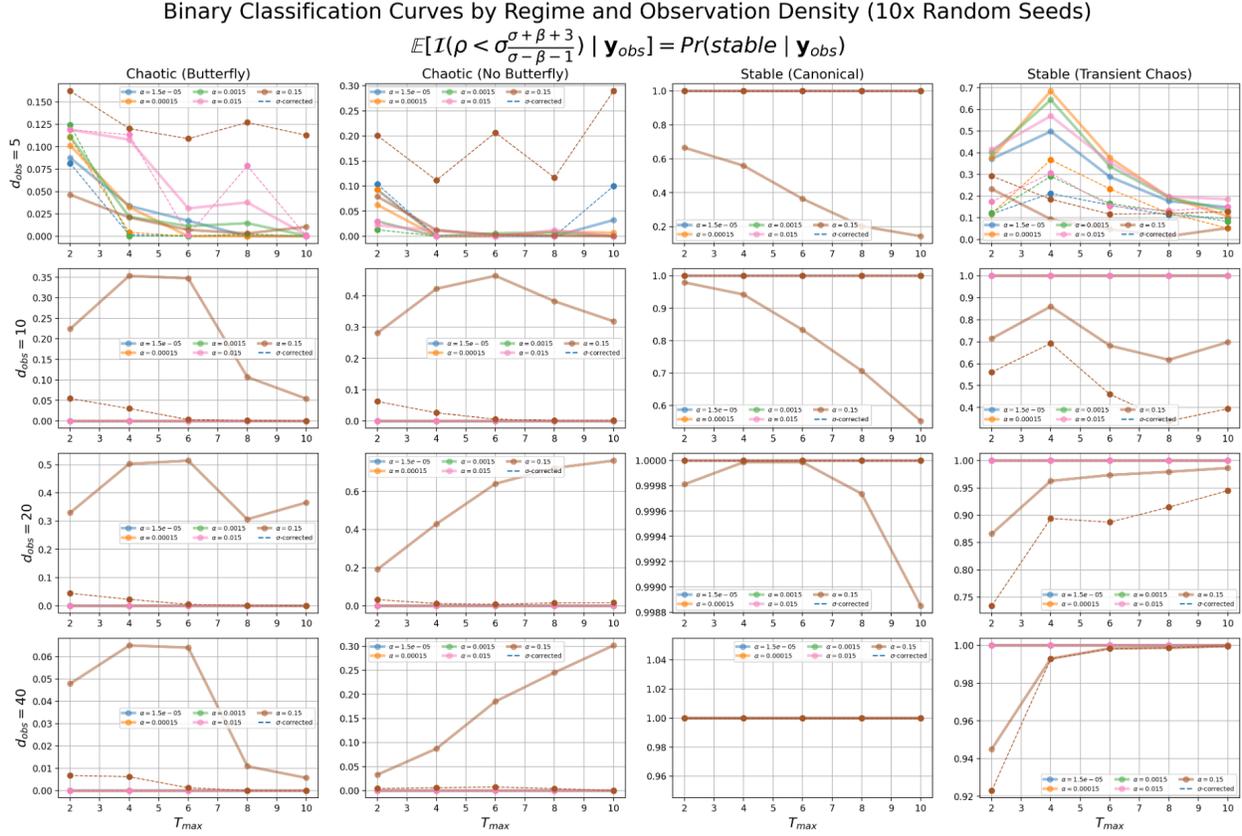

**Figure A.26:** $\sigma$-corrected pMAGI mean estimates of stability probability by regime (column) and $d_{obs}$ (row). The x-axis is the $T_{max}$, the length of our noisy observation interval $t \in [0, T_{max}]$. The y-axis is the mean estimated stability probability averaged across ten randomly-seeded trials. To reduce notational clutter, $(\mathcal{X}_{obs}, \tau_{obs})$ was abbreviated to $\mathbf{y}_{obs}$. $\alpha = 0.15$ corresponds to purple, $\alpha = 0.015$ to red, $\alpha = 1.5 \times 10^{-3}$ to green, $\alpha = 1.5 \times 10^{-4}$ to orange, and $\alpha = 1.5 \times 10^{-5}$ to blue. The dotted curves corresponding to each color represent the $\sigma$-corrected estimated stability probabilities (i.e., manually-encoding $\sigma = 10$ on each sample), while the solid curves represent the uncorrected estimated stability probabilities.

## A.8 Additional Figures for PMSP vs. PINN Prediction (All Regimes)

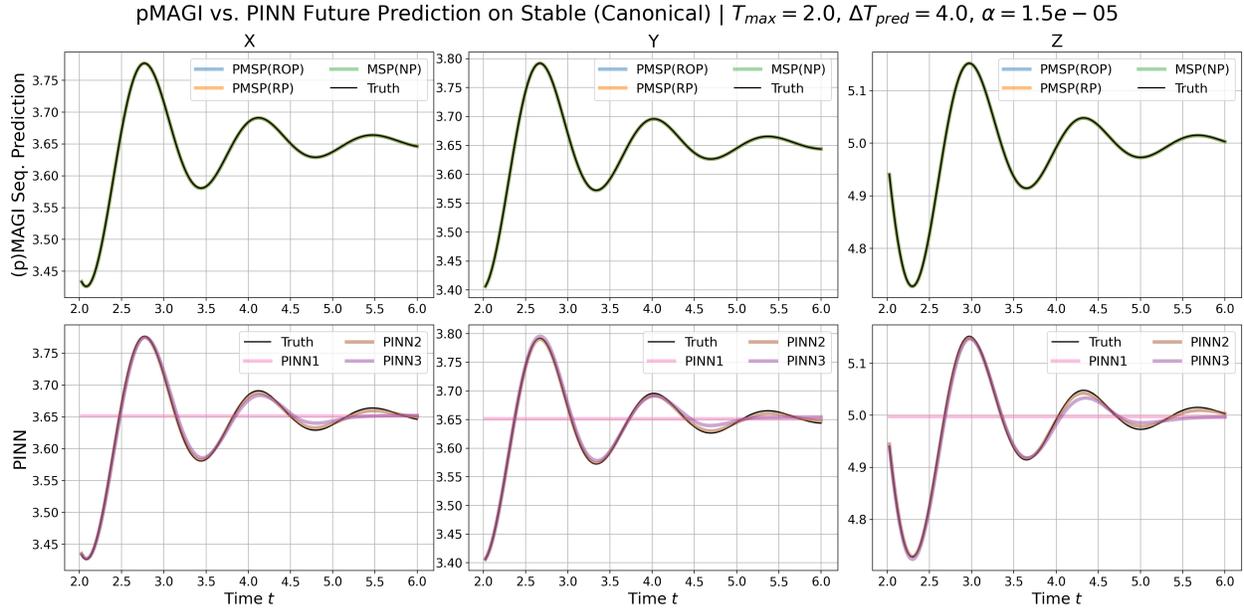

**Figure A.27:** Selected Pilot MAGI Sequential Prediction (PMSP) vs. PINN future prediction trajectories on the Stable (Canonical) regime at $\alpha = 1.5 \times 10^{-5}$, $T_{max} = 2.0$, and $\Delta T_{fc} = 4.0$. PMSP model variants are shown in the first row, while PINN models are shown in the second row, with all prediction curves color-coded by model variant. The x-axis for each subplot is time $t$ on the prediction interval $t \in [2.0, 2.0 + \Delta T_{\text{pred}}]$. Visually, we can estimate model performance by examining how closely the predicted trajectories match the ground truth (in black).

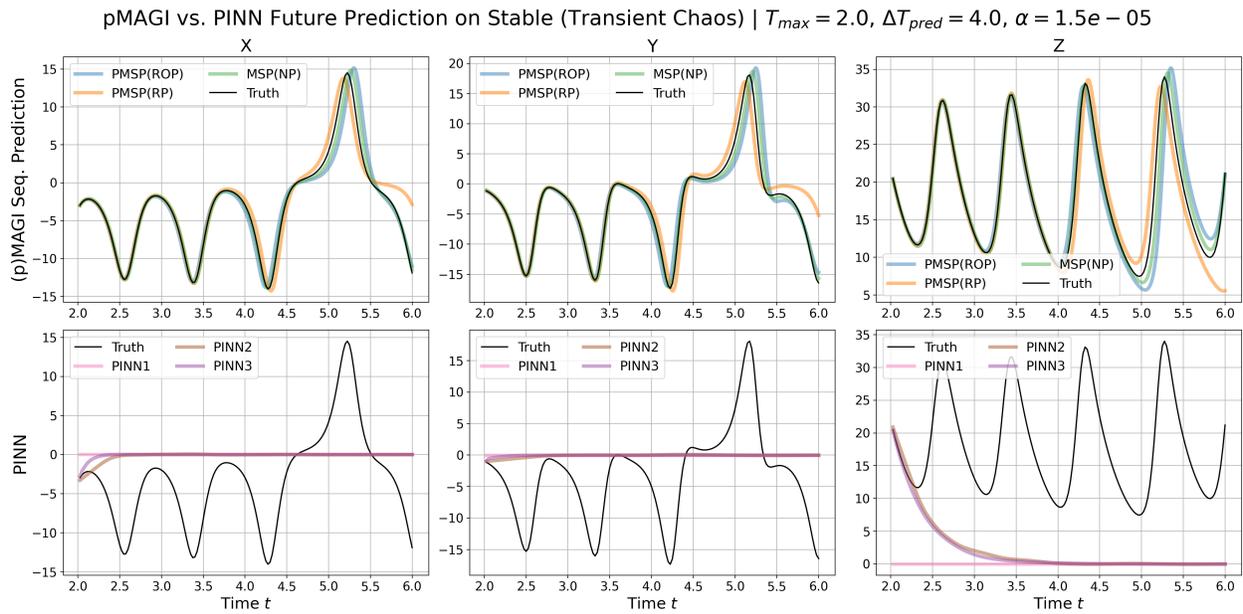

**Figure A.28:** Selected Pilot MAGI Sequential Prediction (PMSP) vs. PINN future prediction trajectories on the Stable (Transient Chaos) regime at $\alpha = 1.5 \times 10^{-5}$, $T_{max} = 2.0$, and $\Delta T_{fc} = 4.0$. PMSP model variants are shown in the first row, while PINN models are shown in the second row, with all prediction curves color-coded by model variant. The x-axis for each subplot is time $t$ on the prediction interval $t \in [2.0, 2.0 + \Delta T_{\text{pred}}]$. Visually, we can estimate model performance by examining how closely the predicted trajectories match the ground truth (in black).

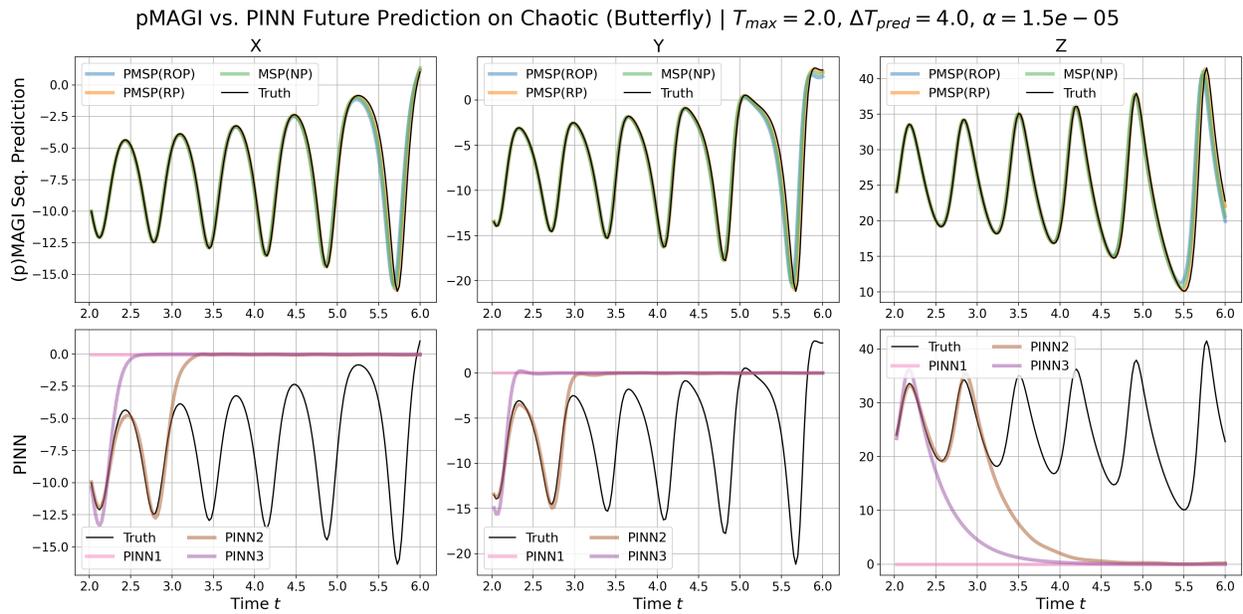

**Figure A.29**: Selected Pilot MAGI Sequential Prediction (PMSP) vs. PINN future prediction trajectories on the Chaotic (Butterfly) regime at $\alpha = 1.5 \times 10^{-5}$, $T_{max} = 2.0$, and $\Delta T_{fc} = 4.0$. PMSP model variants are shown in the first row, while PINN models are shown in the second row, with all prediction curves color-coded by model variant. The x-axis for each subplot is time $t$ on the prediction interval $t \in [2.0, 2.0 + \Delta T_{\text{pred}}]$. Visually, we can estimate model performance by examining how closely the predicted trajectories match the ground truth (in black).

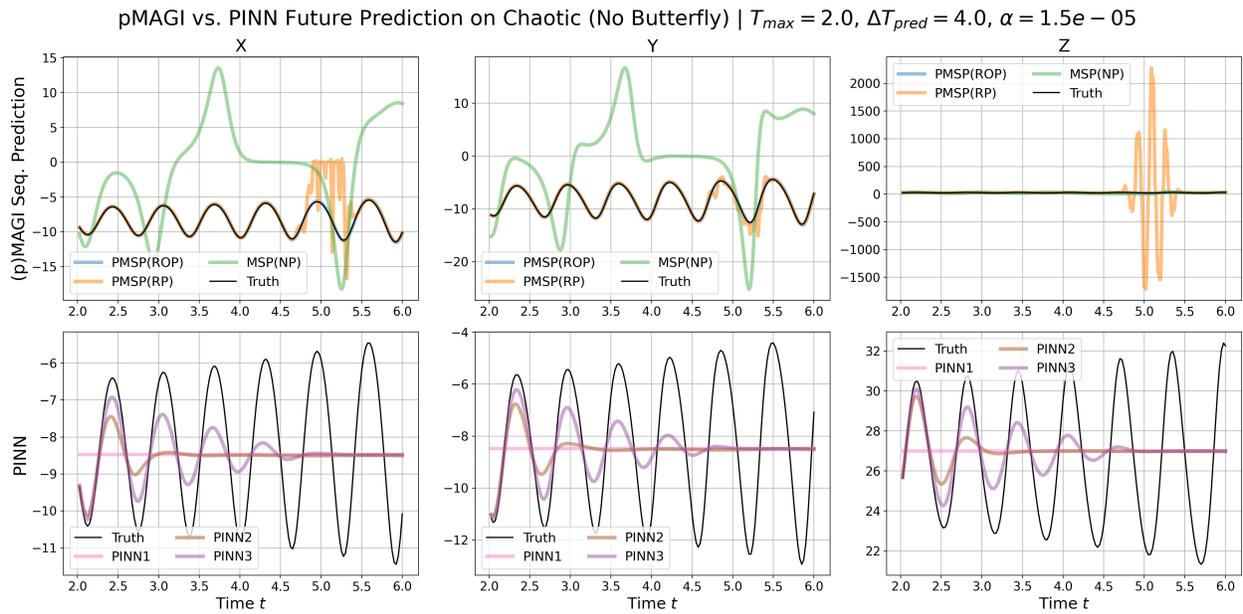

**Figure A.30:** Selected Pilot MAGI Sequential Prediction (PMSP) vs. PINN future prediction trajectories on the Chaotic (No Butterfly) regime at $\alpha = 1.5 \times 10^{-5}$, $T_{max} = 2.0$, and $\Delta T_{fc} = 4.0$. PMSP model variants are shown in the first row, while PINN models are shown in the second row, with all prediction curves color-coded by model variant. The x-axis for each subplot is time $t$ on the prediction interval $t \in [2.0, 2.0 + \Delta T_{\text{pred}}]$. Visually, we can estimate model performance by examining how closely the predicted trajectories match the ground truth (in black).

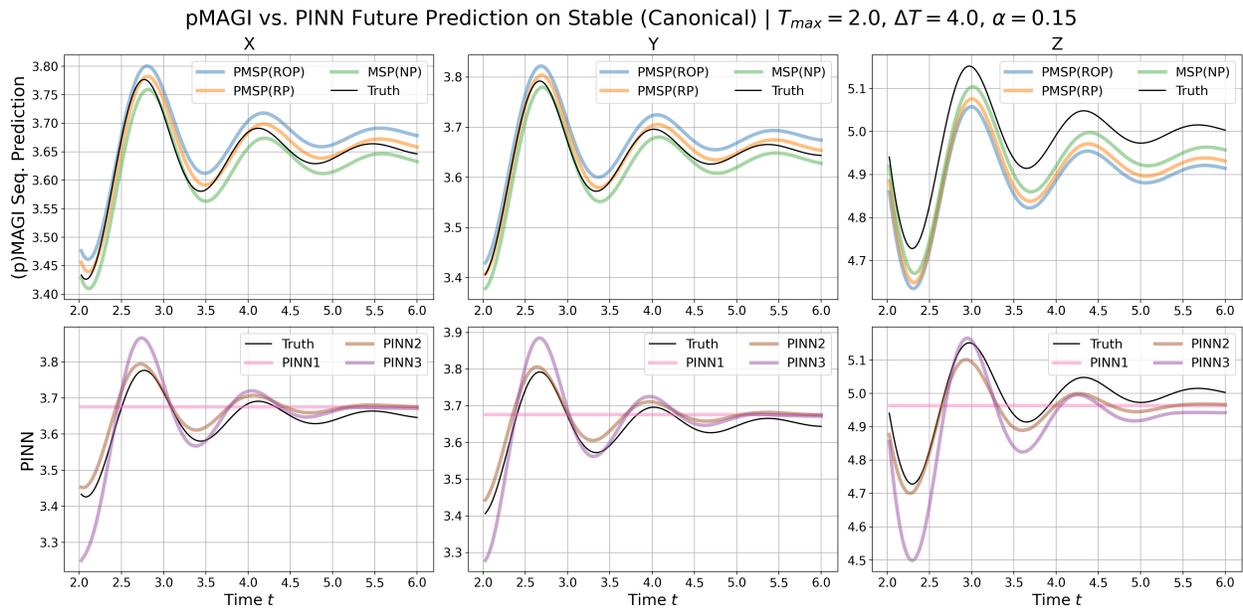

**Figure A.31:** Selected Pilot MAGI Sequential Prediction (PMSP) vs. PINN future prediction trajectories on the Stable (Canonical) regime at $\alpha = 0.15$, $T_{max} = 2.0$, and $\Delta T_{fc} = 4.0$. PMSP model variants are shown in the first row, while PINN models are shown in the second row, with all prediction curves color-coded by model variant. The x-axis for each subplot is time $t$ on the prediction interval $t \in [2.0, 2.0 + \Delta T_{\text{pred}}]$. Visually, we can estimate model performance by examining how closely the predicted trajectories match the ground truth (in black).

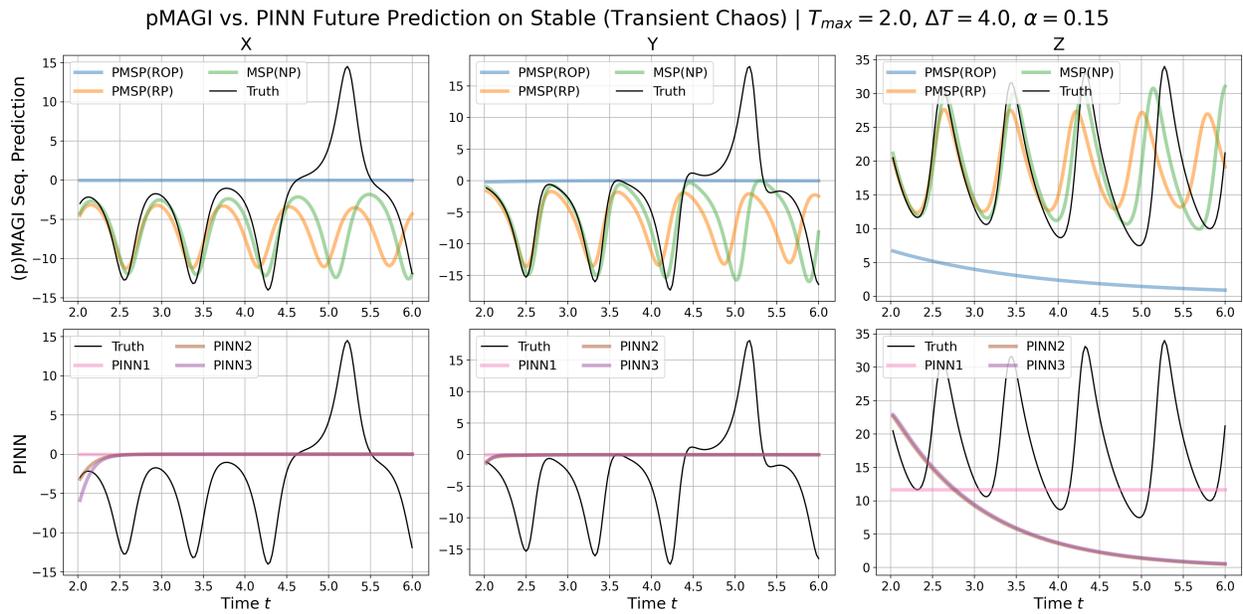

**Figure A.32:** Selected Pilot MAGI Sequential Prediction (PMSP) vs. PINN future prediction trajectories on the Stable (Transient Chaos) regime at $\alpha = 0.15$, $T_{max} = 2.0$, and $\Delta T_{fc} = 4.0$. PMSP model variants are shown in the first row, while PINN models are shown in the second row, with all prediction curves color-coded by model variant. The x-axis for each subplot is time $t$ on the prediction interval $t \in [2.0, 2.0 + \Delta T_{\text{pred}}]$. Visually, we can estimate model performance by examining how closely the predicted trajectories match the ground truth (in black).

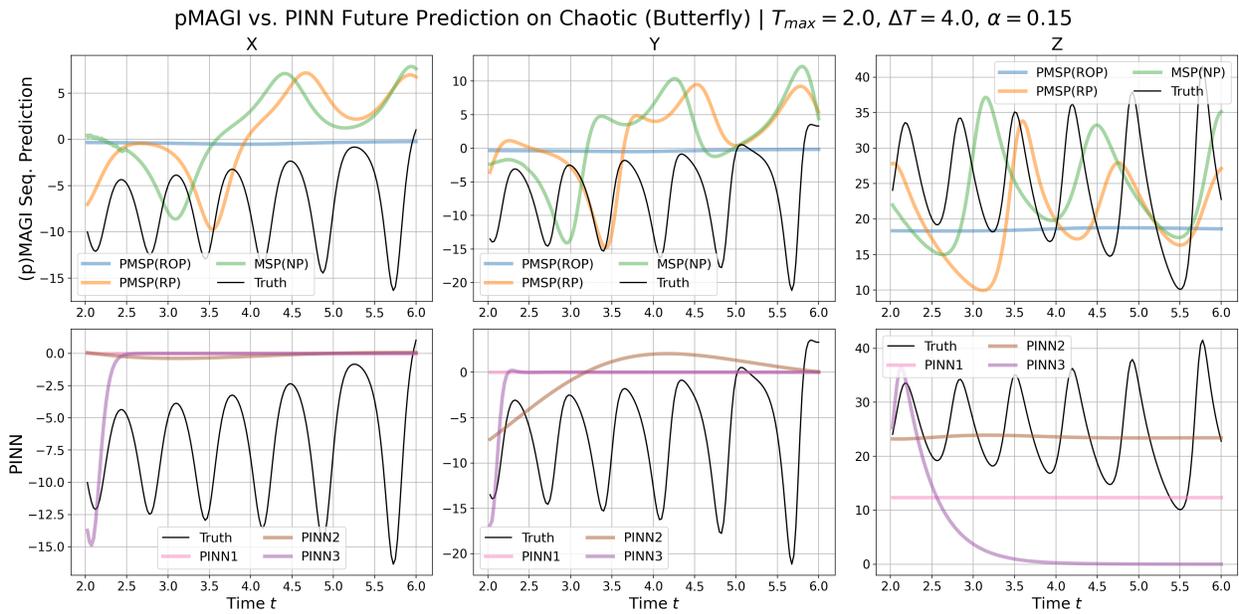

**Figure A.33:** Selected Pilot MAGI Sequential Prediction (PMSP) vs. PINN future prediction trajectories on the Chaotic (Butterfly) regime at $\alpha = 0.15$, $T_{max} = 2.0$, and $\Delta T_{fc} = 4.0$. PMSP model variants are shown in the first row, while PINN models are shown in the second row, with all prediction curves color-coded by model variant. The x-axis for each subplot is time $t$ on the prediction interval $t \in [2.0, 2.0 + \Delta T_{\text{pred}}]$. Visually, we can estimate model performance by examining how closely the predicted trajectories match the ground truth (in black).

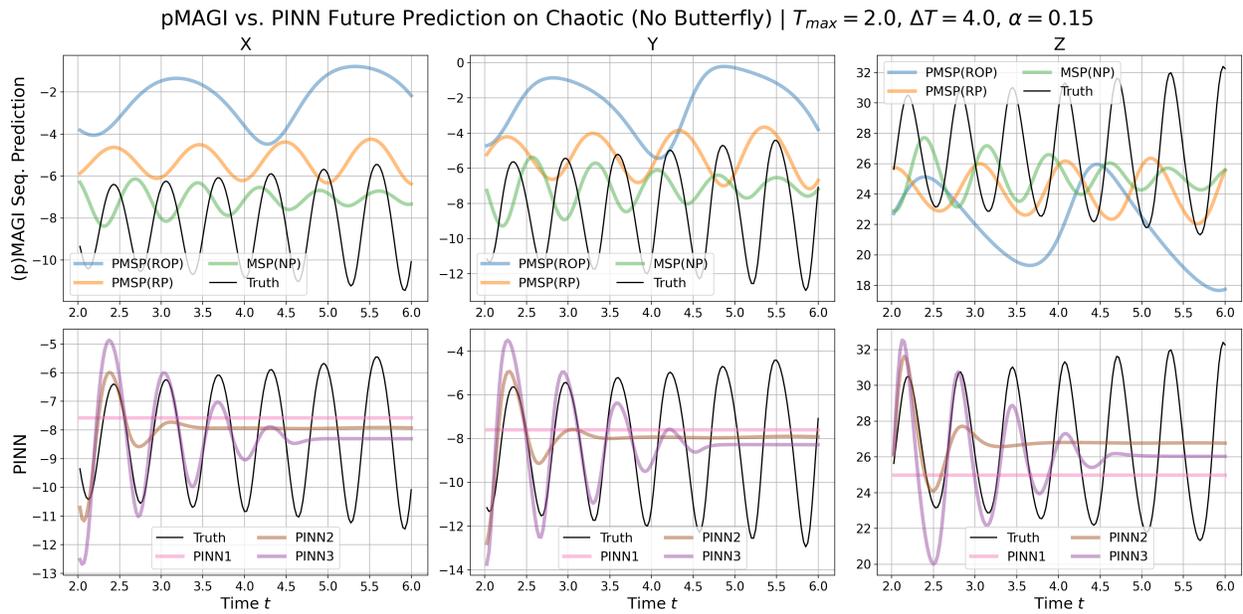

**Figure A.34:** Selected Pilot MAGI Sequential Prediction (PMSP) vs. PINN future prediction trajectories on the Chaotic (No Butterfly) regime at $\alpha = 0.15$, $T_{max} = 2.0$, and $\Delta T_{fc} = 4.0$. PMSP model variants are shown in the first row, while PINN models are shown in the second row, with all prediction curves color-coded by model variant. The x-axis for each subplot is time $t$ on the prediction interval $t \in [2.0, 2.0 + \Delta T_{\text{pred}}]$. Visually, we can estimate model performance by examining how closely the predicted trajectories match the ground truth (in black).